\documentclass[12pt,preprint]{aastex}
\usepackage{pdflscape}
\usepackage{multirow}
\usepackage{booktabs}
\usepackage{graphicx}
\usepackage{amsmath}
\usepackage{longtable}
\usepackage{enumerate}
\usepackage{hyperref}
\usepackage{scrextend}
\usepackage{tablefootnote}
\usepackage{color}

\newcommand{\tdust}{T_{\rm dust}}
\newcommand{\mdisk}{M_{\rm disk}}

\newcommand{\mdust}{M_{\rm dust}}
\newcommand{\mgas}{M_{\rm gas}}

\newcommand{\mdot}{\dot{\mstar}}

\newcommand{\Msun}{M_\odot}
\newcommand{\lacc}{L_{\rm acc}}

\newcommand{\mj}{M_{\rm J}}

\newcommand{\msun}{M_\odot}

\newcommand{\mplanet}{M_{\rm p}}
\newcommand{\mstar}{M_\star}

\def\MJ{\ifmmode {M_J}\else $M_J$\fi}
\def\Mstar{\ifmmode {M_\star}\else $M_\star$\fi}
\def\Msun{\ifmmode {M_\odot}\else $M_\odot$\fi}
\def\Mdust{\ifmmode {M_{\rm dust}}\else $M_{\rm dust}$\fi}
\def\Mdisk{\ifmmode {M_{\rm disk}}\else $M_{\rm disk}$\fi}
\def\Mdiskd{\ifmmode {M_{\rm disk,d}}\else $M_{\rm disk,d}$\fi}
\def\Mdiska{\ifmmode {M_{\rm disk,acc}}\else $M_{\rm disk,acc}$\fi}
\def\MEarth{\ifmmode {M_\oplus}\else $M_\oplus$\fi}
\def\Msunpery{\ifmmode {M_\odot\,{\rm yr}^{-1}}\else $M_\odot\,{\rm yr}^{-1}$\fi}

\begin{document}
\title{Spiral Arms in Disks: Planets or Gravitational Instability?} 
\author{Ruobing Dong\altaffilmark{1}, Joan R. Najita\altaffilmark{2} \& Sean Brittain\altaffilmark{3,2}}
\altaffiltext{1}{Steward Observatory, University of Arizona, 933 N Cherry Ave, Tucson, AZ 85721, USA, rbdong@gmail.com}
\altaffiltext{2}{National Optical Astronomical Observatory, 950 North Cherry Avenue, Tucson, AZ 85719, najita@noao.edu}
\altaffiltext{3}{Department of Physics \& Astronomy, 118 Kinard Laboratory, Clemson University, Clemson, SC 29634-0978, USA, sbritt@clemson.edu}

\clearpage

\begin{abstract}
Spiral arm structures seen in scattered light observations of
protoplanetary disks can potentially serve as signposts of
planetary companions. They can also lend unique insights
into disk masses, which are critical in setting the mass
budget for planet formation but are difficult to determine
directly.
A surprisingly high fraction of disks that have been 
well-studied in scattered light have spiral arms of some kind 
(8/29), as do a high fraction (6/11) of well-studied 
Herbig intermediate mass stars (i.e., Herbig stars $> 1.5\Msun$).
Here we explore the origin of spiral arms in Herbig systems 
by studying their occurrence rates, disk properties, and 
stellar accretion rates. 
We find that two-arm spirals are more common in disks surrounding  
Herbig intermediate mass stars than are 
directly imaged
giant planet companions to mature A and B stars.  
If two-arm spirals are produced by such giant planets, this discrepancy 
suggests that giant planets are much fainter than predicted by hot start models.
In addition, the high stellar accretion rates of Herbig stars, if
sustained over a reasonable fraction of their lifetimes,
suggest that disk masses are much larger than inferred 
from their submillimeter continuum emission.
As a result, gravitational instability is a possible 
explanation for multi-arm spirals. 
Future observations can lend insights into the issues raised
here.
\end{abstract}



\section{Introduction}\label{sec:intro}

As the birthplaces of planets, protoplanetary disks surrounding young stars lend unique insights into the circumstances under which planets form. The initial masses of disks set an upper limit on the mass budget for planet formation, a basic constraint on theories of how planets form.  From the detailed study of disk morphologies, we may also glean evidence of the formation of planets themselves, e.g., from the detection of gaps, rings, and spiral arms created by planets. As a result, numerous studies have examined the protoplanetary disks around T Tauri stars and Herbig stars for clues to their structure and planet formation status. 

Despite their fundamental importance, disk masses have been proven 
difficult to measure. 
Although disk masses are commonly inferred from submillimeter continuum 
fluxes, which probe the dust content of disks,
several arguments strongly suggest that much of the solids in 
T Tauri disks have grown beyond the sizes probed by submillimeter 
continua. 
As one simple example, submillimeter 
continuum disk masses are too low to even account for the
solids that are known to be locked up in exoplanet populations and
debris disks, indicating that T Tauri disks harbor more massive
reservoirs of solids than those probed by continuum measurements 
\citep{najita14},
and disk masses are underestimated as a result. 

More graphically, 
high resolution submillimeter continuum images of disks made 
with ALMA commonly show stunning rings and gaps \citep{brogan15, andrews16, cieza16, isella16hd163296, cieza17, loomis17, fedele18, huang18, dipierro18}, 
structures that signal that disk solids have 
grown beyond centimeter sizes and possibly into planetary mass objects 
\citep[e.g.,][]{dong15gap, dipierro15hltau, dong17doublegap, bae17}.
These results exclaim that disks are highly evolved and 
call into question disk mass estimates made under the usual 
assumption of simple, unevolved, optically thin disks. 
Given the difficulty of measuring disk masses through proxies such 
as dust continuum emission, other approaches are welcome. 

Alongside these developments, high contrast images of disks 
made in scattered light also reveal remarkable features. 
Near-infrared scattered light imaging observations 
probe the spatial distribution of small grains,  
approximately micron-sized or smaller,
that are well-coupled to the gas \citep{zhu12, pinilla12dusttrapping}. 
These studies 
have revealed spiral arms in a growing number of disks (Figure~\ref{fig:spiralarmexamples}), 
both prominent
near-symmetric two-arm spirals \citep[e.g.,][]{grady13,
benisty15} and more flocculent multi-arm spirals on smaller
scales and at lower contrast \citep[e.g.,][]{fukagawa04,
hashimoto11}.
Spiral arms can be produced by either 
gravitational instability \citep[e.g.,][]{rice03, lodato04, stamatellos08, 
kratter10gidisks, kratter16} or the presence of massive companions
\citep[gas giant to stellar masses; e.g.,][]{kley12, zhu15densitywaves, munoz16}.

The presence or absence of spiral arms in disks, if induced by 
gravitational instability, can place a dynamical constraint 
on disk masses. Gravitational instability occurs when the 
destabilizing self-gravity of a disk dominates over the restoring
forces of gas pressure and differential rotation 
\citep{toomre64, goldreich65}, a requirement that roughly translates into
$\mdisk/\mstar \gtrsim 0.1$ under typical conditions \citep{kratter16}.
Spiral arms induced by gravitational instability 
are expected to appear nearly symmetric in scattered light imaging,
with the number of arms roughly scaling as $n \approx\mstar/\mdisk$
\citep{lodato04, lodato05, cossins09, dong15giarm}. 

Planetary companions can also drive spiral
arms \citep{goldreich79, ogilvie02}. A massive companion 
(above a few $\times$ $10^{-3}\Mstar$) generates a set of
near-symmetric two-arm spirals interior to its orbit, 
similar to the morphology of observed two-arm spirals 
\citep[][see also Fig.\ 2a,b]{dong15spiralarm}, 
with the companion located at the tip of the primary arm
\citep[e.g.,][]{zhu15densitywaves, fung15, bae16sao206462}.
Lower mass companions ($\sim10^{-3}\Mstar$)
may generate additional arms interior to their orbits, weaker than
the primary and secondary arms in scattered 
light \citep[e.g.,][see also Fig.\ 2d,e]{juhasz15,
fung15, lee16, bae17, bae18theory}. 
If spiral arms are driven by planets, arm morphologies can 
provide clues to the location and masses of the planets. 

Here we explore these hypotheses for the origin of the spiral arms 
detected in scattered light by comparing
({\it i}) the demographics of arm-bearing disks (disk masses and stellar
accretion rates) with those of other young stars and ({\it ii}) the
incidence rate of arms compared to that of resolved planetary
companions to mature stars. Although we begin by considering the
detection rates of spiral arms in disks surrounding both T Tauri
stars and Herbig stars, we end up focusing more closely on the
properties of the higher mass Herbig stars, where the census of
scattered light structures is more complete. 

In \S\ref{sec:sample} we characterize the occurrence rate of 
spiral arm structures in scattered light imaging of disks. 
To explore whether spiral arms could be produced by orbiting 
companions, we compare in \S\ref{sec:planet} the occurrence rates 
of spiral arms in disks to that of giant planetary companions to 
mature stars. 
In \S\ref{sec:gi} we explore the alternative possibility that 
spiral arms are driven by gravitational instaibility by examining 
the masses of disks that show spiral arm structures, as probed by 
submillimeter continuum emission and stellar accretion rates. 
In \S\ref{sec:discussion} we examine the planet and gravitational 
instability hypotheses in greater detail before suggesting 
scenarios that are potentially consistent with our results. 
We summarize our results in \S\ref{sec:summary}, where we also 
describe future studies 
that can lend insights into the issues raised here.


\section{Sample Selection and Statistics}\label{sec:sample}

In near-infrared scattered light imaging of disks, spiral arms have
been found in both embedded Class 0/I objects surrounded by infalling
envelopes \citep{canovas15zcma, liu16fuori} and revealed Class II
objects (i.e., Herbig and T Tauri stars;
Figure~\ref{fig:spiralarmexamples}). While gravitational instability
may drive the arms in Class 0/I disks 
\citep[e.g.,][]{vorobyov05, vorobyov10burst, tobin16, perez16, dong16protostellar, tomida17, meru17}, which tend to be massive 
and are fed by infall from the envelope,
the origin of the arms detected in Class II disks is unclear.  

\begin{figure}
\begin{center}
\includegraphics[trim=0 0 0 0, clip,width=\textwidth,angle=0]{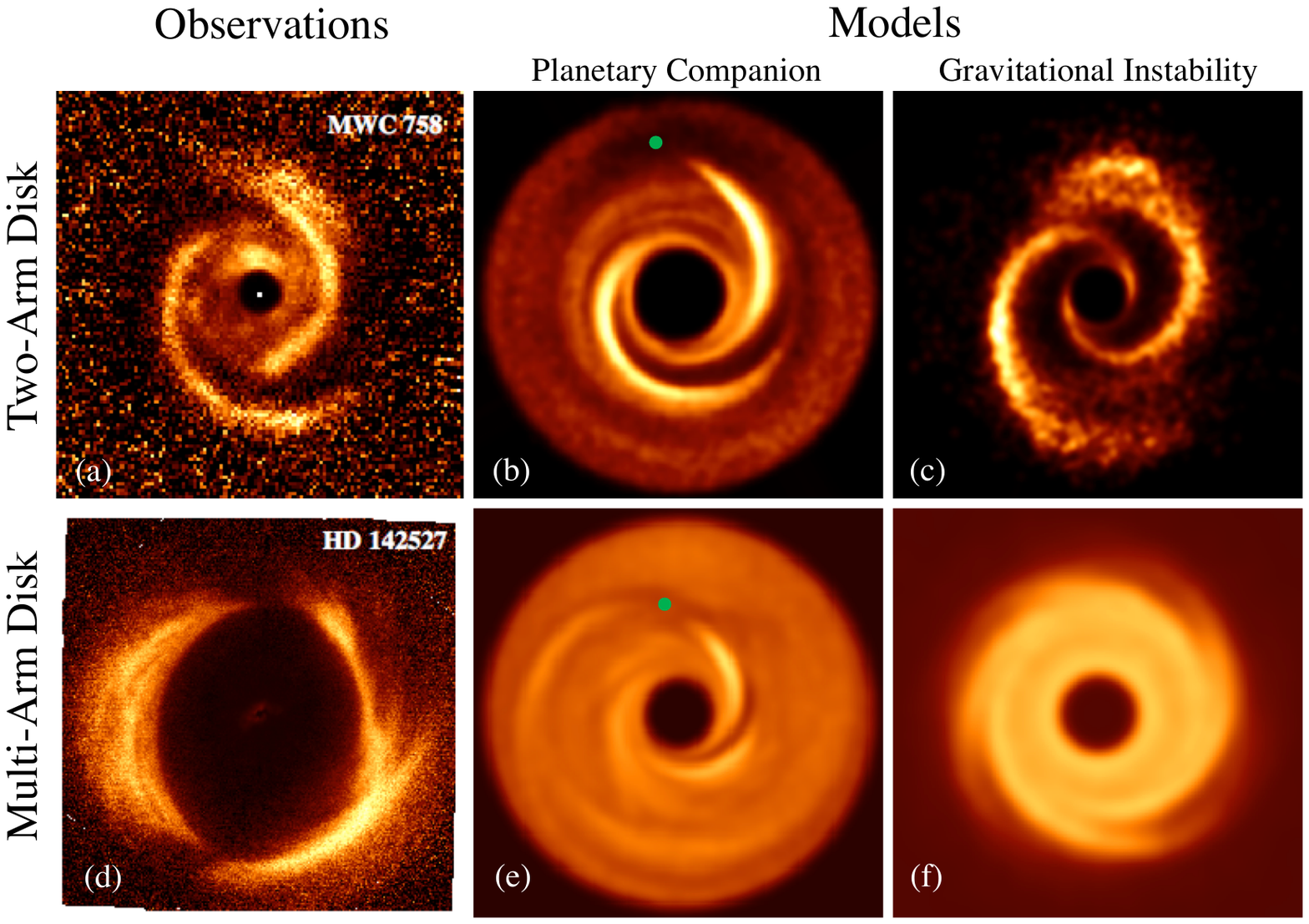}
\end{center}
\figcaption{Examples of observed and modeled spiral structure in disks. 
Observed two-arm and multi-arm
spirals (left column, Panels a and d) are compared with images from 
simulations of spiral arms created by orbiting companions (center column) with
$\mplanet/\mstar=0.006$ \citep[green dot; Panel b]{dong15spiralarm}
and $\mplanet/\mstar=0.001$ \citep[green dot; Panel
e]{dong17spiralarm} and gravitationally unstable disks (right column) 
with  $\mdisk/\mstar=0.5$
\citep[Panel c]{dong15giarm} and $\mdisk/\mstar=0.1$ \citep[Panel
f]{dong15giarm}.  All images have been scaled by $r^2$ to enhance
the faint outer disk, and the color stretch is linear in arbitrary
units.  
\label{fig:spiralarmmechanisms}} 
\end{figure}

Figure~\ref{fig:spiralarmmechanisms} shows how in principle spiral
arms could serve as signposts of either an orbiting planet or a
gravitationally unstable disk. The left column shows observed images
of two representative disks: MWC~758, with its prominent near-symmetric
two-arm spirals (Fig.~\ref{fig:spiralarmmechanisms}a); and HD~142527, with multiple small-scale, 
lower contrast arms (Fig.~\ref{fig:spiralarmmechanisms}d). The middle and right columns show synthetic 
observations from the literature, produced in combined hydrodynamic 
and radiative transfer simulations, which illustrate how 
morphologically similar spiral structures can occur in disks with orbiting 
planets (middle column) or when a disk is gravitational unstable
(right column). 
As discussed in
\S\ref{sec:intro}, both a multi-Jupiter mass giant companion and a
gravitationally unstable disk with $\mdisk \sim 0.5 M_\star$ can
produce a set of prominent two-arm spirals (Fig.~\ref{fig:spiralarmmechanisms}b,c), 
while lower mass planets ($\mplanet\sim\mj$) and marginally
gravitationally unstable disks 
with $\mdisk\sim 0.1 M_\star$ can produce multiple small-scale
arms at low contrast (Fig.~\ref{fig:spiralarmmechanisms}e, f). 

To attempt to distinguish between these two scenarios for the 
origin of spiral structure in disks, we first characterize, in 
this section, the occurrence rate of spiral arms in scattered light 
imaging of disks. \S\ref{sec:nirsample} describes the data available  
and our process of selecting a sample of disks 
well studied in scattered light that are
suitable for our study. 
The occurrence rate of arms within this sample is described 
in \S\ref{sec:spiralarmnir}. 
Because the sample of well-studied disks is small in size, 
in \S\ref{sec:herbig200pc} we examine the occurrence rate 
of arms within in a volume-limited sample of Herbig stars, 
which provides a lower limit on the true occurrence rate.

\subsection{Disk Imaging Sample}\label{sec:nirsample}

We collected from the literature a sample of protoplanetary disks
that have been spatially resolved in scattered light at optical to
NIR wavelengths (Appendix Table~\ref{tab:nirsample}.1) and that meet four
main criteria. First, the disk is gas-rich (i.e., debris disks 
are excluded). Second, the disk was observed with an 
inner working angle $< 1\arcsec$, small enough to reveal
the inner $\sim$100 AU planet forming region 
(most disks imaged to date are 100--200 pc away). 
Third, the disk surface was revealed with sufficient signal-to-noise 
to identify possible structures.
Finally, the disk was observed with an angular resolution 
$\lesssim 0\farcs1$ in order to detect and resolve spiral arms. 

The third condition largely
restricts the sample to Class II objects without a substantial
envelope, as the disk surface of more embedded sources 
is often not accessible in scattered light.
It also excludes disks
that are completely flat or shadowed \citep{dullemond04shadowing}.
The last criterion restricts the sample to those observed at 
$L$-band (3.8~$\micron$)
or shorter wavelengths, because $\lambda/D\lesssim 0\farcs1$ requires
$\lambda\lesssim3.8$~$\micron$ for observations made with $D\approx8$\,m
diameter telescopes. Systems that have been directly imaged, both
with and without a coronagraphic mask, are included, whereas
interferometric observations are not.
The resulting sample of 49 objects (Table~\ref{tab:nirsample}.1) derives mainly from a few major
exoplanet and disk surveys, such as the Subaru-based SEEDS
\citep{tamura09, uyama17} and the VLT-based SHINE surveys, supplemented by
studies of smaller samples that used a variety of ground-based instruments
and {\it HST}. Roughly half of the sources are transitional disks (i.e.,
disks whose inner region is optically thin in the continuum; 
\citealt{espaillat14}).

We further removed sources viewed at high inclination ($\gtrsim 70^\circ$) 
whose scattered light structures may be difficult to
assess (AK~Sco, T~Cha, AA~Tau, RY~Lup, HH~30, HK~Tau, HV~Tau~C, 
LkH$\alpha$~263C, and PDS~144N). 
It is possible to identify spiral arms in disks
at such high inclinations \citep[e.g., RY
Lup;][]{langlois18}; however it is usually a challenge because of 
the resulting geometric effect \citep{dong16armviewing}.
We also excluded any remaining Class 0/I 
sources and sources undergoing accretion outbursts (FU Ori, HL
Tau, R Mon, V 1057 Cyg, V1735 Cyg, and Z CMa) in order to probe a
restricted evolutionary state and to avoid scattered light contamination
from an infalling envelope.

Sources with known stellar companions within 5\arcsec\ 
(HD~100453, GG~Tau, HD~150193, and T~Tau) were also
excluded to avoid systems in which a stellar companion
induces disk structure. Whereas a massive planet exterior to the arms can
generate the observed two-arm spirals, a stellar companion closer
than $\sim 5\arcsec$ could produce similar spiral structure in
a disk with a typical size of $\sim$1$-$2\arcsec;
(e.g., as in the case of the HD~100453 disk; \citealt{wagner15hd100453, dong16hd100453, wagner18}).
The companion to HD~142527, which is located well within the gap 
in the disk, 
is commonly interpreted as having too low a mass \citep[$0.13 \Msun$;][]{lacour16} 
and eccentricity to generate the observed spiral arms, so this 
object is included in our sample (although see \citealt{price18}). 
Lastly, we removed HD~141569 from the sample, because its low dust
mass \citep[on the order of $1\,M_\oplus$ or less;][]{flaherty16,
white16} makes the system prone to radiation pressure and photoelectric
instability, which can also produce spiral arm structures
\citep[e.g.,][]{richert17}.

Finally, we arrive at a sample of 29 ``well-studied NIR disks'', 
whose stellar and disk properties are listed in
Table~\ref{tab:nirsamplegood}. 
The tabulated $M_\star$ values were obtained from $T_{\rm eff}$ 
and $L_\star$ using the \citet{siess00} pre-main-sequence evolutionary 
tracks. 
Notes on individual sources are provided in Appendix~\S\ref{sec:individual}.
The final sample contains 11 Herbig sources
(the first 11 rows in the table) and 18 T Tauri stars.  
Of the 29 sources, 8 have spiral arms: 
the MWC~758, SAO~206462, LkH$\alpha$~330, and DZ~Cha disks show 
prominent two-arm spirals (Fig.\ 1, top),
the AB~Aur, HD~142527, and HD~100546 disks show multiple small arms, 
and the V1247~Ori disk has one arm (Fig.\ 1, bottom).  

\subsection{Spiral Arm Statistics Among Well-studied NIR Disks }\label{sec:spiralarmnir}

From the results shown in Table~\ref{tab:nirsamplegood}, we find 
that the incidence rate of spiral arms in disks is
surprisingly high among young well-studied A and B stars 
(3/7), well-studied FGK stars (5/22), and well-studied
disks overall (8/29; Table~\ref{tab:armfractionnir}).

We can make a similar accounting of spiral fraction by stellar mass 
rather than by spectral type. Anticipating our comparison of the spiral arm
incidence rate to that of massive companions to mature AB stars
(\S\ref{sec:planet}), we define a Herbig intermediate mass star (IMS) 
sample over a broader range of spectral type. A main sequence F0 
star has a mass of $\sim 1.5 \Msun$ \citep[e.g.,][]{pecaut13}. 
So we consider as
Herbig IMS all Herbig systems with spectral types between F6 and
B9, and stellar masses $>1.5\Msun$.

Table~\ref{tab:armfractionnir} shows that the spiral fraction 
is 6/11 among well-studied Herbig IMS and 
2/18 among the low mass stars 
(LMS) that make up the rest of the sample.  
The incidence rate of two-arm spirals is also high: 
2/11 among the well-studied Herbig IMS and 
2/18 among the LMS. 
Although the samples are small, the data can nevertheless lend 
insight into whether there is a significant difference 
between the occurrence rates of spiral structure 
in LMS and Herbig IMS disks.
For example, using Fisher's exact test (a contingency table analysis), 
we find that 
there is a 3\% probability (two-tailed p-value) that the 
spiral occurrence rate among Herbig IMS 
is the same as that among low mass stars.
Thus the apparent low occurrence rate of spirals among 
low mass stars compared to Herbig IMS 
is statistically significant 
(see also \citealt{avenhaus18, isella18}). 

There are a couple of biases associated with these results. 
Firstly, scattered light imaging studies often target sources with
interesting known structures (e.g., transitional disks) and 
sources that appear to show interesting structures at modest 
signal-to-noise receive more extensive follow up. 
Secondly, scattered light imaging studies are only sensitive to 
disk structures that can be seen in scattered light. If 
a disk's surface is not sufficiently flared, its spiral structure 
cannot be detected with this technique. 

Despite these difficulties, we can obtain a lower limit 
on the true spiral fraction of disks by examining
the fraction of disks with known spiral arms 
within a volume-limited sample of comparable young stellar 
objects. 
We therefore
focus our study on the Herbig IMS population, because 
a fair fraction of such sources within the local volume 
have already been surveyed in scattered light. 
In contrast, only a small fraction of the (lower mass) 
T Tauri population has been studied, 
and as a result, a corresponding lower limit on the true 
spiral fraction for LMS would
not be a significant constraint.

\subsection{A Volume-limited Sample of Herbig Stars}\label{sec:herbig200pc}

Excluding V1247~Ori, which is very distant (320 pc), the remaining
Herbig IMS in Table~\ref{tab:nirsamplegood} are all within 200 pc. We therefore
focus on a sample of Herbig IMS within this volume, i.e.,
Herbig sources within 200 pc that have spectral type F6 or earlier, $\Mstar
> 1.5\msun$, and no known stellar companion within $5\arcsec$.

Table~\ref{tab:herbig} compiles a list of sources from the literature meeting these criteria \citep{walker88, the94, malfait98, vieira03, erickson11, chen12spitzer}. 
Generally, these stars are identified by their association with star forming regions or a reflection nebula and the presence of hydrogen 
emission lines (particularly H$\alpha$).
Because the sample could be useful for studies other than the one carried out here, the list also includes, for completeness, sources that we ignore in our study (indicated in Column 4 of Table~\ref{tab:herbig}):  sources with stellar companions 0\farcs3--5\arcsec\ away (HR~811, V892~Tau, PDS~178, CQ~Tau, HD~100453, HR~5999, MWC~863, TY~CrA, T~CrA),
sources with $\Mstar < 1.5 \Msun$ (AK~Sco), sources with low, potentially optically thin dust (HD~141569), and sources with decretion disks (51~Oph). 

Excluding these sources, there are 24 remaining sources, which we carry forward as our ``volume-limited Herbig IMS'' sample. Ten of these 
have been well studied in NIR imaging (i.e., they are in Table~\ref{tab:nirsamplegood}). 
Note that with the above prescription for the Herbig IMS sample, we exclude younger sources such as LkH$\alpha$~330 (Table~\ref{tab:nirsamplegood}) which also has a stellar mass $> 1.5\Msun$ but a later spectral type (G3; Figure~\ref{fig:HRdiagram}), 
These sources are excluded in part because earlier spectral type
sources (F5--B9) have been studied more completely in scattered
light; younger, more embedded disks are more difficult to study in
scattered light due to contamination from the envelope. They are
also more numerous and much less completely studied in scattered
light. The masses of the later spectral type sources are more
uncertain as well.

Within the volume-limited Herbig IMS sample (Table 2), 5 of the 10
well-studied sources display spiral structure (AB~Aur, MWC~758, HD~100546, 
SAO~206462, HD~142527).
The true spiral fraction depends on whether the remaining 14 sources that
have not been well studied (i.e., are not included in Table~\ref{tab:nirsamplegood}) have
arms or not.  Thus, the minimum spiral fraction is
5/24, and the minimum two-arm spiral fraction is 2/24
(Table~\ref{tab:armfractionherbig}).
We conclude that even after correcting for
systems that have not been studied well in scattered light, the
spiral fraction of Herbig IMS is high.
Detailed studies of the additional Herbig IMS within 200 pc would be 
extremely valuable in firming up the spiral fraction.

\section{Comparison with Exoplanet Demographics}\label{sec:planet}

The two-arm spirals observed in scattered light can be produced
by 5--13 $\mj$ planets located 30--300 AU from the star (\citealt{fung15}
and \citealt{dong17spiralarm}).  To explore whether this scenario is 
tenable, we compare the incidence rate
of arms with the incidence rate of planetary companions to 
mature A and B stars in the corresponding mass and separation range.
Planetary companions to mature stars are an advantageous population to 
study because they are better characterized than are 
companions to Herbig stars and T Tauri stars (Bowler 2016). 

Compiling the results of direct imaging 
searches for planetary companions in this mass and separation range, 
\citet{bowler16} reported that only 3
mature (single) AB stars have such a companion out of a total 
sample of 110 stars with spectral types B2 to A9. 
This result assumes that planetary companions have the 
properties predicted by ``hot start'' models 
(see \S\ref{sec:discussion}). 
In comparison, 
the two-arm spiral fraction among well-studied Herbig IMS 
is larger (2/11; \S\ref{sec:spiralarmnir}).
Although the number of detections is small in 
both cases,
there is only a 6\% probability 
of obtaining such discrepant results if the 
the two rates are the same, 
according to Fisher's exact test. 

Further observations are needed to explore this possible 
difference. As noted in \S\ref{sec:herbig200pc}),   
the two-arm spiral fraction in the volume-limited Herbig
IMS sample (Table~\ref{tab:armfractionherbig}) is 
at least 2/24. 
If future observations show that there are no other 
two-arm spirals within 200 pc than the two already known, 
the probability that the two rates are the same would be 21\% 
and the difference in the observed rates would be of little 
statistical significance. 
However, if two additional two-arm spirals are found 
in the sample, 
there would only be a 2\% probability that 
the 4/24 spiral fraction 
and the \citet{bowler16} fraction are the same,
and the difference in the rates would be quite 
significant. 

The occurrence rate of companions to mature FGK stars is lower 
than that for mature AB stars (Bowler 2016). 
To compare with the FGK companion rate, 
we can select sources from Table~\ref{tab:nirsamplegood}
that are ``future FGK stars'', i.e., sources with stellar masses between
$0.5\,\Msun$ and 1.5\,$\Msun$
(TW~Hya, PDS~70, J1604-2130, RX~J1615.3-3255, DoAr~28, V4046~Sgr, GM~Aur, LkCa~15, PDS~66, SR~21, IM~Lup, DZ~Cha). 
The incidence rate of two-arm spirals among the well-studied 
``future FGK stars'' is 1/12 compared with the 0/155 incidence rate 
of 5--13 $\mj$ companions to mature FGK stars in the 30--300 AU 
separation range \citep{bowler16, vigan17}. 
There is a 7\% probability that the two rates are the same, 
similar to the situation for the higher mass stars. 
Further observations are needed to quantify the 
extent to which these rates differ.

With the apparent paucity of massive planetary companions to mature AB stars 
appearing to question the idea that such planetary companions 
drive two-arm spirals in Herbig disks, in the next section we 
explore the alternative possibility that spiral arms are driven by 
gravitational instability. 
We return in \S\ref{sec:discussion} to the planetary companion hypothesis and 
discuss ways in which it may be consistent with observations.

\section{Disk Masses and Gravitational Instability}\label{sec:gi}

To help interpret the origin of the spiral arms detected in scattered
light imaging, we have also compiled disk masses and 
stellar accretion rates 
for the volume-limited Herbig IMS sample (Table~\ref{tab:herbig}).
The latter can provide a complementary estimate of disk (gas)  
mass assuming a typical accretion history for the disk 
\citep{hartmann98}. 
Tables~\ref{tab:nirsamplegood} and \ref{tab:herbig} tabulate the
disk dust masses $\mdust$ calculated from submillimeter continuum fluxes
(Appendix \S\ref{sec:mdisk}). 
Stellar accretion rates $\mdot$ are determined from the veiling of the Balmer
discontinuity or hydrogen emission line fluxes;  
further details are provided in Appendix \S\ref{sec:mdot}.

\begin{figure}
\begin{center}
\text{\large Submillimeter Continuum-Based Disk Masses}\par\smallskip
\includegraphics[trim=30 80 40 120, clip,width=12cm]{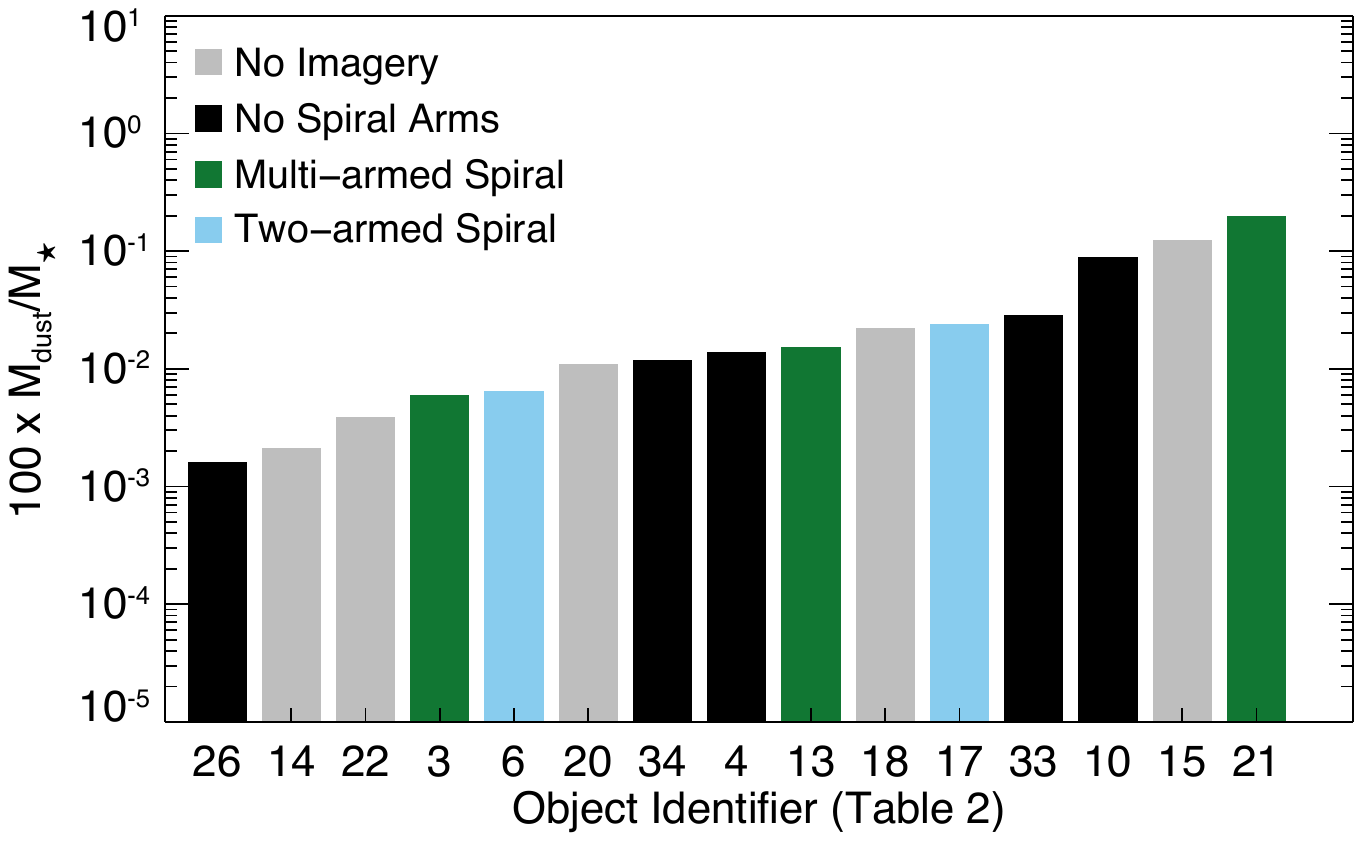}
\end{center}
\figcaption{Ratio of $\mdisk/\mstar$ for the volume-limited Herbig IMS sample (\S\ref{sec:herbig200pc}). The disk mass is estimated from the submillimeter continuum, assuming a gas-to-dust mass ratio of 100. The source name corresponding to each number in the horizontal axis can be found in Table~\ref{tab:herbig}. All disk masses are $< 0.2\mstar,$ with the average ${\mdisk/\mstar}=0.008$. Herbigs with spiral structure are interspersed with other Herbigs. See \S\ref{sec:gi} for details.
\label{fig:diskmass}}
\end{figure}

Turning first to the dust masses, although they may not be a reliable
quantitative measure of disk mass (\S\ref{sec:intro}), they might serve 
as a relative measure:  sources with higher disk gas
masses may also have higher dust masses. 
Assuming $\mdisk= 100\mdust$, 
Figure~\ref{fig:diskmass}
shows the ratio of $\mdisk/\mstar$  
for all sources in the volume-limited Herbig IMS sample 
that have estimated dust masses.
The ratio $\mdisk/\mstar$ ranges from $\sim 10^{-3} - 10^{-1}$. 
Apart from HD~142527, which has a very large dust mass,
the sources with two arms (blue bars) or multiple arms
(green bars) are distributed throughout this range and interspersed
with other sources, both those known to have no arms (black bars)
and those that have not yet been studied in scattered light in
detail (gray bars). 

There is no obvious difference in the
$\mdisk/\mstar$ ratio of disks with and without arms.
The average ratio is 0.008
for the entire volume-limited Herbig IMS sample.
Sources with arms and those known to have no arms
both have the same average ratio, 0.02. 
In other words, 
scattered light surveys have preferentially studied 
sources with higher submillimeter continuum fluxes.  

One indication that these values are flawed as quantitative measures 
of disk mass comes from estimating the remaining disk lifetime
assuming these disk masses,  
$\tau_{\rm life}=\mdisk/\mdot$. Figure~\ref{fig:mdot} shows $\tau_{\rm
life}$ for sources in the volume-limited Herbig IMS sample for
which both measurements are available. If we assume that the remaining
disk lifetime is 2 Myr, comparable to the average age of disk-bearing young 
stars (\citealt{hernandez08, richert18}) and a conservative estimate 
based on the HR diagram (Figure~\ref{fig:HRdiagram}),
half of the sources have lifetimes $\lesssim0.2$ Myr, a mere 10\%
of the average age. The short lifetime is a consequence of the 
high accretion rates. If these values are correct,
half of the Herbigs within 200 pc are in the last 
10\% of their mass-building phase of life, 
which seems implausible. 

We could instead assume that we are viewing Herbigs on average at
middle age, a current age of at least $t_0\sim 2$ Myr. 
Parametrizing the decline of $\mdot$ with time as
$\mdot(t) = \mdot(t_0)(t/t_0)^{-\eta}$ \citep[e.g.,][]{hartmann98},
where $t_0$ is the current age of the star, the total mass that the
star accretes at later times $t> t_0$ is
$\Mdisk = \mdot(t_0)t_0/(\eta -1).$
For $\eta \sim 1.5$ (\citealt{siciliaaguilar10}, \citealt{hartmann98}),
$\Mdisk\sim 2\mdot(t_0)t_0$. 
With this estimate, the average $\Mdisk/\mstar \sim 0.13$, 
about ten times higher than the disk masses estimated from 
submillimeter continua. 
If the stars are older, the implied disk masses are higher.

Figure~\ref{fig:ssdiskmass} shows $\Mdisk/\mstar$ where 
$\Mdisk$ is estimated as above. The sources with two (blue bars)
or multiple arms (green bars) are concentrated in the upper half
of the sample, reaching values as large as 0.6 (AB Aur), with about
one-half to one-third of the sample above the nominal gravitational
instability limit of $\Mdisk/\Mstar = 0.1$. The clustering of
``arm'' sources toward higher $\Mdisk$/\Mstar\ would be consistent
with the arms being generated through or enhanced by gravitational
instability.
It would extremely interesting to measure the scattered light
morphology of the unstudied sources (gray bars). If 
the lower mass disks also show spiral arms,  
it is unlikely that gravitational instability plays
the major role in generating arms. 

\begin{figure}
\begin{center}
\includegraphics[trim=0 80 0 120, clip,width=\textwidth,angle=0]{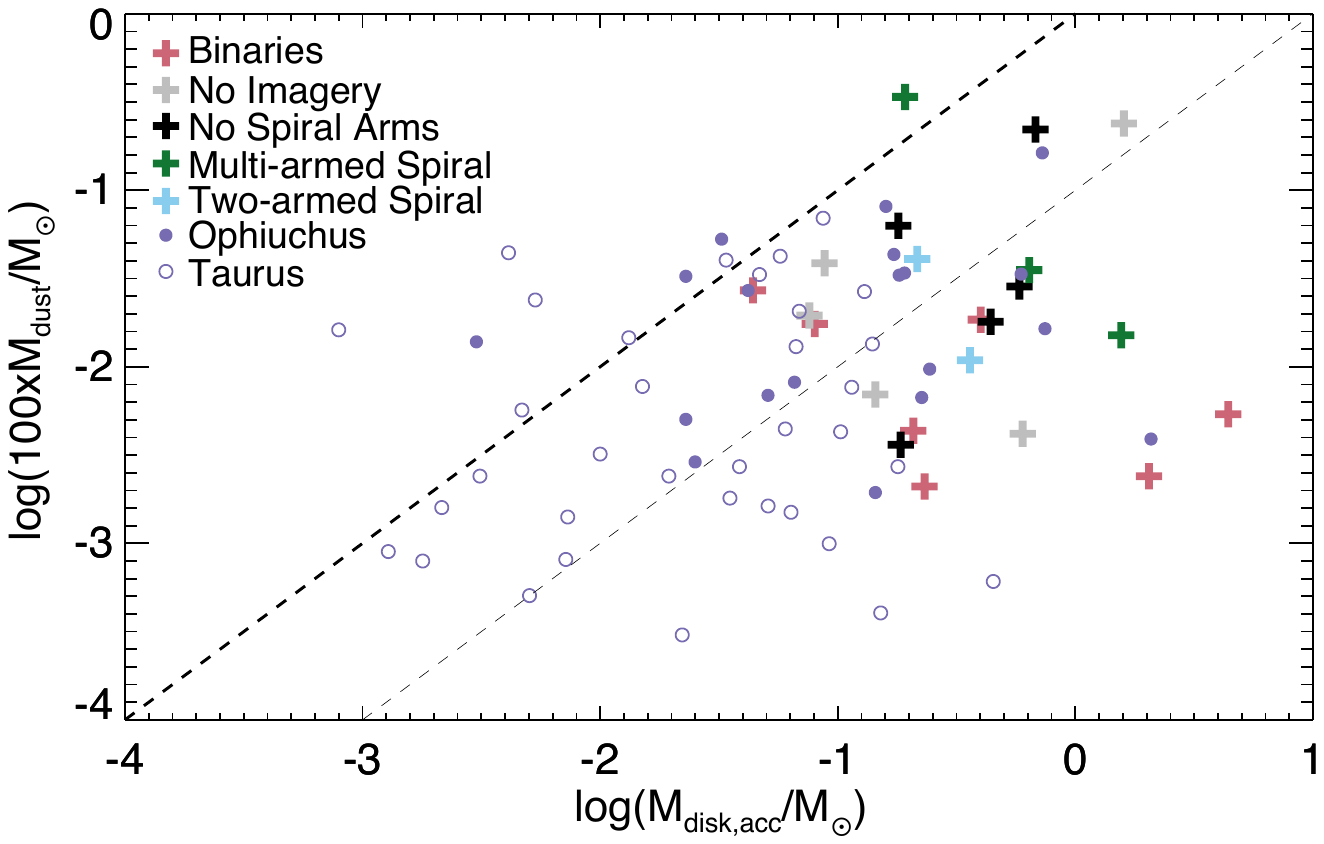}
\end{center}
\figcaption{Disk masses estimated from submillimeter dust continuum
emission (vertical axis) and from stellar accretion rates (horizontal
axis). The thick (thin) dashed lines demarcate continuum-based disk masses
that are equal to (one-tenth of) accretion rate-based disk masses.
For both the Herbig IMS studied here (colored pluses) and T Tauri
stars studied by \citet[Taurus---open circles; 
Ophiuchus---closed circles]{andrews07}, stellar accretion rate-based disk 
masses are a few to tens of times larger than continuum-based 
disk masses. The average ratio of the disk mass inferred from the stellar accretion rate and the disk mass inferred from the flux of the submillimeter continuum is 8.1$\pm$0.4. See \S\ref{sec:gi} and \S\ref{sec:discussion} for details.
\label{fig:ssvsmdust}} 
\end{figure}

Figure~\ref{fig:ssvsmdust} compares the above submillimeter continuum-based and stellar accretion-based disk masses for the volume-limited Herbig IMS sample as well as the T Tauri stars studied by \citet{andrews07}. Both groups of sources have, on average, 3--10 times larger masses estimated from stellar accretion rates than from dust continuum. The discrepancy between the disk masses estimated by these two methods has been previously noted for T Tauri stars by several authors \citep[e.g.,][]{andrews07, rosotti17}. The Herbig IMS studied here show a similar trend.


\section{Discussion}\label{sec:discussion}

Among disks that have been well-studied in scattered light, 
spiral arm structures are surprisingly common (\S\ref{sec:spiralarmnir}), although 
their origin is unclear.
While giant planets and gravitational instability have both 
been put forward as explanations for the observed spiral 
structure, neither explanation is readily confirmed in our study. 
We find that despite the small sample sizes studied to date, 
we can infer that 
two-arm spirals occur more frequently among 
Herbig IMS disks than do massive giant planet companions to 
mature AB stars, challenging the planet hypothesis 
(\S\ref{sec:planet}).
The viability of gravitational instability as an 
explanation for spiral structure is uncertain, 
because submillimeter continuum-based disk masses are low. 
However, as we showed, disk masses may be substantially 
larger if stellar accretion rates are a guide (\S\ref{sec:gi}). 
In this section, we examine the proposed hypotheses in 
greater detail before presenting 
scenarios that are potentially consistent with our results. 

{\bf Shadows and Stellar Flybys.} 
Spiral arms detected in scattered light have also been modeled as
arising from rotating shadows cast by a warped inner disk
\citep[e.g.,][]{montesinos16, montesinos17} and stellar flybys
\citep[e.g.,][]{pfalzner03}. These alternative explanations require
specific conditions that are unlikely to be satisfied in the majority
of cases. The moving shadow scenario is best at explaining grand design
two-arm spirals; yet in the only disk with both a clear $m$=2
shadow pattern and a set of two-arm spirals (HD~100453), the disk's
arms are almost certainly induced by the visible stellar companion
\citep{dong16hd100453, wagner18}.
To catch the spirals induced by a stellar flyby, one needs to image
the system within a few dynamical times (i.e., $<10^4$ years at
typical distances) after the closest encounter before the spirals winded up
due to differential rotation \citep{pfalzner03}. 
The short timescales make it unlikely 
to catch the disk in this situation. 
Because these two options are very unlikely to be the general mechanisms 
responsible for observed spiral arms in scattered light,
we focus more closely on the gravitational instability 
and planetary companion hypotheses that have been considered 
more intently in the recent literature.

{\bf Gravitational Instability.}
Exciting two-arm spirals in a gravitationally unstable disk 
requires $\Mdisk/\mstar\sim 0.5,$ a situation that may 
occur in Herbig disks, especially if they have recently decoupled 
from their infalling envelopes. In such
systems, spiral shocks transport angular momentum and drive rapid
accretion, with $\mdot \gtrsim 10^{-5}\Msunpery$
\citep{kratter08, kratter10gidisks, dong15giarm}. Consequently,
the disk mass decreases quickly and the two-arm phase is brief 
(on the order of $10^4$ yr). 
As the disk mass drops, the disk accretion rate also declines, 
and the disk structure shifts to a multi-arm morphology 
\citep[e.g.,][Hall et al, in prep.]{rice09}.
Eventually the disk is stabilized against gravitational 
instability and the disk settles down to an even lower
accretion rate. 

\begin{figure}
\begin{center}
\includegraphics[trim=80 60 40 40, clip,width=10cm,angle=90]{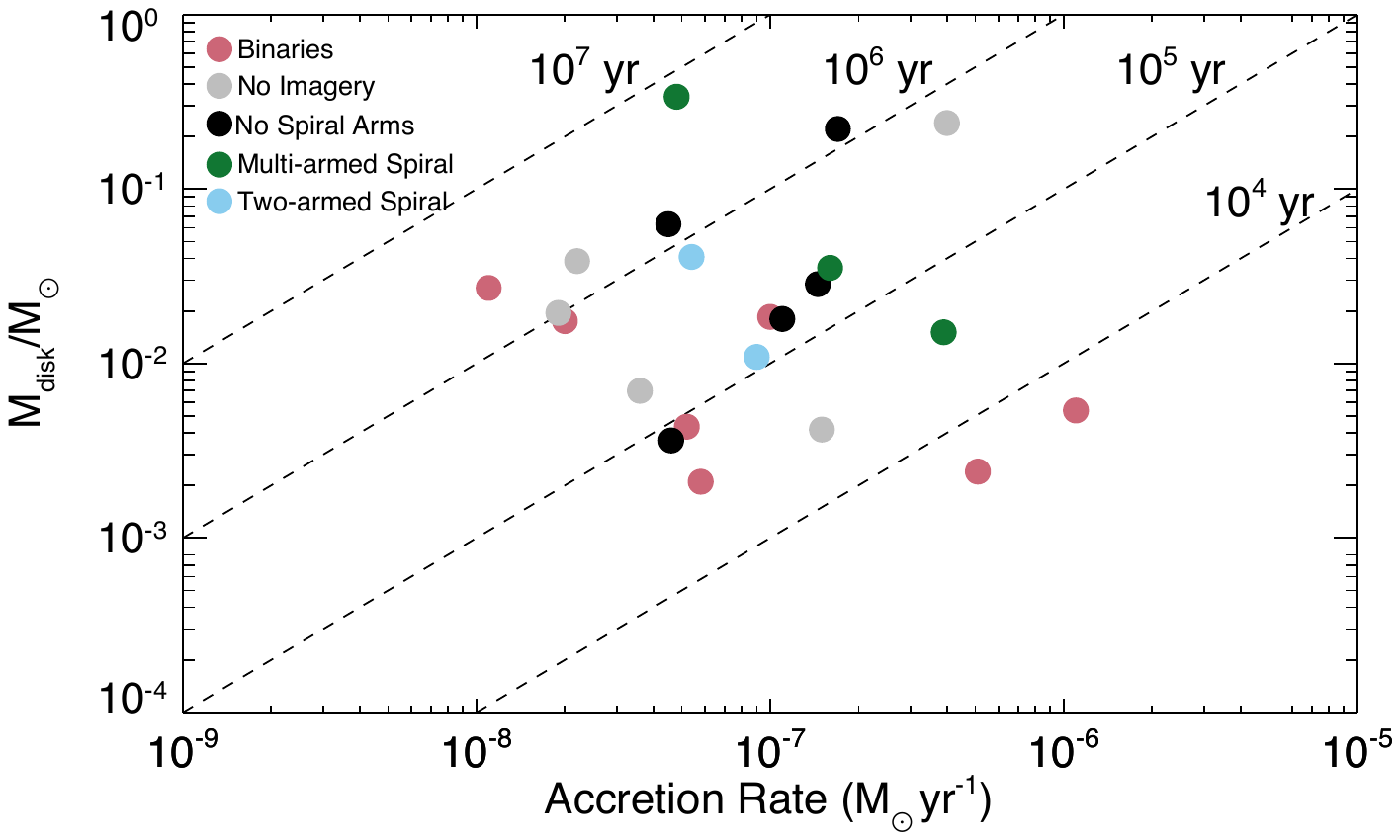}
\end{center}
\figcaption{Submillimeter continuum-based disk mass ($\mdisk$) plotted against 
stellar accretion rate ($\mdot$) for the volume-limited Herbig IMS 
sample. Diagonal dashed lines are lines of constant remaining disk
lifetime ($\tau_{\rm life}=\mdisk/\dot{M_\odot}$). Over half of the sample 
is consistent with a disk lifetime of $\lesssim 2\times10^5$ years. See \S\ref{sec:gi} and \S\ref{sec:discussion} for details.
\label{fig:mdot}} 
\end{figure}

This scenario appears inconsistent with the demographic 
properties of disks with two-arm spirals. 
In the above picture, two-arm spirals should occur in
systems with the largest $\mdot$.  Instead, observed two-arm
spiral disks have middle-of-the-road $\mdot$ values among the
Herbig IMS within 200 pc (Figure~\ref{fig:mdot}). Two-arm
spirals are also expected to be rare, because this phase lasts
for on the order of 1\% of the 2\,Myr average age of disk-bearing stars once envelope infall have ceased. 
In contrast, two-arm spirals are found in 
2/11 (or $\sim$18\%) of the 
well-studied Herbig IMS and $\geq$ 2/24 (or $\gtrsim 8$\%)
of the volume-limited Herbig IMS 
sample (Table~\ref{tab:armfractionherbig}).
Disks with two-arm spirals are also expected to be evolutionarily 
young, and to have just emerged from their infalling envelopes. 
In the HR diagram, disks with two-arm spirals are of 
preferentially later spectral type than other well-studied 
Herbig stars, but they are not preferentially younger 
(Fig.~\ref{fig:HRdiagram}). 

Although it seems unlikely to explain the two-arm spirals, 
gravitational instability is a possible explanation for 
multi-arm spirals. 
Disks with $\mdisk/\mstar\sim 0.1$ have multiple arms and 
are marginally unstable, with disk accretion rates
$\sim10^{-8}$--$10^{-6}\Msunpery$ in the outer disk
\citep[e.g.,][]{dong15giarm, hall16}. They can remain  
in this state for a period $\tau\sim\mdisk/\mdot$, or $10^5$--$10^7$ yr, 
comparable to the
2 Myr average age of disk-bearing stars.
Additional disk accretion mechanisms, such as the 
magnetorotational instability \citep{balbus92, gammie01} 
or magnetized disk winds \citep[e.g.,][]{bai16wind, bai17}, 
will become more dominant as the gravitational instability-driven 
accretion rate declines.

\begin{figure}
\begin{center}
\text{\large Accretion Rate-Based Disk Masses}\par\smallskip
\includegraphics[trim=30 90 30 130, clip,width=\textwidth,angle=0]{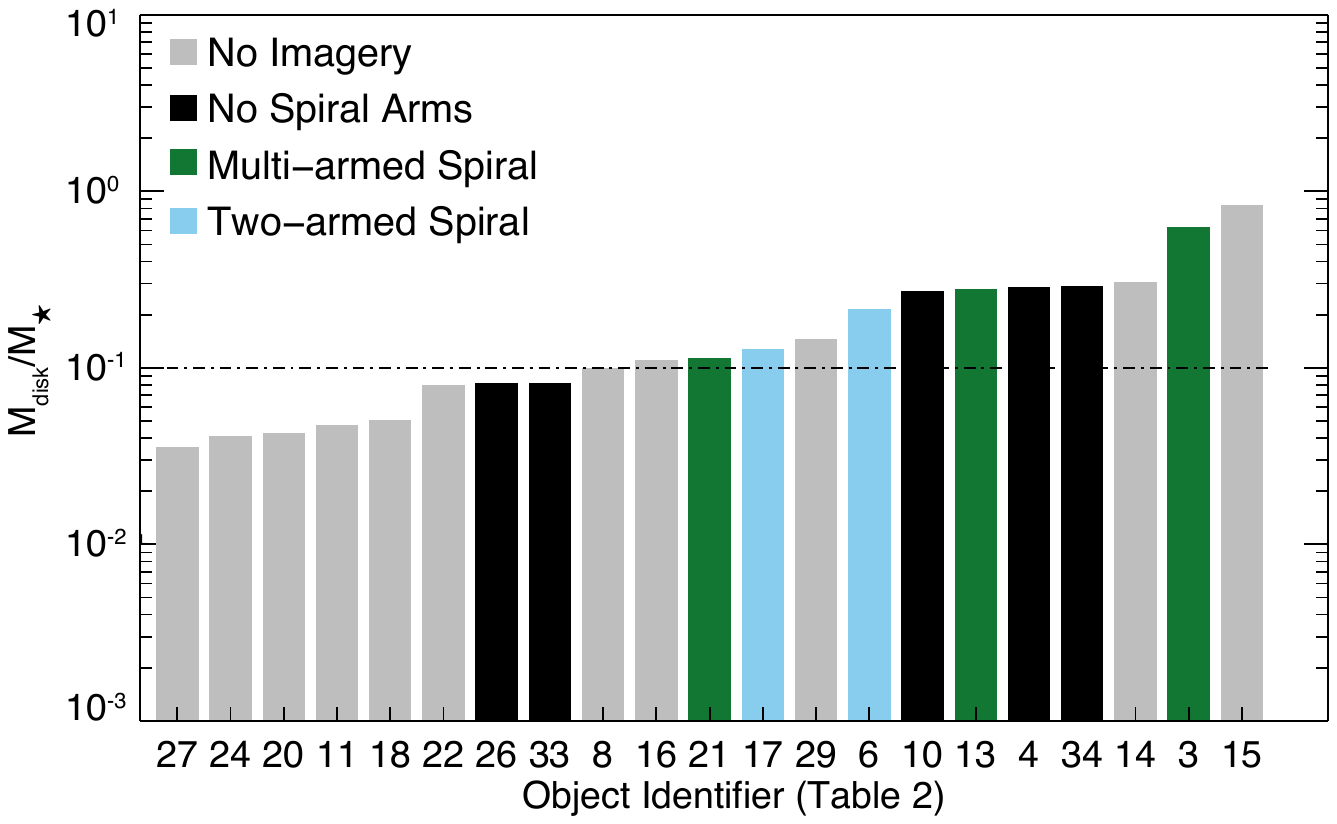}
\end{center}
\figcaption{The ratio $\mdisk/\mstar$ for the volume-limited Herbig IMS sample, where the disk mass is estimated as $2\mdot t_0$ ($t_0$=2 Myr). Approximately one-half of the sources have $\mdisk \gtrsim 0.1 \mstar$. Herbigs with spiral structure have $\Mdisk/\mstar$ as large as $\sim$1 and in the upper half of the distribution, suggesting that disk mass may play a role in generating some of the spiral structures. See \S\ref{sec:gi} and \S\ref{sec:discussion} for details.
\label{fig:ssdiskmass}}
\end{figure}

The relatively long lifetime of 
the multi-arm phase of gravitational instability is possibly 
consistent with the 
3/11 (or 27\%) incidence rate of multi-arm spirals 
among well-studied Herbig IMS and $\geq$ 3/24 (or $\ge 13$\%) 
incidence rate
within the volume-limited Herbig IMS sample
(Table~\ref{tab:armfractionherbig}). 
All three of the multi-arm disks have large disk masses 
based on their stellar accretion rates 
($\Mdisk/\mstar \gtrsim 0.1$; Figure~\ref{fig:ssdiskmass})
as well as stellar accretion rates in the above range 
$\sim$10$^{-8}$--$10^{-6}\Msunpery$ (Figure~\ref{fig:mdot}).
Note that the 
similarity between observed stellar accretion 
rates and the predicted outer disk accretion rates  
suggest that gravitational instability-driven 
accretion in the outer disk could potentially be sustained 
down to small disk radii and onto the star; further theoretical 
work is needed to understand the details of the accretion process.
Numerical simulations that quantify the time that a disk stays 
in each $n$-arm state in a realistic environment 
will be extremely useful to explore this possibility.

{\bf Planetary Companions.}
By studying sources that have been well studied in scattered light,
we have focused on older, revealed Herbig stars that are not obscured by
an infalling envelope. Thus our sample is biased against gravitational
instability, which
is more likely to occur in (younger) massive disks, 
and possibly toward giant
planets, which may take a few Myr to form.  It is therefore surprising
to find that the properties of the sample seem inconsistent with the
reported properties of giant planets around mature A stars: 
two-arm spirals appear to be more common than the 
giant planets massive enough to drive them. 

One possible resolution to this discrepancy is that 
giant planets are fainter than we think, i.e., 
current direct imaging searches for planetary  
companions to mature A stars may underestimate the planetary 
masses to which they are sensitive. 
In converting companion fluxes (and detection upper limits) to
masses,
one can choose between ``hot start'' (e.g., COND and
DUSTY models, \citealt{baraffe03, chabrier00}) and ``cold start''
evolutionary models \citep[e.g.,][]{fortney08, spiegel12}. The former, used
by most investigators, predicts a higher luminosity for a given
star at a given age with a given mass, than the latter. 
Neither set of models has been observationally confirmed for 
planetary mass objects at these young ages. Multi-$\mj$
companions, capable of driving observed two-arm spirals, may not
be detectable by current surveys if the ``cold start'' models 
are appropriate --- at 50 Myr, a 3$\mj$ hot start planet has roughly the 
same $H$-band luminosity
as a 10$\mj$ cold start planet \citep{baraffe03, spiegel12}.
Planets that are fainter than typically assumed would also 
help explain why direct imaging searches for the predicted 
perturbers in two-arm spiral disks
have failed to detect massive giant planets at the 
flux levels predicted by the hot start models
\citep[SAO~206462 and MWC~758;][]{maire17, reggiani18}.

As an alternative way to resolve the discrepancy, one might 
imagine that multi-$\mj$ planets located at 10s to $\sim$100 AU from the
star may be more common at a few Myr than at the 10--100 Myr ages
typically probed by direct imaging studies \citep[e.g.,][]{bowler16},
either because 
planets migrate inward from the original orbital distances via
disk-planet interactions, 
and/or they are scattered out of the system via
planet-planet interactions. 
Both options appear unlikely, however.
While a planet could be driven inward by an outer disk located
beyond the planet \citep[e.g.,][]{kley12}, in all systems with
two-arm spirals the purported planet is located {\it beyond} the
edge of the disk. 
In the planet-planet interaction scenario, 
multiple giant planets that are initially stabilized by the gas disk 
\citep[e.g.,][]{dunhill13} later undergo
planet-planet scattering that removes planets from the system 
once the gas disk dissipates 
\citep[e.g.,][]{dong16td}.
Systems with two-arm spirals, however, are unlikely to have
additional giant planets with similar masses at distances close
enough to interact with the purported arm-driving planet, as
such planets will drive their own sets of two-arm spirals, which
are not seen.  

Thus, the most attractive explanation for two-arm spirals in
disks is that they are driven by massive giant planetary companions
($\gtrsim 5 \mj$) that are fainter than predicted by hot start
evolutionary models.

Planetary companions might also play a role in explaining 
multi-arm spirals.
Numerical simulations have shown that a planet of mass 
$\sim \mj$ usually drives one to three arms on one side 
of its orbit that are detectable in NIR imaging observations
\citep{dong17spiralarm}. In disks with multi-arm spirals
(Figure~\ref{fig:spiralarmexamples}),
the arms are at similar distances.  
To drive them with Jovian planets requires more than one planet, positioned
over a small range of disk radii, potentially in conflict with
theories of giant planet formation.

\section{Summary and Future Prospects}\label{sec:summary}

Spiral arm structures are surprisingly common in NIR scattered light 
imaging of protoplanetary disks, especially among Herbig stars. 
Among the 11 Herbig stars that have been well-studied in scattered light,
6/11 have spiral arms of some kind, and 2/11 have near-symmetric
two-arm spirals. Among the 24 single Herbig stars within 200 pc, 
10 have been imaged in scattered light. 
The spiral arm occurrence rate in this volume-limited
Herbig intermediate mass star (IMS) sample 
is $\ge$ 2/24 for two-arm spirals and $\ge$ 5/24
for spirals of all kinds; if some of the remaining 14 disks in the sample 
turn out to have spiral structures,
the fractions will be higher.
Although the sample studied to date is small and the occurrence
rates of spiral arms poorly determined as a result, the data 
do have the potential to constrain quantities of fundamental 
importance for planet formation.

Two-arm spirals appear to be more common among 
Herbig IMS disks than are 
directly imaged giant planet companions 
to mature AB stars as estimated in the literature. 
This discrepancy is difficult to account for in a simple way 
given our current understanding of the mechanisms that can 
create spiral structure.
Gravitational instability is unlikely to explain the 
two-arm spirals detected in class II disks mainly because of the short lifetime of this 
phase of disk evolution compared with the high fraction of 
two-arm spirals observed. The stellar accretion rates of 
Herbig systems with two-arm spirals are also not particularly high 
compared to other Herbigs. 
To explain the high occurrence rate 
of two-arm spirals, we propose that they may be produced by massive 
giant planets that are fainter than predicted by 
hot start models. Such planets must occur as companions to 
$\gtrsim$8\% of Herbig IMS.

Gravitational instability may drive multi-arm spirals if the disk mass is $\gtrsim$ 10\% of the stellar mass. Submillimeter continuum-based disk mass estimates for disks with multi-arm spirals are generally about one order of magnitude below this threshold.
However, the high stellar accretion rates of Herbig stars, if
sustained  over a reasonable fraction of their lifetimes,
suggest much larger disk masses.
As a result, gravitational instability is a plausible
explanation for multi-arm spirals.
With the multi-arm phase of gravitational instability 
persisting longer than the two-arm phase, it is possibly 
consistent with the 3/11
occurrence rate among well-studied Herbig IMS 
and the $\geq$3/24
occurrence rate in the volume-limited Herbig IMS sample. 

Several future studies can lend insights into the issues 
raised here. 
Our study is limited by the small sample size of disks that have 
been well-studied
in scattered light.  High signal-to-noise scattered light 
imaging of the remainder of the volume-limited Herbig IMS sample 
would be extremely valuable in firming up the occurrence rates 
of spiral features.
ALMA observations of the gas and dust emission structure of disks, 
which do not require illumination by the central star, 
are a valuable way to develop a more complete census of spiral arms. 
To distinguish between the various scenarios that could produce 
the observed spiral structures, it may be useful to compare not 
only the observed and expected number and symmetry of arms 
but also the observed 
and expected surface brightness contrast of spiral features above 
the rest of the disk. 

The current sample is biased toward older sources that lack 
substantial infall and are therefore less likely to have 
gravitationally unstable disks. It would
be very interesting to search for spiral structure among 
the younger precursor population (i.e., among GK-type
intermediate mass T Tauri stars) either in scattered light or
millimeter emission. The time evolution of the occurrence rate of
two-arm spirals may shed light on the formation timescale of giant
planets at large distances, and those of multi-arm spirals may
reveal how disk masses evolve with time.


\section*{Acknowledgments}

\begin{figure}
\begin{center}
\includegraphics[trim=0 0 0 0, clip,width=\textwidth,angle=0]{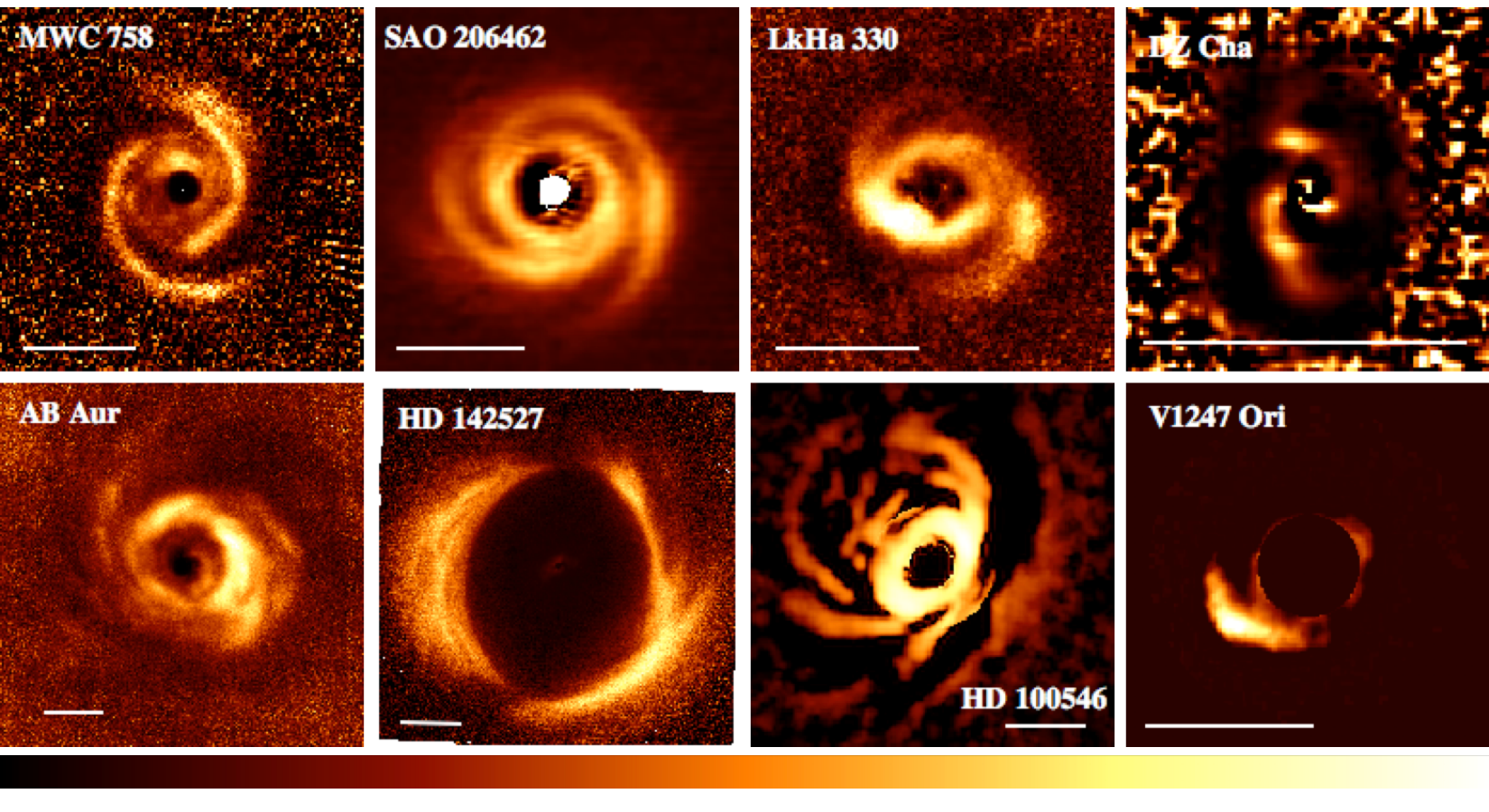}
\end{center}
\figcaption{NIR scattered light images of disks with spiral arms. A set of prominent two-arm spirals are detected in MWC~758 (\citealt{benisty15}), SAO~206462 (\citealt{garufi13}),  LkH$\alpha$~330 (\citealt{uyama18}), and DZ~Cha (\citealt{canovas18}). Multiple weak arms are observed on smaller scales in AB~Aur (\citealt{hashimoto11}), HD~142527 (\citealt{avenhaus17}), and HD~100546 (\citealt{follette17}). A one-arm spiral is observed in V1247~Ori (\citealt{ohta16}). The scale bar in all panels is 0\farcs5. All images show polarized intensity, and are $r^2$-scaled and on a linear stretch, except for HD~100546, which shows total intensity and is on a log stretch because the polarized intensity image does not reveal the spiral arms well (\citealt{follette17}).
\label{fig:spiralarmexamples}}
\end{figure}

We thank Jaehan Bae, Xuening Bai, Josh Eisner, Cassandra Hall, Jun Hashimoto, Scott Kenyon, 
Kaitlin Kratter, Adam Kraus and Ken Rice for 
valuable advice and discussions, and an anonymous referee
for a helpful review of the manuscript.  
We are grateful to Henning Avenhaus, Myriam Benisty, Hector Canovas,
Katherine Follette, Antonio Garufi, Jun Hashimoto, and Taichi Uyama
for making available the observational images used in Figure~\ref{fig:spiralarmexamples}.
This research has made
use of the SIMBAD database, operated at CDS, Strasbourg, France.
SDB acknowledges NASA Agreements No.\ NXX15AD94G and No.\ NNX16AJ81G; and NSF-AST 1517014. JRN acknowledges the stimulating research environment supported by NASA Agreement No.\ NNX15AD94G to the ``Earths in Other Solar Systems'' program. SDB gratefully acknowledges the generous hospitality and stimulating scientific environment provided by the  National Optical Astronomy Observatory.


\clearpage

\setcounter{footnote}{0}

\begin{table}[]
\footnotesize
\centering
\resizebox{\textwidth}{!}{%
\begin{tabular}{@{}ccccccccccccc@{}}
\toprule
\# & Object & Spiral & $L_\star$ & $T_{\rm eff}$ & SpT & D & $M_\star$ & $f_{\rm mm}$ & $\lambda_{\rm mm}$ & $R_{\rm mm}$ & $M_{\rm dust}$ & $M_{\rm disk}/M_\star$ \\
 &  &  & $L_\odot$ & K &  & pc & $M_\odot$ & mJy & $\micron$ & AU & $M_\oplus$ & $\times$1\% \\
 (1)  & (2) & (3) & (4) & (5) & (6) & (7) & (8) & (9) & (10) & (11) & (12) & (13) \\ \midrule
3 & AB Aur & Multiple & 43.8\footref{andrews13} & 9380 & B9.5\footref{valenti03} & 153\footref{gaia16} & 2.50 & 309\footref{andrews13} & 890 & 230\footref{tang12} & 50 & 0.6 \\
4 & MWC 480 & None & 19.6\footref{andrews13} & 8330 & A5 & 142\footref{gaia16} & 2.04 & 331\footref{oberg15} & 1167 & 170 & 95 & 1.4 \\
6 & MWC 758 & $m$$\approx$2 & 8.5\footref{andrews11} & 7600 & A8\footref{beskrovnaya99} & 151\footref{gaia16} & 1.68 & 180\footref{andrews11} & 880 & 151 & 36 & 0.7 \\
\dots & V1247 Ori & One Arm & 15.8\footref{fairlamb15} & 7880 & F0\footref{kraus13} & 320\footref{gaia16} & 1.91 & 300\footref{kraus17} & 870 & 190 & 252 & 4.0 \\
10 & HD 97048 & None & 36.1\footref{fairlamb15} & 10500 & A0 & 179\footref{gaia16} & 2.50 & 2230\footref{walsh16} & 867 & 403 & 737 & 8.8 \\
13 & HD 100546 & Multiple & 24.7\footref{fairlamb15} & 9750 & B9\footref{the94} & 109\footref{gaia16} & 2.30 & 1240\footref{walsh14} & 867 & 230 & 118 & 1.5 \\
17 & SAO 206462 & $m$$\approx$2 & 8.8\footref{fairlamb15} & 6375 & F4 & 156\footref{gaia16} & 1.70 & 620\footref{andrews11} & 880 & 156 & 136 & 2.4 \\
21 & HD 142527 & Multiple & 9.9\footref{fairlamb15} & 6500 & F6 & 156\footref{gaia16} & 1.70 & 3310\footref{boehler17} & 880 & 300 & 1126 & 19.9 \\
26 & IRS 48 & None & 20.4\footref{erickson11} & 9530 & A0 & 120 & 2.25 & 180\footref{vandermarel16whole} & 850 & 60\footref{brown12} & 10 & 0.1 \\
33 & HD 163296 & None & 23.6\footref{fairlamb15} & 9250 & A2\footref{valenti03} & 119\footref{gaia16} & 2.18 & 1820\footref{degregoriomonsalvo13} & 850 & 250 & 210 & 2.9 \\
34 & HD 169142 & None & 6.1\footref{murphy17} & 6700 & F3 & 117\footref{gaia16} & 1.51 & 232\footref{fedele17} & 1288 & 100 & 60 & 1.2 \\
\dots & LkHa 330 & $m$$\approx$2 & 12.8\footref{vandermarel16whole} & 5830 & G3 & 250 & 2.12 & 210\footref{andrews11} & 880 & 170 & 110 & 1.6 \\
\dots & FN Tau & None & 0.5\footref{kudo08} & 3240 & M5 & 140 & 0.30 & 37\footref{momose10} & 882 & \dots & 19 & 1.9 \\
\dots & RY Tau & None & 23.1\footref{andrews13} & 5080 & K1 & 177\footref{gaia16} & 3.30 & 469\footref{andrews13} & 890 & 200\footref{isella09} & 116 & 1.1 \\
\dots & UX Tau & None & 2.0\footref{andrews13} & 4900 & G8 & 158\footref{gaia16} & 1.60 & 150\footref{andrews11} & 880 & 85 & 37 & 0.7 \\
\dots & LkCa 15 & None & 1.0\footref{vandermarel16whole} & 4730 & K5 & 140 & 1.27 & 410\footref{andrews11} & 880 & 182 & 191 & 4.5 \\
\dots & GM Aur & None & 1.2\footref{andrews13} & 4730 & K3 & 140 & 1.16 & 640\footref{andrews11} & 880 & 120 & 193 & 5.0 \\
\dots & SU Aur & None & 10.7\footref{andrews13} & 5520 & G8 & 142\footref{gaia16} & 2.25 & 71\footref{andrews13} & 890 & \dots & 12 & 0.2 \\
\dots & TW Hya & None & 0.4\footref{vandermarel16whole} & 4210 & M2 & 60\footref{gaia16} & 1.00 & 1500\footref{andrews16} & 867 & 65 & 75 & 2.2 \\
\dots & DZ Cha & $m$$\approx$2 & 0.6\footref{canovas18} & 3770 & M0 & 110 & 0.51 & 74\footref{canovas18} & 517 & \dots & 6 & 0.4 \\
\dots & PDS 66 & None & 1.4\footref{vandermarel16whole} & 5080 & K1 & 99\footref{gaia16} & 1.30 & 207\footref{carpenter05} & 1200 & \dots & 73 & 1.7 \\
\dots & PDS 70 & None & 1.2\footref{pecaut13} & 4140 & K5\footref{simbad} & 140 & 0.83 & 38\footref{hashimoto15} & 1300 & 100 & 26 & 0.9 \\
\dots & IM Lupi & None & 0.9\footref{cleeves16} & 3900 & M0 & 161\footref{gaia16} & 0.60 & 590\footref{cleeves16} & 875 & 300 & 614 & 30.7 \\
\dots & J1604-2130 & None & 0.6\footref{vandermarel16whole} & 4350 & K2 & 145 & 1.10 & 226\footref{andrews11} & 880 & 145\footref{dong17j1604} & 116 & 3.2 \\
\dots & Sz 91 & None & 0.3\footref{vandermarel16whole} & 3720 & M0.5 & 200 & 0.48 & 42\footref{zhang14} & 860 & 160 & 60 & 3.7 \\
\dots & RX J1615.3-3255 & None & 1.3\footref{andrews11} & 4350 & K5 & 185 & 1.10 & 430\footref{andrews11} & 880 & 185 & 317 & 8.6 \\
\dots & DoAr 28 & None & 0.7\footref{vandermarel16whole} & 4370 & K5 & 139 & 1.12 & 69\footref{rich15} & 1300 & \dots & 77 & 2.1 \\
\dots & SR 21 & None & 6.5\footref{vandermarel16whole} & 5830 & G3 & 120 & 1.67 & 400\footref{andrews11} & 880 & 84 & 38 & 0.7 \\
\dots & V4046 Sgr & None & 0.7\footref{vandermarel16whole} & 4350 & K5 & 73 & 1.10 & 770\footref{vandermarel16whole} & 850 & 73\footref{rosenfeld13} & 49 & 1.3 \\ \bottomrule
\end{tabular}%
}
\caption{\footnotesize
Stellar and disk properties for the ``well-studied NIR disks'' sample (\S\ref{sec:nirsample}). 
Column (1): Source number from Table~\ref{tab:herbig}. 
Column (2): Source name; RA and Dec available in Appendix Table~\ref{tab:radec}.
Column (3): Type of spiral.
Column (4): Stellar luminosity (rescaled by \citet{gaia16} distance if available). 
Column (5): Effective temperature; references omitted if the same as in Col.\ (4).
Column (6): Spectral type; references omitted if the same as in Col.\ (4). 
Column (7): Distance; references omitted if the same as in Col.\ (4). Distances from \citet{gaia16} unless otherwise noted. 
Column (8): Stellar mass from \citet{siess00} tracks, using listed $T_{\rm eff}$ and $L_\star.$ 
Column (9): Flux density of millimeter continuum emission. 
Column (10): The wavelength at which Col.\ (8) is measured.
Column (11): Disk radius in millimeter continuum emission; references omitted if the same as in Col.\ (8).
Column (12): Dust mass derived from the submillimeter flux; see Appendix \S\ref{sec:mdisk} for details. 
Column (13): Disk mass ($\mdisk$) in units of the stellar mass, where $\mdisk$ is 100$\times$ the dust mass in Col.\ (11).
References can be found at the end of Table~\ref{tab:herbig}.
}
\label{tab:nirsamplegood}
\end{table}

\setcounter{footnote}{0}
\begin{landscape}
\scriptsize
\begin{longtable}[c]{@{}cccccccccccccccccccc@{}}
\toprule
\# & Object & Spiral & Included? & SpT & $L_\star$ & $T_{\rm eff}$ & D  & $M_\star$ & $R_\star$ & $f_{\rm mm}$ & $\lambda_{\rm mm}$ & $R_{\rm mm}$ & $M_{\rm dust}$ &  $\log{L_{\rm acc}}$ & $\log{L_{Br\gamma}}$ & $\log{L_{H \alpha}}$ & $\log{L_{\rm acc}}$ & $\log{\mdot}$  \\
   &        &      &             &     & $L_\odot$ & K & pc & $M_\odot$ & $R_\odot$ & mJy          & $\micron$          & AU           & $M_\oplus$                & $L_\odot$, Lit.     & $L_\odot$, Lit.      & $L_\odot$, Lit.      & $L_\odot$  \\              
(1)  & (2) & (3) & (4) & (5) & (6) & (7) & (8) & (9) & (10) & (11) & (12) & (13) &  (14)  & (15) & (16) & (17) & (18) & (19) \\ \midrule
\endfirsthead
\multicolumn{18}{c}%
{{\bfseries Table \thetable\ continued from previous page}} \\
\toprule
\# & Object & Spiral & Included? & SpT & $L_\star$ & $T_{\rm eff}$ & D  & $M_\star$ & $R_\star$ & $f_{\rm mm}$ & $\lambda_{\rm mm}$ & $R_{\rm mm}$ & $M_{\rm dust}$ &  $\log{L_{\rm acc}}$ & $\log{L_{Br\gamma}}$ & $\log{L_{H \alpha}}$ & $\log{L_{\rm acc}}$ & $\log{\mdot}$  \\
   &        &      &             &     & $L_\odot$ & K & pc & $M_\odot$ & $R_\odot$ & mJy          & $\micron$          & AU           & $M_\oplus$                & $L_\odot$, Lit.     & $L_\odot$, Lit.      & $L_\odot$, Lit.      & $L_\odot$  \\              
(1)  & (2) & (3) & (4) & (5) & (6) & (7) & (8) & (9) & (10) & (11) & (12) & (13) &  (14)  & (15) & (16) & (17) & (18) & (19) \\ \midrule
\endhead
\bottomrule
\endfoot
\endlastfoot
1 & HR811 & \dots & No & B7\footref{malfait98} & 87.1\footref{thiswork} & 14000 & 120\footref{gaia16} & 4.00 & 3.00 & \dots & \dots & \dots & \dots & \dots & \dots & -1.49 & 0.60\footref{valdes04} & -7.02 \\
2 & V892 Tau & \dots & No & A0\footref{the94} & 39.8\footref{hernandez05} & 11220 & 160 & 2.75 & 2.02 & 630\footref{sandell11} & 850 & 100\footref{hamidouche10} & 62 & \dots & \dots & -1.44 & 0.65\footref{hernandez04} & -6.98 \\
3 & AB Aur & Multiple & Yes & B9.5\footref{valenti03} & 43.8\footref{andrews13} & 9380 & 153\footref{gaia16} & 2.50 & 2.53 & 309\footref{andrews13} & 890 & 230 & 50 & \dots & -2.68 & \dots & 1.08\footref{donehew11} & -6.41 \\
4 & MWC 480 & None & Yes & A5\footref{andrews13} & 19.6\footref{andrews13} & 8330 & 142\footref{gaia16} & 2.04 & 2.09 & 331\footref{oberg15} & 1167 & 170 & 95 & \dots & -3.00 & \dots & 0.65\footref{donehew11} & -6.84 \\
5 & PDS 178 & \dots & No & A2/A7\footref{walker88} & 23.4\footref{manoj06} & 8910 & 150 & 2.20 & 1.96 & 20\footref{alonsoalbi09} & 1200 & \dots & 7 & \dots & \dots & -1.78 & 0.31\footref{dunkin98} & -7.23 \\
6 & MWC 758 & $m$$\approx$2 & Yes & A8\footref{vieira03} & 8.5\footref{andrews11} & 7600 & 151\footref{gaia16} & 1.68 & 1.68 & 180\footref{andrews11} & 880 & 151 & 36 & \dots & -3.15 & \dots & 0.45\footref{donehew11} & -7.05 \\
7 & CQ Tau & \dots & No & F5\footref{pecaut13} & 10.7\footref{hernandez05} & 6760 & 160\footref{gaia16} & 1.71 & 2.35 & 143\footref{alonsoalbi09} & 1300 & \dots & 90 & \dots & -4.08 & \dots & -0.61\footref{donehew11} & -7.97 \\
8 & PDS201 & \dots & Yes & A7\footref{vieira03} & 9.5\footref{thiswork} & 7850 & 168\footref{gaia16} & 1.76 & 1.76 & \dots & \dots & \dots & \dots & \dots & \dots & -1.95 & 0.14\footref{vandenancker96} & -7.36 \\
9 & HD56895 & \dots & Yes & F0\footref{malfait98} & 11.7\footref{thiswork} & 7500 & 167\footref{gaia16} & 1.80 & 1.92 & \dots & \dots & \dots & \dots & \dots & \dots & \dots & \dots & \dots \\
10 & HD 97048 & None & Yes & A0\footref{the94} & 36.2\footref{fairlamb15} & 10500 & 179\footref{gaia16} & 2.50 & 1.85 & 2230\footref{walsh16} & 867 & 403 & 736 & \dots & -2.81 & \dots & 0.86\footref{fairlamb17} & -6.77 \\
11 & HIP 54557 & \dots & Yes & B9\footref{malfait98} & 20.9\footref{thiswork} & 10700 & 184\footref{gaia16} & 2.61 & 1.95 &  &  & \dots & \dots & \dots & -3.35 & \dots & 0.11\footref{garcialopez06} & -7.51 \\
12 & HD 100453 & \dots & No & A9\footref{chen12spitzer} & 6.0\footref{fairlamb15} & 7250 & 103\footref{gaia16} & 1.54 & 1.79 & 182\footref{menard_inprep} & 1300 & \dots & 58 & \dots & \dots & -2.35 & -0.26\footref{fairlamb17} & -7.69 \\
13 & HD 100546 & Multiple & Yes & B9\footref{vieira03} & 24.7\footref{fairlamb15} & 9750 & 109\footref{gaia16} & 2.30 & 2.58 & 1240\footref{walsh14} & 867 & 230 & 118 & 0.56 & \dots & \dots & 0.66\footref{fairlamb15} & -6.78 \\
14 & HD104237 & \dots & Yes & A0\footref{the94} & 20.9\footref{fairlamb15} & 8000 & 104\footref{gaia16} & 1.97 & 2.28 & 66\footref{henning94} & 1300 & \dots & 14 & 0.7 & \dots & \dots & 0.61\footref{fairlamb15} & -6.82 \\
15 & PDS 141 & \dots & Yes & F0\footref{spezzi08} & 19.0\footref{spezzi08} & 7200 & 180 & 1.92 & 2.73 & 1470\footref{alcala08} & 1200 & \dots & 796 & \dots & \dots & -1.15 & 0.94\footref{spezzi08} & -6.40 \\
16 & PDS389 & \dots & Yes & A3\footref{vieira03} & 8.9\footref{vieira03} & 8710 & 175 & 1.96 & 1.87 & \dots & \dots & \dots & \dots & \dots & \dots & -1.84 & 0.25\footref{sartori10} & -7.27 \\
17 & SAO 206462 & $m$$\approx$2 & Yes & F4\footref{the94} & 8.8\footref{fairlamb15} & 6375 & 156\footref{gaia16} & 1.70 & 2.54 & 620\footref{andrews11} & 880 & 156 & 136 & -0.04 & \dots & \dots & 0.05\footref{fairlamb15} & -7.27 \\
18 & PDS 395 & \dots & Yes & A8\footref{vieira03} & 5.8\footref{fairlamb15} & 7750 & 131\footref{gaia16} & 1.74 & 1.71 & 242\footref{alonsoalbi09} & 1300 & \dots & 129 & -0.1 & \dots & \dots & -0.16\footref{fairlamb15} & -7.66 \\
19 & HD141569 & \dots & No & A0\footref{vieira03} & 18.8\footref{fairlamb15} & 9750 & 111\footref{gaia16} & 2.26 & 2.58 & 13\footref{white16} & 870 & 60 & 1 & -0.05 & \dots & \dots & -0.06\footref{fairlamb15} & -7.50 \\
20 & PDS 76 & \dots & Yes & A8\footref{vieira03} & 9.8\footref{fairlamb15} & 7500 & 150\footref{gaia16} & 1.78 & 1.78 & 313\footref{sandell11} & 850 & \dots & 65 & \dots & -3.63 & \dots & -0.22\footref{fairlamb17} & -7.72 \\
21 & HD 142527 & Multiple & Yes & F6\footref{malfait98} & 9.9\footref{fairlamb15} & 6500 & 156\footref{gaia16} & 1.70 & 2.52 & 3310\footref{boehler17} & 880 & 300 & 1126 & -0.09 & \dots & \dots & 0.01\footref{fairlamb15} & -7.32 \\
22 & PDS 78 & \dots & Yes & A8\footref{vieira03} & 11.0\footref{fairlamb15} & 7500 & 160\footref{gaia16} & 1.80 & 1.93 & 37\footref{alonsoalbi09} & 1300 & \dots & 23 & 0.02 & \dots & \dots & 0.02\footref{fairlamb15} & -7.44 \\
23 & HR 5999 & \dots & No & A7\footref{the94} & 59.6\footref{fairlamb15} & 8500 & 163\footref{gaia16} & 2.53 & 3.09 & 60\footref{sandell11} & 850 & \dots & 8 & 1.1 & \dots & \dots & 1.12\footref{fairlamb15} & -6.29 \\
24 & PDS 80 & \dots & Yes & A8\footref{vieira03} & 8.2\footref{fairlamb15} & 8000 & 149\footref{gaia16} & 1.76 & 1.75 & \dots & \dots & \dots & \dots & \dots & \dots & -2.34 & -0.25\footref{fairlamb17} & -7.75 \\
25 & HIP80425 & \dots & Yes & A1\footref{chen12spitzer} & 15.9\footref{thiswork} & 9200 & 152\footref{gaia16} & 2.39 & 2.17 & \dots & \dots & \dots & \dots & \dots & \dots & \dots & \dots & \dots \\
26 & IRS 48 & None & Yes & A0\footref{erickson11} & 20.4\footref{erickson11} & 9530 & 120 & 2.25 & 2.52 & 1000\footref{vandermarel13} & 441 & 60 & 12 & \dots & -3.35 & \dots & 0.11\footref{erickson11} & -7.34 \\
27 & WLY 1-53 & \dots & Yes & A7\footref{erickson11} & 17.4\footref{erickson11} & 7850 & 130 & 1.91 & 2.13 & \dots & \dots & \dots & \dots & \dots & -3.67 & \dots & -0.32\footref{greene95} & -7.76 \\
28 & HD148352 & \dots & Yes & F2\footref{erickson11} & 16.2\footref{erickson11} & 6890 & 130 & 1.90 & 2.70 & \dots & \dots & \dots & \dots & \dots & \dots & \dots & \dots & \dots \\
29 & HIP81474 & \dots & Yes & B9.5\footref{malfait98} & 78.1\footref{thiswork} & 10400 & 149\footref{gaia16} & 2.73 & 2.78 & \dots & \dots & \dots & \dots & \dots & -3.04 & \dots & 0.50\footref{garcialopez06} & -6.98 \\
30 & MWC863 & \dots & No & B9.5\footref{malfait98} & 23.8\footref{fairlamb15} & 9000 & 145 & 2.20 & 1.94 & 101\footref{sandell11} & 850 & \dots & 14 & 0.1 & \dots & \dots & 0.27\footref{fairlamb15} & -7.28 \\
31 & AK Sco & \dots & No & F5\footref{simbad} & 4.7\footref{fairlamb15} & 6250 & 143\footref{gaia16} & 1.43 & 1.72 & 33\footref{czekala15} & 1300 & 140 & 18 & \dots & -2.60 & \dots & 1.45\footref{fairlamb17} & -5.96 \\
32 & 51 Oph & \dots & No & A2\footref{malfait98} & 147.9\footref{thiswork} & 9700 & 124\footref{gaia16} & 3.20 & 3.75 & 5\footref{thi13} & 1200 & \dots & 1 & \dots & -1.99 & \dots & 1.87\footref{garcialopez06} & -5.56 \\
33 & HD 163296 & None & Yes & A2\footref{valenti03} & 23.4\footref{fairlamb15} & 9250 & 119\footref{gaia16} & 2.18 & 1.84 & 1820\footref{degregoriomonsalvo13} & 850 & 250 & 210 & 0.08 & \dots & \dots & 0.22\footref{fairlamb15} & -7.35 \\
34 & HD 169142 & None & Yes & F3\footref{murphy17} & 6.0\footref{murphy17} & 6700 & 117\footref{gaia16} & 1.51 & 1.90 & 232\footref{fedele17} & 1288 & 100 & 60 & \dots & -3.08 & \dots & 0.45\footref{garcialopez06} & -6.95 \\
35 & TY CrA & \dots & No & B9\footref{the94} & 37.2\footref{thiswork} & 10700 & 130 & 2.61 & 1.95 & \dots & \dots & \dots & \dots & \dots & -3.78 & \dots & -0.46\footref{finkenzeller84} & -8.08 \\
36 & T CrA & \dots & No & F0\footref{the94} & 1.6\footref{thiswork} & 7200 & 130 & 1.54 & 1.78 & \dots & \dots & \dots & \dots & \dots & -3.94 & \dots & -0.66\footref{finkenzeller84} & -8.09
\\* \bottomrule
\caption{The sample of Herbig IMS within 200 pc introduced in \S\ref{sec:herbig200pc}. Note that this list includes objects that are not studied in this paper, such as multiples.
Column (1): Source number, ordered by RA.
Column (2): Source name; RA and Dec available in Appendix Table~\ref{tab:radec}; notes on individual stars can be found in Appendix~\S\ref{sec:individual}.
Column (3): Type of Spiral; omitted if high sensitivity high resolution scattered light imaging unavailable. 
Column (4): Whether the source is included in our ``volume-limited Herbig IMS'' sample (\S\ref{sec:herbig200pc}). 
Column (5): Spectral Type.
Column (6): Stellar luminosity, rescaled by Gaia distance if available.
Column (7): Effective temperature; references omitted if the same as in Col.\ (6).
Column (8): Distance; references omitted if the same as in Col.\ (6); distances from \citet{gaia16} adopted if available. 
Column (9): Stellar mass.
Column (10): Stellar radius. Cols.\ (9) and (10) are derived from \citet{siess00} tracks and the listed $T_{\rm eff}$ and $L_\star.$ 
Column (11): Flux density of millimeter continuum emission. 
Column (12): Wavelength at which Col.\ (10) is measured. 
Column (13): Disk radius in $\sim$mm continuum emission; references omitted if the same as Col.\ (10).
Column (14): Dust mass from Eqn.~(\ref{eq:mdisk}); see Appendix \S\ref{sec:mdisk} for details.
Column (15): Accretion luminosity from veiling of the Balmer discontinuity; references listed in Col.\ (17).
Column (16): Br$\gamma$ emission line flux from literature; references listed in Col.\ (17).
Column (17): H$\alpha$ emission line flux from literature; references listed in Col.\ (17).
Column (18): Accretion luminosity adopted in this work; see Appendix \S\ref{sec:mdot} for details.
Column (19): Stellar accretion rate from Eqn.~(\ref{eq:mdot}) and Cols.\ (8), (9), and (17).
}
\label{tab:herbig}\\
\end{longtable}
\end{landscape}

References in Tables~\ref{tab:nirsamplegood} and \ref{tab:herbig}. 
\footnote{\label{andrews13}\citet{andrews13}}
\footnote{\label{andrews11}\citet{andrews11}}
\footnote{\label{fairlamb15}\citet{fairlamb15}}
\footnote{\label{erickson11}\citet{erickson11}}
\footnote{\label{murphy17}\citet{murphy17}}
\footnote{\label{vandermarel16whole}\citet{vandermarel16whole}}
\footnote{\label{kudo08}\citet{kudo08}}
\footnote{\label{canovas18}\citet{canovas18}}
\footnote{\label{pecaut13}Stellar temperature estimated from the spectral type using \citet{pecaut13}; luminosity calculated from the photometry reported in SIMBAD.}
\footnote{\label{cleeves16}\citet{cleeves16}}
\footnote{\label{valenti03}\citet{valenti03}}
\footnote{\label{beskrovnaya99}\citet{beskrovnaya99}}
\footnote{\label{kraus13}\citet{kraus13}}
\footnote{\label{the94}\citet{the94}}
\footnote{\label{simbad}SIMBAD}
\footnote{\label{gaia16}\citet{gaia16}}
\footnote{\label{oberg15}\citet{oberg15}}
\footnote{\label{kraus17}\citet{kraus17}}
\footnote{\label{walsh16}\citet{walsh16}}
\footnote{\label{walsh14}\citet{walsh14}}
\footnote{\label{boehler17}\citet{boehler17}}
\footnote{\label{degregoriomonsalvo13}\citet{degregoriomonsalvo13}}
\footnote{\label{fedele17}\citet{fedele17}}
\footnote{\label{momose10}\citet{momose10}}
\footnote{\label{andrews16}\citet{andrews16}}
\footnote{\label{carpenter05}\citet{carpenter05}}
\footnote{\label{hashimoto15}\citet{hashimoto15}}
\footnote{\label{zhang14}\citet{zhang14}}
\footnote{\label{rich15}\citet{rich15}}
\footnote{\label{tang12}\citet{tang12}}
\footnote{\label{brown12}\citet{brown12}}
\footnote{\label{isella09}\citet{isella09}}
\footnote{\label{dong17j1604}\citet{dong17j1604}}
\footnote{\label{rosenfeld13}\citet{rosenfeld13}}
\footnote{\label{malfait98}\citet{malfait98}}
\footnote{\label{walker88}\citet{walker88}}
\footnote{\label{vieira03}\citet{vieira03}}
\footnote{\label{chen12spitzer}\citet{chen12spitzer}}
\footnote{\label{spezzi08}\citet{spezzi08}}
\footnote{\label{thiswork}This work; see Appendix~\ref{sec:individual} for details.}
\footnote{\label{hernandez05}\citet{hernandez05}}
\footnote{\label{manoj06}\citet{manoj06}}
\footnote{\label{sandell11}\citet{sandell11}}
\footnote{\label{alonsoalbi09}\citet{alonsoalbi09}}
\footnote{\label{henning94}\citet{henning94}}
\footnote{\label{menard_inprep}Menard et al. in prep.}
\footnote{\label{alcala08}\citet{alcala08}}
\footnote{\label{white16}\citet{white16}}
\footnote{\label{vandermarel13}\citet{vandermarel13}}
\footnote{\label{czekala15}\citet{czekala15}}
\footnote{\label{thi13}\citet{thi13}}
\footnote{\label{valdes04}\citet{valdes04}}
\footnote{\label{hamidouche10}\citet{hamidouche10}}
\footnote{\label{hernandez04}\citet{hernandez04}}
\footnote{\label{donehew11}\citet{donehew11}}
\footnote{\label{dunkin98}\citet{dunkin98}}
\footnote{\label{vandenancker96}\citet{vandenancker96}}
\footnote{\label{fairlamb17}\citet{fairlamb17}}
\footnote{\label{garcialopez06}\citet{garcialopez06}}
\footnote{\label{sartori10}\citet{sartori10}}
\footnote{\label{greene95}\citet{greene95}}
\footnote{\label{finkenzeller84}\citet{finkenzeller84}}

\clearpage

\begin{deluxetable}{rc|cc|c}
\tablecaption{Incidence Rate of Spiral Arms in the ``Well-Studied NIR Disks'' Sample}
\tablehead{
	      & 
\colhead{All Stars} &  
\colhead{AB Stars} & 
\colhead{FGK Stars} & 
\colhead{Herbig IMS} 
}
\startdata
Two arms      & 4/29 
		& 1/7 
		& 3/22 
		& 2/11 
		\\
Multiple arms & 3/29 
		& 2/7 
		& 1/22 
		& 3/11 
		\\
One arm	      & 1/29           
		&               
		& 1/22 
		& 1/11 
		\\
All arm types & 8/29 
		& 3/7 
		& 5/22 
		& 6/11 
		\\
\enddata
\label{tab:armfractionnir}
\end{deluxetable}

\begin{deluxetable}{rccc}
\tablecaption{Incidence Rate of Spiral Arms in the ``Volume-limited Herbig IMS'' Sample}
\tablehead{
	      & 
\colhead{All Types of Arms} &  
\colhead{Two Arms}  &
\colhead{Multiple Arms} 
}
\startdata
Sources with arms            & 5           & 2    & 3  \\
Arm fraction, well-studied disks  & 5/10 
				& 2/10 
				& 3/10 
				\\
Arm fraction, all disks      & $\geq$ 5/24 
				& $\geq$ 2/24 
				& $\geq$ 3/24 
				\\
\enddata
\label{tab:armfractionherbig}
\end{deluxetable}

\begin{figure}
\begin{center}
\includegraphics[trim=40 130 70 160, clip,width=0.5\textwidth,angle=0]{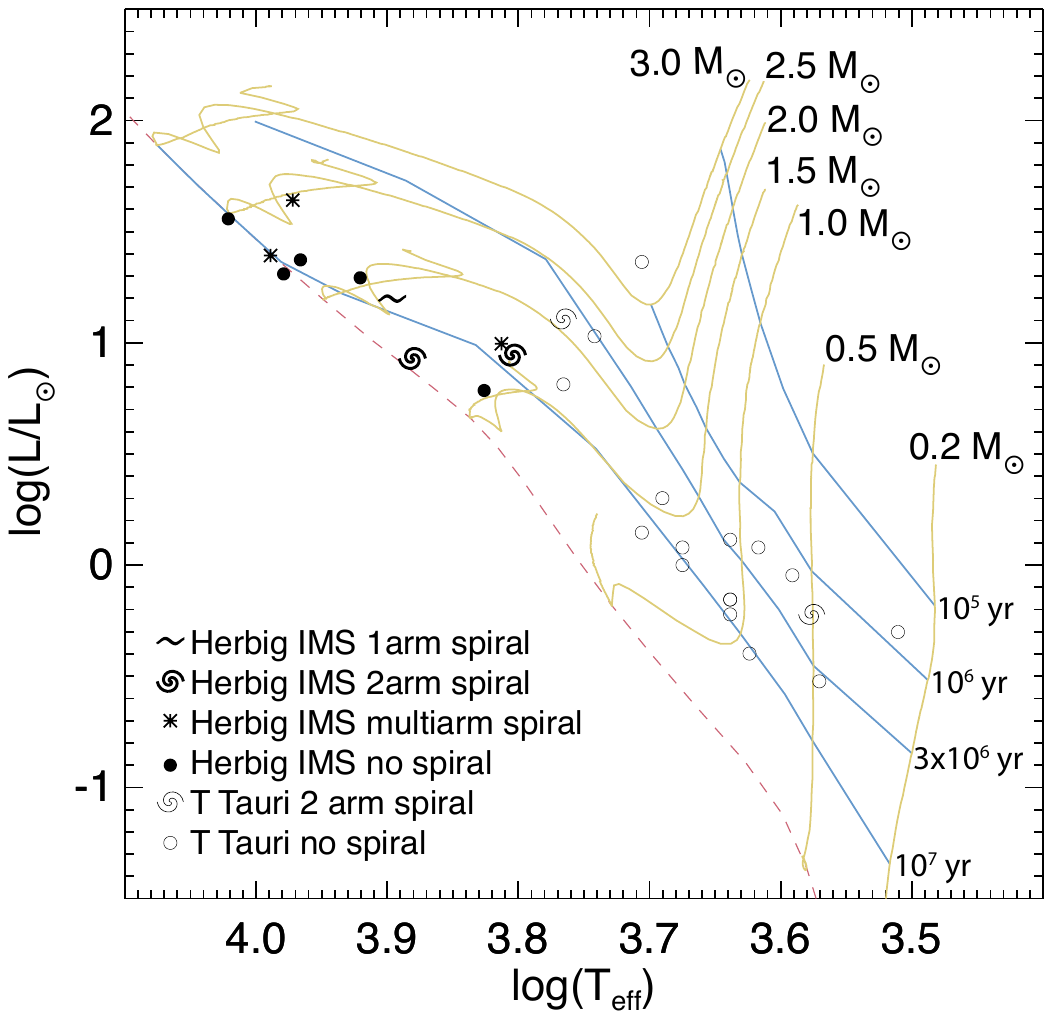}
\end{center}
\figcaption{HR diagram for young stars with disks that have 
been well-studied in 
scattered light. Disks with multi-arm spirals (asterisks), 
two-arm spirals (spiral symbol), one-arm spirals (twiddle symbol), 
and no arms (circles and dots) are shown for Herbig IMS (heavy symbols, dots) 
and T Tauri stars (light symbols, open circles). 
Evolutionary tracks and isochrones from \citet{siess00}
are also shown. 
\label{fig:HRdiagram}}
\end{figure}

\clearpage

\appendix
\renewcommand\thefigure{\thesection.\arabic{figure}}    
\setcounter{figure}{0}    
\renewcommand\thetable{\thesection.\arabic{table}}    
\setcounter{table}{0}    

\section{Scattered Light Disk Sample}\label{sec:app:nirsample}

\begin{deluxetable}{@{}lcllll@{}}
\tabletypesize{\scriptsize}
\tablecaption{Scattered Light Protoplanetary Disk Sample}
\tablehead{
\colhead{Target} & \colhead{Instrument} & \colhead{Wavelength} & \colhead{Intensity} & \colhead{Mode} & \colhead{Reference} \\
\colhead{(1)} & \colhead{(2)} & \colhead{(3)} & \colhead{(4)} & \colhead{(5)} & \colhead{(6)}
}
\startdata
AA Tau & HST / STIS & optical & total & Coro & \citet{cox13} \\
AB Aur & Subaru / CIAO & $H$ & total & Coro & \citet{fukagawa04} \\
\dots & HST / NICMOS & 1.1$\micron$, 2$\micron$ & pol, total & Coro & \citet{perrin09} \\
\dots & Subaru / HiCIAO & $H$ & pol & Coro & \citet{hashimoto11} \\
AK Sco & VLT / SPHERE & $Y$, $J$, $H$ & total & Coro & \citet{janson16} \\
DoAr 28 & Subaru / HiCIAO & $H$ & pol & DI & \citet{rich15} \\
DZ Cha & VLT / SPHERE & $J$ & pol & DI$+$Coro & \citet{canovas18} \\
FN Tau & Subaru / CIAO & $H$ & total & Coro & \citet{kudo08} \\
GG Tau & HST / NICMOS & 1um & pol & DI & \citet{silber00} \\
\dots & HST / WFPC2 & optical & total & DI & \citet{krist02} \\
\dots & HST / NICMOS & 1.1 to 2$\micron$ & total & DI & \citet{mccabe02} \\
\dots & Subaru / CIAO & $H$ & total & Coro & \citet{itoh02} \\
\dots & Keck / NIRC2 & $L^\prime$ & total & DI & \citet{duchene04} \\
\dots & HST / ACS & optical & total & DI & \citet{krist05} \\
\dots & Subaru / HiCIAO & $H$ & pol & Coro & \citet{itoh14} \\
\dots & Subaru / HiCIAO & $H$ & pol & DI & \citet{yang17} \\
GM Aur & Subaru / HiCIAO & $H$ & pol & DI & \citet{oh16gmaur} \\
HD 100453 & VLT / SPHERE & $Y$, $J$, $H$, $K$ & total & DI & \citet{wagner15hd100453} \\
\dots & VLT / SPHERE & optical, $J$ & pol & DI$+$Coro & \citet{benisty17} \\
\dots & Gemini / GPI & $Y$, $J$, $K_1$ & pol & DI & \citet{long17} \\
HD 100546 & HST / STIS & 1.1 to 2$\micron$ & total & Coro & \citet{grady01} \\
\dots & HST / NICMOS2 & $H$ & total & Coro & \citet{augereau01} \\
\dots & VLT / NACO & $H$, $K_{s}$ & pol & DI & \citet{quanz11} \\
\dots & Gemini / NICI & $K_{s}$ & total & Coro & \citet{boccaletti13} \\
\dots & VLT / NACO & $H$, $K_{s}$, $L^\prime$ & pol & DI & \citet{avenhaus14} \\
\dots & Gemini / NICI & $K_{s}$, $L^\prime$ & total & DI & \citet{currie14} \\
\dots & Gemini / GPI & $H$ & pol, total & DI$+$Coro & \citet{currie15} \\
\dots & VLT / SPHERE & optical, $H$, $K$ & pol, total & DI$+$Coro & \citet{garufi16} \\
\dots & Gemini / GPI, Magellan / MagAO & optical, $H$ & total & DI$+$Coro & \citet{rameau17} \\
\dots & Gemini / GPI, Magellan / MagAO & optical, $Y$, $H$ & pol, total & Coro & \citet{follette17} \\
HD 141569 & HST / NICMOS & 1.1 $\micron$ & total & Coro & \citet{weinberger99}  \\
\dots & HST / STIS & optical & total & Coro & \citet{mouillet01} \\
\dots & HST / ACS & optical & total & Coro & \citet{clampin03} \\
\dots & Gemini / NICI & $H$ & total & Coro & \citet{biller15} \\
\dots & HST / STIS & optical & total & Coro & \citet{konishi16} \\
\dots & Keck / NIRC2 & $L^\prime$ & total & DI & \citet{currie16hd141569} \\
\dots & VLT / SPHERE & $Y$, $J$, $H$, $K$ & total & Coro & \citet{perrot16} \\
HD 142527 & Subaru / CIAO & $H$, $K$ & total & Coro & \citet{fukagawa06} \\
\dots & Subaru / CIAO & 3$\micron$, $L^\prime$ & total & DI$+$Coro & \citet{honda09} \\
\dots & VLT / NACO & $L^\prime$ & total & DI & \citet{rameau12} \\
\dots & VLT / NACO & $H$, $K_{s}$ & pol & DI & \citet{canovas13} \\
\dots & Gemini / GPI & $Y$ & pol & DI & \citet{rodigas14} \\
\dots & VLT / NACO & $H$, $K_{s}$ & pol & DI & \citet{avenhaus14} \\
\dots & VLT / SPHERE & optical & pol & DI & \citet{mendigutia17hd100546} \\
\dots & VLT / SPHERE & optical & pol & DI & \citet{avenhaus17} \\
HD 150193 & Subaru / CIAO & $H$ & total & Coro & \citet{fukagawa03} \\
HD 163296 & VLT / NACO & $H$, $K_{s}$ & pol & DI & \citet{garufi14} \\
\dots & Gemini / GPI & $J$ & pol & Coro & \citet{monnier17} \\
\dots & VLT / NACO & $K_{s}$ & pol & DI & \citet{garufi17} \\
HD 169142 & HST / NICMOS & 1.1 $\micron$ & total & Coro & \citet{grady07} \\
\dots & VLT / NACO & $H$ & pol & DI & \citet{quanz13gap} \\
\dots & Subaru / HiCIAO & $H$ & pol & Coro & \citet{momose15} \\
\dots & VLT / SPHERE & optical, $Y$, $J$, $H$ & pol & Coro & \citet{pohl17hd169142} \\
\dots & Gemini / GPI & $J$ & pol & Coro & \citet{monnier17} \\
\dots & VLT / SPHERE & $Y$, $J$, $H$, $K$ & pol, total & DI$+$Coro & \citet{ligi18} \\
\dots & VLT / SPHERE & optical & pol & DI & \citet{bertrang18} \\
HD 97048 & VLT / NACO & $H$, $K_{s}$ & pol & DI & \citet{quanz12} \\
\dots & VLT / SPHERE & $J$, $H$ & pol, total & Coro & \citet{ginski16} \\
IM Lupi & HST / WFPC2 / NICMOS & optical, $H$ & total & DI$+$Coro & \citet{pinte08} \\
IRS 48 & Subaru / HiCIAO & $H$, $K_{s}$ & pol & DI & \citet{follette15} \\
J1604-2130 & Subaru / HiCIAO & $H$ & pol & DI & \citet{mayama12} \\
\dots & VLT / SPHERE & optical & pol & DI & \citet{pinilla15j1604} \\
\dots & VLT / SPHERE & $Y$, $J$, $H$, $K$ & total & Coro & \citet{canovas17} \\
LkCa 15 & Subaru / HiCIAO & $H$, $K_{s}$ & total & DI & \citet{thalmann10} \\
\dots & Gemini / NIRI & $K_{s}$ & total & DI & \citet{thalmann14} \\
\dots & VLT / SPHERE & optical & pol & DI & \citet{thalmann15} \\
\dots & Subaru / HiCIAO & $H$ & pol & DI & \citet{oh16lkca15} \\
\dots & VLT / SPHERE & $J$ & pol & DI$+$Coro & \citet{thalmann16} \\
LkH$\alpha$ 330 & Subaru / HiCIAO & $H$ & pol & DI & \citet{akiyama16} \\
\dots  & Subaru / HiCIAO & $H$, $K$ & pol & DI & \citet{uyama18} \\
MWC 480 & HST / NICMOS & 1.1, 1.6 $\micron$ & total & Coro & \citet{grady10} \\
\dots & Subaru / HiCIAO & $H$ & pol & Coro & \citet{kusakabe12} \\
MWC 758 & Subaru / CIAO & $H$ & pol & DI$+$Coro & \citet{grady13} \\
\dots & VLT / SPHERE & $Y$ & pol & Coro & \citet{benisty15} \\
\dots & Keck / NIRC2 & $L^\prime$ & total & Coro & \citet{reggiani18} \\
\dots & Keck / NIRC2 & $L^\prime$ & total & Coro & \citet{ren18} \\
PDS 66 & HST / STIS & 1.1$\micron$ & total & Coro & \citet{schneider14} \\
\dots & Gemini / GPI & $H$, $K_1$ & pol & Coro & \citet{wolff16} \\
PDS 70 & Subaru / HiCIAO & $H$ & pol & DI & \citet{hashimoto12} \\
RX J1615.3-3255 & Subaru / HiCIAO & $H$ & pol & DI & \citet{kooistra17} \\
\dots & VLT / SPHERE & optical, $Y$, $J$, $H$ & pol, total & DI$+$Coro & \citet{deboer16} \\
RY Lup & VLT / SPHERE & $Y$, $J$, $H$ & pol, total & DI$+$Coro & \citet{langlois18} \\
RY Tau & Subaru / HiCIAO & $H$ & pol & Coro & \citet{takami13} \\
SAO 206462 & HST / NICMOS & optical, 1.1$\micron$, 1.6 $\micron$ & total & DI$+$Coro & \citet{grady09} \\
\dots & Subaru / HiCIAO & $H$ & pol & DI & \citet{muto12} \\
\dots & VLT / NACO & $H$, $K_{s}$ & pol & DI & \citet{garufi13} \\
\dots & Gemini / GPI & $J$ & total & Coro & \citet{wahhaj15} \\
\dots & VLT / SPHERE & optical, $Y$, $J$ & pol & DI$+$Coro & \citet{stolker16sao206462} \\
\dots & VLT / SPHERE & $J$, $H$, $K$ & pol & Coro & \citet{stolker17} \\
\dots & VLT / SPHERE & $Y$, $J$, $H$, $K$ & total & Coro & \citet{maire17} \\
SR 21 & Subaru / HiCIAO & $H$ & pol & DI & \citet{follette13} \\
SU Aur & Subaru / HiCIAO & $H$ & pol & DI & \citet{deleon15} \\
Sz 91 & Subaru / HiCIAO & $K_{s}$ & pol & DI & \citet{tsukagoshi14} \\
T Cha & VLT / SPHERE & optical, $Y$, $J$, $H$ & pol, total & Coro & \citet{pohl17tcha} \\
T Tau & VLT / SPHERE & $J$, $H$ & total & Coro & \citet{kasper16} \\
TW Hya & HST / WFPC2 & optical & total & DI & \citet{krist00} \\
\dots & HST / NICMOS & 1.1$\micron$, 1.6$\micron$ & total & Coro & \citet{weinberger02} \\
\dots & HST / STIS & optical, 1.1-2.2 $\micron$ & total & Coro & \citet{debes13} \\
\dots & Subaru / HiCIAO & $H$ & pol & DI & \citet{akiyama15} \\
\dots & Gemini / GPI & $J$, $K_1$ & pol & Coro & \citet{rapson15twhya} \\
\dots & VLT / SPHERE & optical, $J$, $H$ & pol & DI$+$Coro & \citet{vanboekel17} \\
\dots & HST / STIS & optical & total & Coro & \citet{debes17} \\
UX Tau & Subaru / HiCIAO & $H$ & pol & Coro & \citet{tanii12} \\
V1247 Ori & Subaru / HiCIAO & $H$ & pol & Coro & \citet{ohta16} \\
V4046 Sgr & Gemini / GPI & $J$, $K_2$ & pol & Coro & \citet{rapson15v4046} \\
 \hline
 &  &  &  &  &  \\
\multicolumn{6}{c}{Edge On Sources (Central Source Invisible)}\\
 &  &  &  &  &  \\
HH 30 & HST / WFPC2 & optical & total & DI & \citet{burrows96} \\
\dots & HST / WFPC2 & optical & total & DI & \citet{stapelfeldt99} \\
\dots & HST / NICMOS & 1-2 $\micron$ & total & DI & \citet{cotera01} \\
\dots & HST / WFPC2 & optical & total & DI & \citet{watson04} \\
HK Tau & HST / WFPC2 & optical & total & DI & \citet{stapelfeldt98} \\
HV Tau C & HST / WFPC2 & optical & total & DI & \citet{stapelfeldt03} \\
LkH$\alpha$ 263C & Gemini / Hokupa & $H$ & total & DI & \citet{jayawardhana02} \\
PDS 144 N & Keck / NIRC2 & $H$, $K$, $L^\prime$ & total & DI & \citet{perrin06} \\
 \hline
 &  &  &  &  &  \\
\multicolumn{6}{c}{Partially Embedded / Class 0 / Class I Sources}\\
 &  &  &  &  &  \\
FU Ori & Subaru / HiCIAO & $H$ & pol & Coro & \citet{liu16} \\
HL Tau & Subaru / CIAO & $J$, $H$, $K$ & pol & DI & \citet{lucas04} \\
\dots & Subaru / CIAO & $J$, $H$, $K$ & pol & DI & \citet{murakawa08} \\
R Mon & HST / WFPC2 & $J$, $H$, $K$ & pol, total & DI & \citet{close97} \\
V1057 Cyg & Subaru / HiCIAO & $H$ & pol & Coro & \citet{liu16} \\
V1735 Cyg & Subaru / HiCIAO & $H$ & pol & Coro & \citet{liu16} \\
Z CMa & VLT / SPHERE & $H$, $K$ & pol & DI & \citet{canovas15zcma} \\
\dots & VLT / SPHERE & optical & total & DI & \citet{antoniucci16} \\
\dots & Subaru / HiCIAO & $H$ & pol & Coro & \citet{liu16} \\
\enddata
\tablecomments{
Column (2): the instrument used in each observation, including STIS, NICMOS, ACS, and WFPC2 on HST, CIAO \citep{tamura00} and HiCIAO \citep{tamura06} on Subaru, NICI \citep{chun08} and GPI \citep{macintosh08} on Gemini, NACO \citep{lenzen03} and SPHERE \citep{beuzit08} on VLT, NIRC2 on Keck, and MagAO on Magellan \citep{morzinski14}. Column (4): ``pol'' -- polarized intensity; ``total'' -- total intensity. Column (5): observing mode; ``Coro'' -- coronagraphic imaging; ``DI'' -- direct imaging with no coronagraph. See \S\ref{sec:sample} for details.}
\label{tab:nirsample}
\end{deluxetable}
\clearpage

\section{Disk Mass Measurements}\label{sec:mdisk}

The direct measurement of the mass of an accretion disk around a young star is almost impossible. Most of the disk mass is comprised of molecular hydrogen which lacks a dipole moment and whose lowest energy emission line has an upper state energy level of 510~K \citep{mandy93}. Thus emission lines from this molecule only trace relatively warm gas. Additionally, disks are very optically thick. For a disk that is 1\% of the minimum mass solar nebula, the surface density at 1~au is $\sim$20 g cm$^{-2}$ \citep{hayashi81}. For a solar abundance and \citet{mathis77} dust distribution, the disk is optically thick at the mid-infrared wavelengths at which molecular hydrogen emits. Thus only the optically thin atmosphere can be observed. HD is a promising gas tracer owing to its lower upper energy state, far infrared emission lines, and allowed dipole transitions \citep{bergin13, mcclure16}, however, the conversion to a total gas mass is still model dependent. Numerous other gas tracers have been explored for measuring disk masses, but these values are also model dependent and have not been independently validated \citep[e.g., CO;][]{williams14}. 

Because of the difficulty in measuring gas in disks, the most common avenue to estimate $\mdisk$ is to convert the observed continuum flux density, $f_\nu$, at a submillimeter frequency $\nu$, to the dust (solids) mass, $\mdust$, then scale $\mdust$ by a constant gas-to-dust mass ratio $\xi$, often assumed to be 100:1, to obtain the gas disk mass, $\mgas$:
\begin{equation}
\mdisk\approx\mgas=\xi\mdust=\xi\frac{f_\nu d^2}{B_\nu(\tdust) \ \kappa_\nu},
\label{eq:mdisk}
\end{equation}
where $d$ is the distance to the source, $B_\nu(\tdust)$ is Planck function at the dust temperature $\tdust$, and $\kappa_\nu$ is the dust opacity. Unfortunately, the midplane temperature of the dust has not been measured directly, the opacity of the dust is poorly constrained \citep{draine06}, massive disks may be optically thick in mm continuum observations (e.g., \citealt{evans17}), and the gas to dust ratio in the disk has not been robustly measured (cf. \citealt{forgan13l1527}, \citealt{eisner16}, and \citealt{eisner18}). With these caveats in mind, we have calculated an estimate of the disk mass following this approach. 

We provide a submillimeter flux-based $\mdisk$ estimate (Eqn~\ref{eq:mdisk}) for the targets listed in Tables~\ref{tab:nirsamplegood} and \ref{tab:herbig} in the following steps. First, we obtain the submillimeter flux density of the sources $f_\nu$ from literature (Columns 8 and 10 in Tables~\ref{tab:nirsamplegood} and \ref{tab:herbig}, respectively). We adopt measurements at $\lambda=880$ $\micron$ whenever possible. For all ``Well studied NIR disks'' we are able to find flux density measurements at wavelengths within a factor of 2 of 880$\micron$.
Next, we convert all submillimeter flux density measurements to the flux density at 880 $\micron$, assuming a spectral index $\alpha=2.4$ \citep{andrews13}. For the dust opacity at 880 $\micron$, we adopt $\kappa=3$ cm$^2$ g$^{-1}$ \citep[e.g., ][]{beckwith90}. 

For the dust temperature $\tdust$, \citet{andrews05} advocated $\tdust\approx20$K for T Tauri disks, and \citet{henning94} advocated 50~K for Herbigs. \citet{sandell11} performed simple model fitting to the SED of a sample of Herbig disks and found $\tdust\sim$ 30$-$70 K. \citet{andrews13} pointed out the $\tdust$ dependence on the stellar luminosity $L_\star$, and suggested scaling $\tdust$ as $L_\star^{1/4}$. 
\citet{pascucci16} suggested that the size of the dust 
disk should also be taken into consideration when estimating $\tdust$.

Here, we introduce another correction factor on $\tdust$ based on the outer radius of the disk in submillimeter continuum observations, $R_{\rm mm}$ (Column 12 in Table~\ref{tab:herbig}), as where the dust is emitting is also important --- a dust ring further from the star is colder than a dust ring closer in, everything else being equal. We scale $\tdust$ as 
\begin{equation}
T_{\rm dust} = 30\left( \frac{L_\star}{38L_\odot} \right)^{1/4} \left(\frac{0.5R_{\rm mm}}{100\rm AU} \right)^{-1/2},
\label{eq:tdust}
\end{equation}
The factor of 0.5 in $0.5R_{\rm mm}$ is an approximate correction factor meant to identify the radius where the dust temperature is the global average. This is to account for the fact that dust at different radii has different temperature. If the surface density follows a $\Sigma\sim1/R$ radial profile and the disk extends from $R=0$ to $R=R_{\rm mm}$, $0.5R_{\rm mm}$ is the half-mass radius, i.e., the radius inside which half of the disk mass is enclosed. Equation~\ref{eq:tdust} gives the specific temperature at $R=0.5R_{\rm mm}$, not $R_{\rm mm}$, in a disk; this temperature is considered as a proxy for the average dust temperature in a disk whose outer edge is at $R_{\rm mm}$.
For unresolved submillimeter sources, we adopted the average $R_{\rm mm}$ in resolved sources: 191 AU for Herbig disks and 144 AU for T Tauri disks. This relation is calibrated against a model for HD~163296 \citep{facchini17}, and reproduces the midplane temperature produced by several radiative transfer models of both T Tauri and Herbig disks in the literature to within 40\% \citep{dalessio98, dullemond04shadowing, dong15shadow, andrews16}.

Another simple scaling for estimating the disk mass comes from the stellar accretion rate \citep{hartmann98}. The stellar accretion rate declines roughly as 
$t^{-3/2}$; thus the mass remaining in the disk is $M_{\rm disk} = 2\mdot t_{\rm age}$. 
Because stellar accretion rates are time variable 
by a factor of $\sim$3, the mass estimate of an individual 
disk is uncertain by at least this factor.
However, for an unbiased ensemble of disks, the mean mass of the 
disks is a reasonable estimate. 
For targets in our volume-limited Herbig IMS sample (Table~\ref{tab:herbig}), we conservatively estimate $t_{\rm age}$ to be at least
2 Myr based on the HR diagram (Figure~\ref{fig:HRdiagram}).

\section{Accretion Rate of Herbig Stars in Table~\ref{tab:herbig}}\label{sec:mdot}

Measuring the accretion rate on to Herbig Ae/Be (HAeBe) stars is complicated by the fact that the stellar photosphere is bright at ultraviolet wavelengths. Thus the contrast between the accretion luminosity and stellar photosphere is low \citep[e.g.,][]{muzerolle04}. To address this challenge, several groups have calibrated the veiling of the Balmer jump against various emission line diagnostics for stars with the greatest ultraviolet contrast. In an X-Shooter survey of 91 Herbig Ae/Be stars, \citet{fairlamb15, fairlamb17} have calibrated the luminosity of 32 emission lines against the accretion luminosity as inferred from the veiling of the Balmer jump. 

We recalculated the accretion rates for all sources to obtain a self-consistent set of values. We adopted the accretion luminosity for our sources from the measurement of the Balmer discontinuity when this measurement was available as it is the most direct measure of stellar accretion. When the Balmer discontinuity was not available, we used the luminosity of Br$\gamma$ or H$\alpha$ ($L_{\rm Br\gamma}$, $L_{\rm H\alpha}$) to calculate the accretion luminosity \citep[Table 2]{fairlamb17}:
\begin{align}
\log{\lacc}=(4.46\pm 0.23)+(1.30\pm 0.09)\log{L_{\rm Br\gamma}},\\
\log{\lacc}=(2.09\pm 0.06)+(1.00\pm 0.05)\log{L_{\rm H\alpha}}.
\end{align}
We prioritized Br$\gamma$ over H$\alpha$ because it is less sensitive to 
corrections for both reddening and photospheric absorption. 

There are two caveats to keep in mind concerning the measurement of the stellar accretion rate of HAeBes. Firstly, it is not exactly clear how HAeBes accrete. While young A/B stars do not have the convective envelopes believed to play a key role in sustaining the kG magnetic fields observed among their later type counterparts, there is some evidence of magnetospheric accretion \citep{guimaraes06, cauley14}. If magnetospheric accretion controls the star/disk interface, it is likely through higher order field components. However, the conversion of the veiling of the Balmer discontinuity to an accretion luminosity relies on an assumption about the energy in the accretion column ($F$) and the filling factor ($f$; \citealt{muzerolle04}). Neither is well constrained directly for HAeBes. Adopting values that reproduce the observations for T Tauris stars may not be appropriate for these higher mass analogs. Secondly, early type stars with metal poor photospheres, in particular the $\lambda$~Boo stars, show a UV excess \citep{murphy17}. \citet{folsom12} and \citet{kama15} found a relatively high fraction of HAeBes in their samples (up to $\sim$50\%) showing the $\lambda$~Boo pattern to varying degrees. If this effect is common among the targets in our sample, it is possible that $\mdot$ have been systematically overestimated. Future work to account for the uncertainty in $F$ and $f$, as well as the effect of the photospheric abundance, will improve our understanding of the stellar accretion rate of these systems.

The luminosities collected from the literature were rescaled using the distances available from \citet{gaia16} when they were available. Conversion of equivalent width measurements of the HI lines to line luminosities requires correction for the veiled equivalent width of photospheric absorption, 
\begin{equation}
W_{\rm circ} = W_{\rm obs} - W_{\rm phot}10^{-0.4\Delta m}
\end{equation}
where $W_{\rm obs}$ is the equivalent width of the emission line, $W_{\rm phot}$ is the equivalent width of the photospheric line, $W_{\rm circ}$ is the equivalent width of the circumstellar emission, and $\Delta m$ is the excess broadband emission above the stellar photosphere \citep{rodgers01}. The equivalent widths of the photospheric features were taken from \citet{fairlamb17} when available. For stars not included in their study, stars with matching temperatures and gravities were used to infer the photospheric equivalent width. The veiling in the K-band was determined by the dereddened color excess, $E(V-Ks)$. The intrinsic colors were taken from \citet{pecaut13}. For the sources where H$\alpha$ was used, the extinction at 6563$\AA$ was inferred from $E(B-V)$, $R_V$=5, and the reddening law presented in \citet{mathis90}. 

The stellar accretion rate is proportional the accretion luminosity so that,
\begin{equation}
\dot{M} = \frac{\lacc R_\star} {GM_\star},
\label{eq:mdot}
\end{equation}
where $R_\star$ is the stellar radius and $M_\star$ is the stellar mass. The stellar parameters were taken from the \citet{siess00} pre-main sequence evolutionary tracks and tabulated in Table~\ref{tab:herbig}.

\section{Comment on Individual Systems}\label{sec:individual}
Stars for which the determination of the stellar parameters were not straight forward are discussed in detail. We also comment on the scattered light imaging observations of the AB~Aur and HD~100546 disks.

\paragraph{AB Aur} ALMA revealed a two-arm spiral structure inside
$0\farcs 4$ in the AB Aur disk in CO emission \citep{tang17}.
This structure was however not well revealed in NIR scattered light
\citep{hashimoto11}. We conservatively consider the disk to
have only multiple weak arms on $\sim 1\arcsec$ scale in scattered light.

\paragraph{HR~811} This star was included in a survey of the SED of Herbig Ae/Be stars by \citet{malfait98}. The SED only reveals an infrared excess longward of 20$\micron$ leading them to conclude that this is a Vega-type system. \citet{folsom12} included this source in their survey of abundances among Herbig Ae/Be stars, noted the lack of strong optical emission lines, and found that the star had a solar abundance pattern (see also \citealt{fossati09}). We infer the luminosity for this source from the published photometry \citep{hog00} adopting the distance from \citet{gaia16} and arrive at $L_{\star}$= 87$L_{\sun}$. This value for the luminosity places the star below the early-MS (127~$L_{\sun}$), thus we adopt the stellar mass and radius for a star of this temperature on the zero age main sequence (ZAMS). 

\paragraph{V892~Tau} There is not parallax measurement for this heavily extinguished source, thus we adopt the average distance to Taurus \citep{vieira03}. We adopted the effective temperature (T$\sim 11,000$~K). and luminosity ($L_{\star}$=40~$L_{\sun}$) from \citet{hernandez04}, and found that the star is slightly below the ZAMS (49~$L\sun$). Given the uncertainty in the distance and reddening correction, this is consistent with a location on the ZAMS, and the mass and radius we adopt for this star reflect that position.

\paragraph{MWC~758} To calculate the accretion rate of this source, we adopted the observed equivalent width of Br$\gamma$, stellar temperature, and luminosity reported in the literature \citep[respectively]{donehew11, beskrovnaya99, andrews11}. We rescaled the luminosity reported by \citet{andrews11} from the distance adopted in that work (200~pc) to the updated distance provided by \citet{gaia16} (151~pc). 

\paragraph{PDS~201} We infer the luminosity for this source from the published photometry \citep{ducati02} adopting the distance from \citet{gaia16} and arrive at $L_{\star}$= 9.5$L_{\sun}$.

\paragraph{HD~56895} We infer the luminosity for this source from the published photometry \citep{hog00} adopting the distance from \citet{gaia16} and arrive at $L_{\star}$= 11.7$L_{\sun}$.

\paragraph{HD~97048} We adopt the effective temperature determined by \citet{fairlamb15} and rescale the luminosity using the distance provided by \citet{gaia16}. We find $L_{\star}$=36~$L_{\sun}$ which falls below the zero age main sequence (ZAMS) for a temperature of 10,500~K for the Siess evolutionary tracks. We assume the temperature is the more reliable parameter, so we scale the luminosity so that the star falls on the ZAMS. We adopt $M_{\star}$=2.5$M_{\sun}$ and $R_{\star}$=1.8$R_{\sun}$.

\paragraph{HIP 54557} We infer the luminosity for this source from the published photometry \citep{hog00} adopting the distance from \citet{gaia16} and arrive at $L_{\star}$= 20.9$L_{\sun}$. This source is under luminous by a factor of two. The spectral type was taken from \citet{houk75}. They assign a quality flag of 2 (one a 1-4 scale where 1 is the best).  This rating indicates that ``the spectra may be slightly under or over-exposed, or slightly overlapped". It is not clear what uncertainty is appropriate for the inferred stellar temperature. To remain consistent with our other targets, we adopt the stellar mass and radius for a source on the ZAMS. 

\paragraph{HD~100546} We adopt the effective temperature determined by \citet{fairlamb15} and rescale the stellar luminosity using the updated distance provided by \citet{gaia16}. We find $L_{\star}$=25~$L_{\sun}$ which falls below the ZAMS for a temperature of 9,750~K for the Siess evolutionary tracks. We assume the temperature is the more reliable parameter, so we scale the luminosity so that the star falls on the ZAMS. We adopt $M_{\star}$=2.3$M_{\sun}$ and $R_{\star}$=2.6$R_{\sun}$.
\citet{follette17}
showed that processing a total intensity image of a
modestly inclined disk with a small number of spiral arms with
Angular Differential Imaging (ADI) techniques may increase the number
of arms by breaking them into pieces. The arms seen in the HD 100546
disk may be susceptible to this effect.

\paragraph{PDS~389} There is no parallax measurement reported for this source, so we adopt the distance used by \citet{vieira03} based on its cluster membership (d=175~pc). We find $L_{\star}$=9~$L_{\sun}$ which falls below the ZAMS (14~$L_{\sun}$) for a star with this effective temperature (8700~K). The reddening for this source is significant (B-V=1.87; \citealt{vieira03}), and uncertainty in the reddening correction likely accounts for our underestimate. We adopt the stellar mass and radius for a source on the ZAMS. 

\paragraph{PDS~395} We adopt the effective temperature from \citet{fairlamb15}, and rescale the luminosity based on the updated distance found from \citet{gaia16}. This star falls 30\% below the ZAMS. The origin of this discrepancy is not clear. 

\paragraph{HD~141569} Historically, HD~141569 has been included among Herbig Ae/Be stars \citep[e.g.,][]{the94}. It shows \ion{H}{1} emission lines, has an extended disk in scattered light \citep[e.g.,][]{weinberger99}, and a MIR excess \citep[e.g.,][]{malfait98}. However this IR excess is with a factor of a few of $\beta$~Pic \citep{clampin03} indicating that it may straddle the transition from gas rich accretion disk to debris disk \citep[e.g.,][]{mawet17}. This makes the stellar accretion rate inferred from the H$\alpha$ emission, $10^{-8} M_{\rm \sun} \rm yr^{-1}$ surprising \citep{mendigutia17}. One solution to this puzzle is comes from the fact that HD~141569 is rotating near its break up velocity and the \ion{H}{1} emission is double peaked \citep{brittain07}. This situation is consistent with a picture of the \ion{H}{1} emission arising from a decretion disk rather than accretion disk. Because the dust disk around the star is optically thin, stellar radiation may sculpt the morphology of the dust leading to the observed rings and spiral structure \citep[e.g.,][]{richert17}. Thus, we exclude this star from our sample. 

\paragraph{PDS~80} We adopt the effective temperature from \citet{fairlamb15}, and rescale the luminosity they report based on the updated distance found from \citet{gaia16}. This star falls  15\% below the ZAMS. The origin of this discrepancy is not clear. 

\paragraph{HIP~80452} We infer the stellar temperature from the reported spectral type and the luminosity from reported photometry \citep{hog00}. The reddening indicates that this source lies behind 1~mag of extinction. We find  $L_{\star}$=16~$L_{\sun}$ which is 16~\% below the ZAMS. This discrepancy is likely due to uncertainty in the conversion from a color excess to extinction measure. We adopt the stellar parameters for a star on the ZAMS.

\paragraph{IRS~48} For the effective temperature and stellar luminosity adopted for this source, we find that it falls far below the ZAMS. The extinction towards this source is quite high, (A$_{\rm V}$=11.5; \citealt{brown12}), so the uncertainty in the extinction correction dominates the uncertainty in the stellar luminosity. We assume the the effective temperature is correct and adopt the stellar mass and radius that places this star on the ZAMS. 

\paragraph{HIP~81474} We infer the luminosity for this source from the published photometry \citep{hog00} adopting the distance from \citet{gaia16} and arrive at $L_{\star}$= 78.1~$L_{\sun}$.

\paragraph{51~Oph} We infer the luminosity for this source from the published photometry \citep{ducati02} adopting the distance from \citet{gaia16} and arrive at $L_{\star}$= 147.9$L_{\sun}$. 51~Oph is a rapidly rotating A0 star \citep{jamialahmadi15} with a minimal FIR excess \citep{thi13}. It is not clear whether this small excess is due to a very flat and thus cold disk or a very compact ($R\rm _{out}\sim 10-15$~au disk; \citealt{thi13}). Interferometric observations of 51~Oph indicate that the photosphere is highly flattened due to the large rotational velocity of the star and that  H$\alpha$ arises from a geometrically thin Keplerian disk. As it is possible that this star is a post-Main Sequence object with a decretion disk rather than accretion disk, we exclude it from our sample. 

\paragraph{HD~163296} We adopt the effective temperature from \citet{fairlamb15}, and rescale the luminosity they report based on the updated distance found from \citet{gaia16}.

\paragraph{HD~169142} The spectral type designation of this source has ranged from B9-mid-F in the literature. This is a face-on star rotating near its break-up velocity (Grady et al. 2007) resulting in a significant deformation of the star and resultant temperature gradient from the stellar pole to the equatorial region of the star. As part of their study of $\lambda$~Boo stars, \citet{murphy17} find T=6700$\pm$322~K and log $L_{\star}$/$L_{\sun}$=0.78$\pm$0.03. We adopt these values to estimate the stellar mass and radius from the \citet{siess00} evolutionary tracks.  

\paragraph{TY~CrA} This is a heavily extinguished B9 star (A$\rm _V$=1.9-3.0; \citealt{ducati02}). The star falls 14\% below the ZAMS which can be accounted for by uncertainty in the reddening correction. We adopt the stellar parameters for a star of this temperature on the ZAMS. 

\paragraph{T CrA} There is no parallax measurement of this source, so we adopt the distance to this star forming region (130~pc; \citealt{vieira03}). 
The spectral type is adopted from \citet{the94}. The uncertainty in the spectral type and luminosity is uncertain. Additionally, the proper reddening correction is uncertain. The star is highly variable, thus we select the photometry for which the B and V were acquired contemporaneously \citep{zacharias12} and arrive at B-V=0.975. For a spectral type of F0, this implies E(B-V)=0.66. The extinction ranges from 2.1 - 3.3 for $R\rm_V$=3.1-5.0. For a temperature of 7200 K, the stellar luminosity is underestimated by a factor of 4. The origin of this discrepancy is unclear, though it is likely due to underestimating the extinction to the source. We adopt the stellar parameters for this star assuming that it falls on the ZAMS. 

\section{Coordinates of Sources}

\begin{longtable}[c]{@{}llll@{}}
\toprule
\#    & Object          & RA          & Dec          \\* \midrule
\endfirsthead
\multicolumn{4}{c}%
{{\bfseries Table \thetable\ continued from previous page}} \\
\toprule
\#    & Object          & RA          & Dec          \\* \midrule
\endhead
\bottomrule
\endfoot
\endlastfoot
1     & HR811           & 02 44 07.35 & -13 51 31.31 \\
\dots & LkH$\alpha$ 330        & 03 45 48.28 & +32 24 11.87 \\
\dots & FN Tau          & 04 14 14.59 & +28 27 58.06 \\
2     & V892 Tau        & 04 18 40.62 & +28 19 15.51 \\
\dots & RY Tau          & 04 21 57.41 & +28 26 35.54 \\
\dots & UX Tau          & 04 30 04.00 & +18 13 49.44 \\
\dots & LkCa 15         & 04 39 17.80 & +22 21 03.48 \\
\dots & GM Aur          & 04 55 10.98 & +30 21 59.54 \\
3     & AB Aur          & 04 55 45.85 & +30 33 04.29 \\
\dots & SU Aur          & 04 55 59.39 & +30 34 01.50 \\
4     & MWC 480         & 04 58 46.27 & +29 50 36.99 \\
5     & PDS 178         & 05 24 01.17 & +24 57 37.58 \\
6     & MWC 758         & 05 30 27.53 & +25 19 57.08 \\
7     & CQ Tau          & 05 35 58.47 & +24 44 54.09 \\
\dots & V1247 Ori       & 05 38 05.25 & -01 15 21.70 \\
8     & PDS 201          & 05 44 18.79 & +00 08 40.40 \\
9     & HD 56895         & 07 18 31.79 & -11 11 34.94 \\
\dots & TW Hya          & 11 01 51.91 & -34 42 17.03 \\
10    & HD 97048        & 11 08 03.31 & -77 39 17.49 \\
11    & HIP 54557       & 11 09 50.02 & -76 36 47.72 \\
12    & HD 100453        & 11 33 05.58 & -54 19 28.55 \\
13    & HD 100546       & 11 33 25.44 & -70 11 41.24 \\
\dots & DZ Cha          & 11 49 31.84 & -78 51 01 1  \\
14    & HD 104237       & 12 00 05.09 & -78 11 34.57 \\
15    & PDS 141         & 12 53 17.22 & -77 07 10.61 \\
\dots & PDS 66          & 13 22 07.54 & -69 38 12.22 \\
\dots & PDS 70          & 14 08 10.15 & -41 23 52.5  \\
16    & PDS 389          & 15 14 47.05 & -62 16 59.73 \\
17    & SAO 206462      & 15 15 48.45 & -37 09 16.04 \\
18    & PDS 395         & 15 40 46.38 & -42 29 53.54 \\
19    & HD 141569        & 15 49 57.75 & -03 55 16.34 \\
\dots & IM Lupi         & 15 56 09.18 & -37 56 06.12 \\
20    & PDS 76          & 15 56 40.02 & -22 01 40.00 \\
21    & HD 142527 A     & 15 56 41.89 & -42 19 23.27 \\
\dots & J1604-2130      & 16 04 21.66 & -21 30 28.4  \\
22    & PDS 78          & 16 06 57.95 & -27 43 09.79 \\
\dots & Sz 91           & 16 07 11.59 & -39 03 47.54 \\
23    & HR 5999         & 16 08 34.29 & -39 06 18.33 \\
24    & PDS 80          & 16 13 11.59 & -22 29 06.62 \\
\dots & RX J1615.3-3255 & 16 15 20.23 & -32 55 05.10 \\
25    & HIP 80425        & 16 24 59.15 & -25 21 17.98 \\
\dots & DoAr 28         & 16 26 47.42 & -23 14 52.2  \\
\dots & SR 21           & 16 27 10.28 & -24 19 12.74 \\
26    & IRS 48          & 16 27 37.19 & -24 30 35.03 \\
27    & WLY 1-53        & 16 27 49.87 & -24 25 40.2  \\
28    & HD 148352        & 16 28 25.16 & -24 45 01.00 \\
29    & HIP 81474        & 16 38 28.65 & -18 13 13.71 \\
30    & MWC 863          & 16 40 17.92 & -23 53 45.18 \\
31    & AK Sco          & 16 54 44.85 & -36 53 18.56 \\
32    & 51 Oph       & 17 31 24.95 & -23 57 45.51 \\
33    & HD 163296       & 17 56 21.29 & -21 57 21.87 \\
\dots & V4046 Sgr       & 18 14 10.47 & -32 47 34.50 \\
34    & HD 169142       & 18 24 29.78 & -29 46 49.33 \\
35    & TY CrA          & 19 01 40.83 & -36 52 33.88 \\
36    & T CrA           & 19 01 58.79 & -36 57 49.93 \\* \bottomrule
\caption{SIMBAD coordinates (ICRS coord. (ep=J2000)) of the sources in Tables~\ref{tab:nirsamplegood} and \ref{tab:herbig}, ranked by RA. The first column is the number of Herbig objects in Table~\ref{tab:herbig}, if applicable.}
\label{tab:radec}\\
\end{longtable}


\begin{thebibliography}{}
\expandafter\ifx\csname natexlab\endcsname\relax\def\natexlab#1{#1}\fi

\bibitem[{{Akiyama} {et~al.}(2015){Akiyama}, {Muto}, {Kusakabe}, {Kataoka},
  {Hashimoto}, {Tsukagoshi}, {Kwon}, {Kudo}, {Kandori}, {Grady}, {Takami},
  {Janson}, {Kuzuhara}, {Henning}, {Sitko}, {Carson}, {Mayama}, {Currie},
  {Thalmann}, {Wisniewski}, {Momose}, {Ohashi}, {Abe}, {Brandner}, {Brandt},
  {Egner}, {Feldt}, {Goto}, {Guyon}, {Hayano}, {Hayashi}, {Hayashi}, {Hodapp},
  {Ishi}, {Iye}, {Knapp}, {Matsuo}, {Mcelwain}, {Miyama}, {Morino},
  {Moro-Martin}, {Nishimura}, {Pyo}, {Serabyn}, {Suenaga}, {Suto}, {Suzuki},
  {Takahashi}, {Takato}, {Terada}, {Tomono}, {Turner}, {Watanabe}, {Yamada},
  {Takami}, {Usuda}, \& {Tamura}}]{akiyama15}
{Akiyama}, E., {Muto}, T., {Kusakabe}, N., {et~al.} 2015, \apjl, 802, L17

\bibitem[{{Akiyama} {et~al.}(2016){Akiyama}, {Hashimoto}, {Liu}, {Li},
  {Bonnefoy}, {Dong}, {Hasegawa}, {Henning}, {Sitko}, {Janson}, {Feldt},
  {Wisniewski}, {Kudo}, {Kusakabe}, {Tsukagoshi}, {Momose}, {Muto}, {Taki},
  {Kuzuhara}, {Satoshi}, {Takami}, {Ohashi}, {Grady}, {Kwon}, {Thalmann},
  {Abe}, {Brandner}, {Brandt}, {Carson}, {Egner}, {Goto}, {Guyon}, {Hayano},
  {Hayashi}, {Hayashi}, {Hodapp}, {Ishii}, {Iye}, {Knapp}, {Kandori}, {Matsuo},
  {Mcelwain}, {Miyama}, {Morino}, {Moro-Martin}, {Nishimura}, {Pyo}, {Serabyn},
  {Suenaga}, {Suto}, {Suzuki}, {Takahashi}, {Takato}, {Terada}, {Tomono},
  {Turner}, {Watanabe}, {Yamada}, {Takami}, {Usuda}, \& {Tamura}}]{akiyama16}
{Akiyama}, E., {Hashimoto}, J., {Liu}, H.~B., {et~al.} 2016, \aj, 152, 222

\bibitem[{{Alcal{\'a}} {et~al.}(2008){Alcal{\'a}}, {Spezzi}, {Chapman},
  {Evans}, {Huard}, {J{\o}rgensen}, {Mer{\'{\i}}n}, {Stapelfeldt}, {Covino},
  {Frasca}, {Gandolfi}, \& {Oliveira}}]{alcala08}
{Alcal{\'a}}, J.~M., {Spezzi}, L., {Chapman}, N., {et~al.} 2008, \apj, 676, 427

\bibitem[{{ALMA Partnership} {et~al.}(2015){ALMA Partnership}, {Brogan},
  {P{\'e}rez}, {Hunter}, {Dent}, {Hales}, {Hills}, {Corder}, {Fomalont},
  {Vlahakis}, {Asaki}, {Barkats}, {Hirota}, {Hodge}, {Impellizzeri}, {Kneissl},
  {Liuzzo}, {Lucas}, {Marcelino}, {Matsushita}, {Nakanishi}, {Phillips},
  {Richards}, {Toledo}, {Aladro}, {Broguiere}, {Cortes}, {Cortes}, {Espada},
  {Galarza}, {Garcia-Appadoo}, {Guzman-Ramirez}, {Humphreys}, {Jung}, {Kameno},
  {Laing}, {Leon}, {Marconi}, {Mignano}, {Nikolic}, {Nyman}, {Radiszcz},
  {Remijan}, {Rod{\'o}n}, {Sawada}, {Takahashi}, {Tilanus}, {Vila Vilaro},
  {Watson}, {Wiklind}, {Akiyama}, {Chapillon}, {de Gregorio-Monsalvo}, {Di
  Francesco}, {Gueth}, {Kawamura}, {Lee}, {Nguyen Luong}, {Mangum}, {Pietu},
  {Sanhueza}, {Saigo}, {Takakuwa}, {Ubach}, {van Kempen}, {Wootten},
  {Castro-Carrizo}, {Francke}, {Gallardo}, {Garcia}, {Gonzalez}, {Hill},
  {Kaminski}, {Kurono}, {Liu}, {Lopez}, {Morales}, {Plarre}, {Schieven},
  {Testi}, {Videla}, {Villard}, {Andreani}, {Hibbard}, \&
  {Tatematsu}}]{brogan15}
{ALMA Partnership}, {Brogan}, C.~L., {P{\'e}rez}, L.~M., {et~al.} 2015, \apjl,
  808, L3

\bibitem[{{Alonso-Albi} {et~al.}(2009){Alonso-Albi}, {Fuente}, {Bachiller},
  {Neri}, {Planesas}, {Testi}, {Bern{\'e}}, \& {Joblin}}]{alonsoalbi09}
{Alonso-Albi}, T., {Fuente}, A., {Bachiller}, R., {et~al.} 2009, \aap, 497, 117

\bibitem[{{Andrews} {et~al.}(2013){Andrews}, {Rosenfeld}, {Kraus}, \&
  {Wilner}}]{andrews13}
{Andrews}, S.~M., {Rosenfeld}, K.~A., {Kraus}, A.~L., \& {Wilner}, D.~J. 2013,
  \apj, 771, 129

\bibitem[{{Andrews} \& {Williams}(2005)}]{andrews05}
{Andrews}, S.~M., \& {Williams}, J.~P. 2005, \apj, 631, 1134

\bibitem[{{Andrews} \& {Williams}(2007)}]{andrews07}
---. 2007, \apj, 671, 1800

\bibitem[{{Andrews} {et~al.}(2011){Andrews}, {Wilner}, {Espaillat}, {Hughes},
  {Dullemond}, {McClure}, {Qi}, \& {Brown}}]{andrews11}
{Andrews}, S.~M., {Wilner}, D.~J., {Espaillat}, C., {et~al.} 2011, \apj, 732,
  42

\bibitem[{{Andrews} {et~al.}(2016){Andrews}, {Wilner}, {Zhu}, {Birnstiel},
  {Carpenter}, {P{\'e}rez}, {Bai}, {{\"O}berg}, {Hughes}, {Isella}, \&
  {Ricci}}]{andrews16}
{Andrews}, S.~M., {Wilner}, D.~J., {Zhu}, Z., {et~al.} 2016, \apjl, 820, L40

\bibitem[{{Antoniucci} {et~al.}(2016){Antoniucci}, {Podio}, {Nisini},
  {Bacciotti}, {Lagadec}, {Sissa}, {La Camera}, {Giannini}, {Schmid},
  {Gratton}, {Turatto}, {Desidera}, {Bonnefoy}, {Chauvin}, {Dougados},
  {Bazzon}, {Thalmann}, \& {Langlois}}]{antoniucci16}
{Antoniucci}, S., {Podio}, L., {Nisini}, B., {et~al.} 2016, \aap, 593, L13

\bibitem[{{Augereau} {et~al.}(2001){Augereau}, {Lagrange}, {Mouillet}, \&
  {M{\'e}nard}}]{augereau01}
{Augereau}, J.~C., {Lagrange}, A.~M., {Mouillet}, D., \& {M{\'e}nard}, F. 2001,
  \aap, 365, 78

\bibitem[{{Avenhaus} {et~al.}(2014){Avenhaus}, {Quanz}, {Schmid}, {Meyer},
  {Garufi}, {Wolf}, \& {Dominik}}]{avenhaus14}
{Avenhaus}, H., {Quanz}, S.~P., {Schmid}, H.~M., {et~al.} 2014, \apj, 781, 87

\bibitem[{{Avenhaus} {et~al.}(2017){Avenhaus}, {Quanz}, {Schmid}, {Dominik},
  {Stolker}, {Ginski}, {de Boer}, {Szul{\'a}gyi}, {Garufi}, {Zurlo},
  {Hagelberg}, {Benisty}, {Henning}, {M{\'e}nard}, {Meyer}, {Baruffolo},
  {Bazzon}, {Beuzit}, {Costille}, {Dohlen}, {Girard}, {Gisler}, {Kasper},
  {Mouillet}, {Pragt}, {Roelfsema}, {Salasnich}, \& {Sauvage}}]{avenhaus17}
---. 2017, \aj, 154, 33

\bibitem[{{Avenhaus} {et~al.}(2018){Avenhaus}, {Quanz}, {Garufi}, {Perez},
  {Casassus}, {Pinte}, {Bertrang}, {Caceres}, {Benisty}, \&
  {Dominik}}]{avenhaus18}
{Avenhaus}, H., {Quanz}, S.~P., {Garufi}, A., {et~al.} 2018, ArXiv e-prints,
  arXiv:1803.10882

\bibitem[{{Bae} \& {Zhu}(2018)}]{bae18theory}
{Bae}, J., \& {Zhu}, Z. 2018, \apj, 859, 118

\bibitem[{{Bae} {et~al.}(2016){Bae}, {Zhu}, \& {Hartmann}}]{bae16sao206462}
{Bae}, J., {Zhu}, Z., \& {Hartmann}, L. 2016, \apj, 819, 134

\bibitem[{{Bae} {et~al.}(2017){Bae}, {Zhu}, \& {Hartmann}}]{bae17}
---. 2017, \apj, 850, 201

\bibitem[{{Bai}(2017)}]{bai17}
{Bai}, X.-N. 2017, \apj, 845, 75

\bibitem[{{Bai} {et~al.}(2016){Bai}, {Ye}, {Goodman}, \& {Yuan}}]{bai16wind}
{Bai}, X.-N., {Ye}, J., {Goodman}, J., \& {Yuan}, F. 2016, \apj, 818, 152

\bibitem[{{Balbus} \& {Hawley}(1992)}]{balbus92}
{Balbus}, S.~A., \& {Hawley}, J.~F. 1992, \apj, 400, 610

\bibitem[{{Baraffe} {et~al.}(2003){Baraffe}, {Chabrier}, {Barman}, {Allard}, \&
  {Hauschildt}}]{baraffe03}
{Baraffe}, I., {Chabrier}, G., {Barman}, T.~S., {Allard}, F., \& {Hauschildt},
  P.~H. 2003, \aap, 402, 701

\bibitem[{{Beckwith} {et~al.}(1990){Beckwith}, {Sargent}, {Chini}, \&
  {Guesten}}]{beckwith90}
{Beckwith}, S.~V.~W., {Sargent}, A.~I., {Chini}, R.~S., \& {Guesten}, R. 1990,
  \aj, 99, 924

\bibitem[{{Benisty} {et~al.}(2015){Benisty}, {Juhasz}, {Boccaletti},
  {Avenhaus}, {Milli}, {Thalmann}, {Dominik}, {Pinilla}, {Buenzli}, {Pohl},
  {Beuzit}, {Birnstiel}, {de Boer}, {Bonnefoy}, {Chauvin}, {Christiaens},
  {Garufi}, {Grady}, {Henning}, {Huelamo}, {Isella}, {Langlois}, {M{\'e}nard},
  {Mouillet}, {Olofsson}, {Pantin}, {Pinte}, \& {Pueyo}}]{benisty15}
{Benisty}, M., {Juhasz}, A., {Boccaletti}, A., {et~al.} 2015, \aap, 578, L6

\bibitem[{{Benisty} {et~al.}(2017){Benisty}, {Stolker}, {Pohl}, {de Boer},
  {Lesur}, {Dominik}, {Dullemond}, {Langlois}, {Min}, {Wagner}, {Henning},
  {Juhasz}, {Pinilla}, {Facchini}, {Apai}, {van Boekel}, {Garufi}, {Ginski},
  {M{\'e}nard}, {Pinte}, {Quanz}, {Zurlo}, {Boccaletti}, {Bonnefoy}, {Beuzit},
  {Chauvin}, {Cudel}, {Desidera}, {Feldt}, {Fontanive}, {Gratton}, {Kasper},
  {Lagrange}, {LeCoroller}, {Mouillet}, {Mesa}, {Sissa}, {Vigan}, {Antichi},
  {Buey}, {Fusco}, {Gisler}, {Llored}, {Magnard}, {Moeller-Nilsson}, {Pragt},
  {Roelfsema}, {Sauvage}, \& {Wildi}}]{benisty17}
{Benisty}, M., {Stolker}, T., {Pohl}, A., {et~al.} 2017, \aap, 597, A42

\bibitem[{{Bergin} {et~al.}(2013){Bergin}, {Cleeves}, {Gorti}, {Zhang},
  {Blake}, {Green}, {Andrews}, {Evans}, {Henning}, {{\"O}berg}, {Pontoppidan},
  {Qi}, {Salyk}, \& {van Dishoeck}}]{bergin13}
{Bergin}, E.~A., {Cleeves}, L.~I., {Gorti}, U., {et~al.} 2013, \nat, 493, 644

\bibitem[{{Bertrang} {et~al.}(2018){Bertrang}, {Avenhaus}, {Casassus},
  {Montesinos}, {Kirchschlager}, {Perez}, {Cieza}, \& {Wolf}}]{bertrang18}
{Bertrang}, G.~H.-M., {Avenhaus}, H., {Casassus}, S., {et~al.} 2018, \mnras,
  474, 5105

\bibitem[{{Beskrovnaya} {et~al.}(1999){Beskrovnaya}, {Pogodin},
  {Miroshnichenko}, {Th{\'e}}, {Savanov}, {Shakhovskoy}, {Rostopchina},
  {Kozlova}, \& {Kuratov}}]{beskrovnaya99}
{Beskrovnaya}, N.~G., {Pogodin}, M.~A., {Miroshnichenko}, A.~S., {et~al.} 1999,
  \aap, 343, 163

\bibitem[{{Beuzit} {et~al.}(2008){Beuzit}, {Feldt}, {Dohlen}, {Mouillet},
  {Puget}, {Wildi}, {Abe}, {Antichi}, {Baruffolo}, {Baudoz}, {Boccaletti},
  {Carbillet}, {Charton}, {Claudi}, {Downing}, {Fabron}, {Feautrier},
  {Fedrigo}, {Fusco}, {Gach}, {Gratton}, {Henning}, {Hubin}, {Joos}, {Kasper},
  {Langlois}, {Lenzen}, {Moutou}, {Pavlov}, {Petit}, {Pragt}, {Rabou}, {Rigal},
  {Roelfsema}, {Rousset}, {Saisse}, {Schmid}, {Stadler}, {Thalmann}, {Turatto},
  {Udry}, {Vakili}, \& {Waters}}]{beuzit08}
{Beuzit}, J.-L., {Feldt}, M., {Dohlen}, K., {et~al.} 2008, in Society of
  Photo-Optical Instrumentation Engineers (SPIE) Conference Series, Vol. 7014,
  Society of Photo-Optical Instrumentation Engineers (SPIE) Conference Series

\bibitem[{{Biller} {et~al.}(2015){Biller}, {Liu}, {Rice}, {Wahhaj}, {Nielsen},
  {Hayward}, {Kuchner}, {Close}, {Chun}, {Ftaclas}, \& {Toomey}}]{biller15}
{Biller}, B.~A., {Liu}, M.~C., {Rice}, K., {et~al.} 2015, \mnras, 450, 4446

\bibitem[{{Boccaletti} {et~al.}(2013){Boccaletti}, {Pantin}, {Lagrange},
  {Augereau}, {Meheut}, \& {Quanz}}]{boccaletti13}
{Boccaletti}, A., {Pantin}, E., {Lagrange}, A.-M., {et~al.} 2013, \aap, 560,
  A20

\bibitem[{{Boehler} {et~al.}(2017){Boehler}, {Weaver}, {Isella}, {Ricci},
  {Grady}, {Carpenter}, \& {Perez}}]{boehler17}
{Boehler}, Y., {Weaver}, E., {Isella}, A., {et~al.} 2017, \apj, 840, 60

\bibitem[{{Bowler}(2016)}]{bowler16}
{Bowler}, B.~P. 2016, \pasp, 128, 102001

\bibitem[{{Brittain} {et~al.}(2007){Brittain}, {Simon}, {Najita}, \&
  {Rettig}}]{brittain07}
{Brittain}, S.~D., {Simon}, T., {Najita}, J.~R., \& {Rettig}, T.~W. 2007, \apj,
  659, 685

\bibitem[{{Brown} {et~al.}(2012){Brown}, {Rosenfeld}, {Andrews}, {Wilner}, \&
  {van Dishoeck}}]{brown12}
{Brown}, J.~M., {Rosenfeld}, K.~A., {Andrews}, S.~M., {Wilner}, D.~J., \& {van
  Dishoeck}, E.~F. 2012, \apjl, 758, L30

\bibitem[{{Burrows} {et~al.}(1996){Burrows}, {Stapelfeldt}, {Watson}, {Krist},
  {Ballester}, {Clarke}, {Crisp}, {Gallagher}, {Griffiths}, {Hester},
  {Hoessel}, {Holtzman}, {Mould}, {Scowen}, {Trauger}, \&
  {Westphal}}]{burrows96}
{Burrows}, C.~J., {Stapelfeldt}, K.~R., {Watson}, A.~M., {et~al.} 1996, \apj,
  473, 437

\bibitem[{{Canovas} {et~al.}(2013){Canovas}, {M{\'e}nard}, {Hales},
  {Jord{\'a}n}, {Schreiber}, {Casassus}, {Gledhill}, \& {Pinte}}]{canovas13}
{Canovas}, H., {M{\'e}nard}, F., {Hales}, A., {et~al.} 2013, \aap, 556, A123

\bibitem[{{Canovas} {et~al.}(2015){Canovas}, {Perez}, {Dougados}, {de Boer},
  {M{\'e}nard}, {Casassus}, {Schreiber}, {Cieza}, {Caceres}, \&
  {Girard}}]{canovas15zcma}
{Canovas}, H., {Perez}, S., {Dougados}, C., {et~al.} 2015, \aap, 578, L1

\bibitem[{{Canovas} {et~al.}(2017){Canovas}, {Hardy}, {Zurlo}, {Wahhaj},
  {Schreiber}, {Vigan}, {Villaver}, {Olofsson}, {Meeus}, {M{\'e}nard},
  {Caceres}, {Cieza}, \& {Garufi}}]{canovas17}
{Canovas}, H., {Hardy}, A., {Zurlo}, A., {et~al.} 2017, \aap, 598, A43

\bibitem[{{Canovas} {et~al.}(2018){Canovas}, {Montesinos}, {Schreiber},
  {Cieza}, {Eiroa}, {Meeus}, {de Boer}, {M{\'e}nard}, {Wahhaj},
  {Riviere-Marichalar}, {Olofsson}, {Garufi}, {Rebollido}, {van Holstein},
  {Caceres}, {Hardy}, \& {Villaver}}]{canovas18}
{Canovas}, H., {Montesinos}, B., {Schreiber}, M.~R., {et~al.} 2018, \aap, 610,
  A13

\bibitem[{{Carpenter} {et~al.}(2005){Carpenter}, {Wolf}, {Schreyer},
  {Launhardt}, \& {Henning}}]{carpenter05}
{Carpenter}, J.~M., {Wolf}, S., {Schreyer}, K., {Launhardt}, R., \& {Henning},
  T. 2005, \aj, 129, 1049

\bibitem[{{Cauley} \& {Johns-Krull}(2014)}]{cauley14}
{Cauley}, P.~W., \& {Johns-Krull}, C.~M. 2014, \apj, 797, 112

\bibitem[{{Chabrier} {et~al.}(2000){Chabrier}, {Baraffe}, {Allard}, \&
  {Hauschildt}}]{chabrier00}
{Chabrier}, G., {Baraffe}, I., {Allard}, F., \& {Hauschildt}, P. 2000, \apj,
  542, 464

\bibitem[{{Chen} {et~al.}(2012){Chen}, {Pecaut}, {Mamajek}, {Su}, \&
  {Bitner}}]{chen12spitzer}
{Chen}, C.~H., {Pecaut}, M., {Mamajek}, E.~E., {Su}, K.~Y.~L., \& {Bitner}, M.
  2012, \apj, 756, 133

\bibitem[{{Chun} {et~al.}(2008){Chun}, {Toomey}, {Wahhaj}, {Biller}, {Artigau},
  {Hayward}, {Liu}, {Close}, {Hartung}, {Rigaut}, \& {Ftaclas}}]{chun08}
{Chun}, M., {Toomey}, D., {Wahhaj}, Z., {et~al.} 2008, in \procspie, Vol. 7015,
  Adaptive Optics Systems, 70151V

\bibitem[{{Cieza} {et~al.}(2016){Cieza}, {Casassus}, {Tobin}, {Bos},
  {Williams}, {Perez}, {Zhu}, {Caceres}, {Canovas}, {Dunham}, {Hales},
  {Prieto}, {Principe}, {Schreiber}, {Ruiz-Rodriguez}, \& {Zurlo}}]{cieza16}
{Cieza}, L.~A., {Casassus}, S., {Tobin}, J., {et~al.} 2016, \nat, 535, 258

\bibitem[{{Cieza} {et~al.}(2017){Cieza}, {Casassus}, {P{\'e}rez}, {Hales},
  {C{\'a}rcamo}, {Ansdell}, {Avenhaus}, {Bayo}, {Bertrang}, {C{\'a}novas},
  {Christiaens}, {Dent}, {Ferrero}, {Gamen}, {Olofsson}, {Orcajo}, {Osses},
  {Pe{\~n}a-Ramirez}, {Principe}, {Ru{\'{\i}}z-Rodr{\'{\i}}guez}, {Schreiber},
  {van der Plas}, {Williams}, \& {Zurlo}}]{cieza17}
{Cieza}, L.~A., {Casassus}, S., {P{\'e}rez}, S., {et~al.} 2017, \apjl, 851, L23

\bibitem[{{Clampin} {et~al.}(2003){Clampin}, {Krist}, {Ardila}, {Golimowski},
  {Hartig}, {Ford}, {Illingworth}, {Bartko}, {Ben{\'{\i}}tez}, {Blakeslee},
  {Bouwens}, {Broadhurst}, {Brown}, {Burrows}, {Cheng}, {Cross}, {Feldman},
  {Franx}, {Gronwall}, {Infante}, {Kimble}, {Lesser}, {Martel}, {Menanteau},
  {Meurer}, {Miley}, {Postman}, {Rosati}, {Sirianni}, {Sparks}, {Tran},
  {Tsvetanov}, {White}, \& {Zheng}}]{clampin03}
{Clampin}, M., {Krist}, J.~E., {Ardila}, D.~R., {et~al.} 2003, \aj, 126, 385

\bibitem[{{Cleeves} {et~al.}(2016){Cleeves}, {{\"O}berg}, {Wilner}, {Huang},
  {Loomis}, {Andrews}, \& {Czekala}}]{cleeves16}
{Cleeves}, L.~I., {{\"O}berg}, K.~I., {Wilner}, D.~J., {et~al.} 2016, \apj,
  832, 110

\bibitem[{{Close} {et~al.}(1997){Close}, {Roddier}, {Hora}, {Graves},
  {Northcott}, {Roddier}, {Hoffman}, {Dayal}, {Fazio}, \& {Deutsch}}]{close97}
{Close}, L.~M., {Roddier}, F., {Hora}, J.~L., {et~al.} 1997, \apj, 489, 210

\bibitem[{{Cossins} {et~al.}(2009){Cossins}, {Lodato}, \& {Clarke}}]{cossins09}
{Cossins}, P., {Lodato}, G., \& {Clarke}, C.~J. 2009, \mnras, 393, 1157

\bibitem[{{Cotera} {et~al.}(2001){Cotera}, {Whitney}, {Young}, {Wolff}, {Wood},
  {Povich}, {Schneider}, {Rieke}, \& {Thompson}}]{cotera01}
{Cotera}, A.~S., {Whitney}, B.~A., {Young}, E., {et~al.} 2001, \apj, 556, 958

\bibitem[{{Cox} {et~al.}(2013){Cox}, {Grady}, {Hammel}, {Hornbeck}, {Russell},
  {Sitko}, \& {Woodgate}}]{cox13}
{Cox}, A.~W., {Grady}, C.~A., {Hammel}, H.~B., {et~al.} 2013, \apj, 762, 40

\bibitem[{{Currie} {et~al.}(2015){Currie}, {Cloutier}, {Brittain}, {Grady},
  {Burrows}, {Muto}, {Kenyon}, \& {Kuchner}}]{currie15}
{Currie}, T., {Cloutier}, R., {Brittain}, S., {et~al.} 2015, \apjl, 814, L27

\bibitem[{{Currie} {et~al.}(2014){Currie}, {Muto}, {Kudo}, {Honda}, {Brandt},
  {Grady}, {Fukagawa}, {Burrows}, {Janson}, {Kuzuhara}, {McElwain}, {Follette},
  {Hashimoto}, {Henning}, {Kandori}, {Kusakabe}, {Kwon}, {Mede}, {Morino},
  {Nishikawa}, {Pyo}, {Serabyn}, {Suenaga}, {Takahashi}, {Wisniewski}, \&
  {Tamura}}]{currie14}
{Currie}, T., {Muto}, T., {Kudo}, T., {et~al.} 2014, \apjl, 796, L30

\bibitem[{{Currie} {et~al.}(2016){Currie}, {Grady}, {Cloutier}, {Konishi},
  {Stassun}, {Debes}, {van der Marel}, {Muto}, {Jayawardhana}, \&
  {Ratzka}}]{currie16hd141569}
{Currie}, T., {Grady}, C.~A., {Cloutier}, R., {et~al.} 2016, \apjl, 819, L26

\bibitem[{{Czekala} {et~al.}(2015){Czekala}, {Andrews}, {Jensen}, {Stassun},
  {Torres}, \& {Wilner}}]{czekala15}
{Czekala}, I., {Andrews}, S.~M., {Jensen}, E.~L.~N., {et~al.} 2015, \apj, 806,
  154

\bibitem[{{D'Alessio} {et~al.}(1998){D'Alessio}, {Cant{\"o}}, {Calvet}, \&
  {Lizano}}]{dalessio98}
{D'Alessio}, P., {Cant{\"o}}, J., {Calvet}, N., \& {Lizano}, S. 1998, \apj,
  500, 411

\bibitem[{{de Boer} {et~al.}(2016){de Boer}, {Salter}, {Benisty}, {Vigan},
  {Boccaletti}, {Pinilla}, {Ginski}, {Juhasz}, {Maire}, {Messina}, {Desidera},
  {Cheetham}, {Girard}, {Wahhaj}, {Langlois}, {Bonnefoy}, {Beuzit}, {Buenzli},
  {Chauvin}, {Dominik}, {Feldt}, {Gratton}, {Hagelberg}, {Isella}, {Janson},
  {Keller}, {Lagrange}, {Lannier}, {Menard}, {Mesa}, {Mouillet}, {Mugrauer},
  {Peretti}, {Perrot}, {Sissa}, {Snik}, {Vogt}, {Zurlo}, \& {SPHERE
  Consortium}}]{deboer16}
{de Boer}, J., {Salter}, G., {Benisty}, M., {et~al.} 2016, \aap, 595, A114

\bibitem[{{de Gregorio-Monsalvo} {et~al.}(2013){de Gregorio-Monsalvo},
  {M{\'e}nard}, {Dent}, {Pinte}, {L{\'o}pez}, {Klaassen}, {Hales},
  {Cort{\'e}s}, {Rawlings}, {Tachihara}, {Testi}, {Takahashi}, {Chapillon},
  {Mathews}, {Juhasz}, {Akiyama}, {Higuchi}, {Saito}, {Nyman}, {Phillips},
  {Rod{\'o}n}, {Corder}, \& {Van Kempen}}]{degregoriomonsalvo13}
{de Gregorio-Monsalvo}, I., {M{\'e}nard}, F., {Dent}, W., {et~al.} 2013, \aap,
  557, A133

\bibitem[{{de Leon} {et~al.}(2015){de Leon}, {Takami}, {Karr}, {Hashimoto},
  {Kudo}, {Sitko}, {Mayama}, {Kusakabe}, {Akiyama}, {Liu}, {Usuda}, {Abe},
  {Brandner}, {Brandt}, {Carson}, {Currie}, {Egner}, {Feldt}, {Follette},
  {Grady}, {Goto}, {Guyon}, {Hayano}, {Hayashi}, {Hayashi}, {Henning},
  {Hodapp}, {Ishii}, {Iye}, {Janson}, {Kandori}, {Knapp}, {Kuzuhara}, {Kwon},
  {Matsuo}, {McElwain}, {Miyama}, {Morino}, {Moro-Martin}, {Nishimura}, {Pyo},
  {Serabyn}, {Suenaga}, {Suto}, {Suzuki}, {Takahashi}, {Takato}, {Terada},
  {Thalmann}, {Tomono}, {Turner}, {Watanabe}, {Wisniewski}, {Yamada}, {Takami},
  \& {Tamura}}]{deleon15}
{de Leon}, J., {Takami}, M., {Karr}, J.~L., {et~al.} 2015, \apjl, 806, L10

\bibitem[{{Debes} {et~al.}(2013){Debes}, {Jang-Condell}, {Weinberger},
  {Roberge}, \& {Schneider}}]{debes13}
{Debes}, J.~H., {Jang-Condell}, H., {Weinberger}, A.~J., {Roberge}, A., \&
  {Schneider}, G. 2013, \apj, 771, 45

\bibitem[{{Debes} {et~al.}(2017){Debes}, {Poteet}, {Jang-Condell}, {Gaspar},
  {Hines}, {Kastner}, {Pueyo}, {Rapson}, {Roberge}, {Schneider}, \&
  {Weinberger}}]{debes17}
{Debes}, J.~H., {Poteet}, C.~A., {Jang-Condell}, H., {et~al.} 2017, \apj, 835,
  205

\bibitem[{{Dipierro} {et~al.}(2015){Dipierro}, {Price}, {Laibe}, {Hirsh},
  {Cerioli}, \& {Lodato}}]{dipierro15hltau}
{Dipierro}, G., {Price}, D., {Laibe}, G., {et~al.} 2015, \mnras, 453, L73

\bibitem[{{Dipierro} {et~al.}(2018){Dipierro}, {Ricci}, {P{\'e}rez}, {Lodato},
  {Alexander}, {Laibe}, {Andrews}, {Carpenter}, {Chandler}, {Greaves}, {Hall},
  {Henning}, {Kwon}, {Linz}, {Mundy}, {Sargent}, {Tazzari}, {Testi}, \&
  {Wilner}}]{dipierro18}
{Dipierro}, G., {Ricci}, L., {P{\'e}rez}, L., {et~al.} 2018, \mnras, 475, 5296

\bibitem[{{Donehew} \& {Brittain}(2011)}]{donehew11}
{Donehew}, B., \& {Brittain}, S. 2011, \aj, 141, 46

\bibitem[{{Dong}(2015)}]{dong15shadow}
{Dong}, R. 2015, \apj, 810, 6

\bibitem[{{Dong} \& {Dawson}(2016)}]{dong16td}
{Dong}, R., \& {Dawson}, R. 2016, \apj, 825, 77

\bibitem[{{Dong} \& {Fung}(2017)}]{dong17spiralarm}
{Dong}, R., \& {Fung}, J. 2017, \apj, 835, 38

\bibitem[{{Dong} {et~al.}(2016{\natexlab{a}}){Dong}, {Fung}, \&
  {Chiang}}]{dong16armviewing}
{Dong}, R., {Fung}, J., \& {Chiang}, E. 2016{\natexlab{a}}, \apj, 826, 75

\bibitem[{{Dong} {et~al.}(2015{\natexlab{a}}){Dong}, {Hall}, {Rice}, \&
  {Chiang}}]{dong15giarm}
{Dong}, R., {Hall}, C., {Rice}, K., \& {Chiang}, E. 2015{\natexlab{a}}, \apjl,
  812, L32

\bibitem[{{Dong} {et~al.}(2017{\natexlab{a}}){Dong}, {Li}, {Chiang}, \&
  {Li}}]{dong17doublegap}
{Dong}, R., {Li}, S., {Chiang}, E., \& {Li}, H. 2017{\natexlab{a}}, \apj, 843,
  127

\bibitem[{{Dong} {et~al.}(2016{\natexlab{b}}){Dong}, {Vorobyov},
  {Pavlyuchenkov}, {Chiang}, \& {Liu}}]{dong16protostellar}
{Dong}, R., {Vorobyov}, E., {Pavlyuchenkov}, Y., {Chiang}, E., \& {Liu}, H.~B.
  2016{\natexlab{b}}, \apj, 823, 141

\bibitem[{{Dong} {et~al.}(2016{\natexlab{c}}){Dong}, {Zhu}, {Fung}, {Rafikov},
  {Chiang}, \& {Wagner}}]{dong16hd100453}
{Dong}, R., {Zhu}, Z., {Fung}, J., {et~al.} 2016{\natexlab{c}}, \apjl, 816, L12

\bibitem[{{Dong} {et~al.}(2015{\natexlab{b}}){Dong}, {Zhu}, {Rafikov}, \&
  {Stone}}]{dong15spiralarm}
{Dong}, R., {Zhu}, Z., {Rafikov}, R.~R., \& {Stone}, J.~M. 2015{\natexlab{b}},
  \apjl, 809, L5

\bibitem[{{Dong} {et~al.}(2015{\natexlab{c}}){Dong}, {Zhu}, \&
  {Whitney}}]{dong15gap}
{Dong}, R., {Zhu}, Z., \& {Whitney}, B. 2015{\natexlab{c}}, \apj, 809, 93

\bibitem[{{Dong} {et~al.}(2017{\natexlab{b}}){Dong}, {van der Marel},
  {Hashimoto}, {Chiang}, {Akiyama}, {Liu}, {Muto}, {Knapp}, {Tsukagoshi},
  {Brown}, {Bruderer}, {Koyamatsu}, {Kudo}, {Ohashi}, {Rich}, {Satoshi},
  {Takami}, {Wisniewski}, {Yang}, {Zhu}, \& {Tamura}}]{dong17j1604}
{Dong}, R., {van der Marel}, N., {Hashimoto}, J., {et~al.} 2017{\natexlab{b}},
  \apj, 836, 201

\bibitem[{{Draine}(2006)}]{draine06}
{Draine}, B.~T. 2006, \apj, 636, 1114

\bibitem[{{Ducati}(2002)}]{ducati02}
{Ducati}, J.~R. 2002, VizieR Online Data Catalog, 2237

\bibitem[{{Duch{\^e}ne} {et~al.}(2004){Duch{\^e}ne}, {McCabe}, {Ghez}, \&
  {Macintosh}}]{duchene04}
{Duch{\^e}ne}, G., {McCabe}, C., {Ghez}, A.~M., \& {Macintosh}, B.~A. 2004,
  \apj, 606, 969

\bibitem[{{Dullemond} \& {Dominik}(2004)}]{dullemond04shadowing}
{Dullemond}, C.~P., \& {Dominik}, C. 2004, \aap, 417, 159

\bibitem[{{Dunhill} {et~al.}(2013){Dunhill}, {Alexander}, \&
  {Armitage}}]{dunhill13}
{Dunhill}, A.~C., {Alexander}, R.~D., \& {Armitage}, P.~J. 2013, \mnras, 428,
  3072

\bibitem[{{Dunkin} \& {Crawford}(1998)}]{dunkin98}
{Dunkin}, S.~K., \& {Crawford}, I.~A. 1998, \mnras, 298, 275

\bibitem[{{Eisner} {et~al.}(2016){Eisner}, {Bally}, {Ginsburg}, \&
  {Sheehan}}]{eisner16}
{Eisner}, J.~A., {Bally}, J.~M., {Ginsburg}, A., \& {Sheehan}, P.~D. 2016,
  \apj, 826, 16

\bibitem[{{Eisner} {et~al.}(2018){Eisner}, {Arce}, {Ballering}, {Bally},
  {Andrews}, {Boyden}, {Di Francesco}, {Fang}, {Johnstone}, {Kim}, {Mann},
  {Matthews}, {Pascucci}, {Ricci}, {Sheehan}, \& {Williams}}]{eisner18}
{Eisner}, J.~A., {Arce}, H.~G., {Ballering}, N.~P., {et~al.} 2018, \apj, 860,
  77

\bibitem[{{Erickson} {et~al.}(2011){Erickson}, {Wilking}, {Meyer}, {Robinson},
  \& {Stephenson}}]{erickson11}
{Erickson}, K.~L., {Wilking}, B.~A., {Meyer}, M.~R., {Robinson}, J.~G., \&
  {Stephenson}, L.~N. 2011, \aj, 142, 140

\bibitem[{{Espaillat} {et~al.}(2014){Espaillat}, {Muzerolle}, {Najita},
  {Andrews}, {Zhu}, {Calvet}, {Kraus}, {Hashimoto}, {Kraus}, \&
  {D'Alessio}}]{espaillat14}
{Espaillat}, C., {Muzerolle}, J., {Najita}, J., {et~al.} 2014, Protostars and
  Planets VI, 497

\bibitem[{{Evans} {et~al.}(2017){Evans}, {Ilee}, {Hartquist}, {Caselli}, {Sz{\H
  u}cs}, {Purser}, {Boley}, {Durisen}, \& {Rawlings}}]{evans17}
{Evans}, M.~G., {Ilee}, J.~D., {Hartquist}, T.~W., {et~al.} 2017, \mnras, 470,
  1828

\bibitem[{{Facchini} {et~al.}(2017){Facchini}, {Birnstiel}, {Bruderer}, \& {van
  Dishoeck}}]{facchini17}
{Facchini}, S., {Birnstiel}, T., {Bruderer}, S., \& {van Dishoeck}, E.~F. 2017,
  \aap, 605, A16

\bibitem[{{Fairlamb} {et~al.}(2015){Fairlamb}, {Oudmaijer},
  {Mendigut{\'{\i}}a}, {Ilee}, \& {van den Ancker}}]{fairlamb15}
{Fairlamb}, J.~R., {Oudmaijer}, R.~D., {Mendigut{\'{\i}}a}, I., {Ilee}, J.~D.,
  \& {van den Ancker}, M.~E. 2015, \mnras, 453, 976

\bibitem[{{Fairlamb} {et~al.}(2017){Fairlamb}, {Oudmaijer}, {Mendigutia},
  {Ilee}, \& {van den Ancker}}]{fairlamb17}
{Fairlamb}, J.~R., {Oudmaijer}, R.~D., {Mendigutia}, I., {Ilee}, J.~D., \& {van
  den Ancker}, M.~E. 2017, \mnras, 464, 4721

\bibitem[{{Fedele} {et~al.}(2017){Fedele}, {Carney}, {Hogerheijde}, {Walsh},
  {Miotello}, {Klaassen}, {Bruderer}, {Henning}, \& {van Dishoeck}}]{fedele17}
{Fedele}, D., {Carney}, M., {Hogerheijde}, M.~R., {et~al.} 2017, \aap, 600, A72

\bibitem[{{Fedele} {et~al.}(2018){Fedele}, {Tazzari}, {Booth}, {Testi},
  {Clarke}, {Pascucci}, {Kospal}, {Semenov}, {Bruderer}, {Henning}, \&
  {Teague}}]{fedele18}
{Fedele}, D., {Tazzari}, M., {Booth}, R., {et~al.} 2018, \aap, 610, A24

\bibitem[{{Finkenzeller} \& {Mundt}(1984)}]{finkenzeller84}
{Finkenzeller}, U., \& {Mundt}, R. 1984, \aaps, 55, 109

\bibitem[{{Flaherty} {et~al.}(2016){Flaherty}, {Hughes}, {Andrews}, {Qi},
  {Wilner}, {Boley}, {White}, {Harney}, \& {Zachary}}]{flaherty16}
{Flaherty}, K.~M., {Hughes}, A.~M., {Andrews}, S.~M., {et~al.} 2016, \apj, 818,
  97

\bibitem[{{Follette} {et~al.}(2013){Follette}, {Tamura}, {Hashimoto},
  {Whitney}, {Grady}, {Close}, {Andrews}, {Kwon}, {Wisniewski}, {Brandt},
  {Mayama}, {Kandori}, {Dong}, {Abe}, {Brandner}, {Carson}, {Currie}, {Egner},
  {Feldt}, {Goto}, {Guyon}, {Hayano}, {Hayashi}, {Hayashi}, {Henning},
  {Hodapp}, {Ishii}, {Iye}, {Janson}, {Knapp}, {Kudo}, {Kusakabe}, {Kuzuhara},
  {McElwain}, {Matsuo}, {Miyama}, {Morino}, {Moro-Martin}, {Nishimura}, {Pyo},
  {Serabyn}, {Suto}, {Suzuki}, {Takami}, {Takato}, {Terada}, {Thalmann},
  {Tomono}, {Turner}, {Watanabe}, {Yamada}, {Takami}, \& {Usuda}}]{follette13}
{Follette}, K.~B., {Tamura}, M., {Hashimoto}, J., {et~al.} 2013, \apj, 767, 10

\bibitem[{{Follette} {et~al.}(2015){Follette}, {Grady}, {Swearingen}, {Sitko},
  {Champney}, {van der Marel}, {Takami}, {Kuchner}, {Close}, {Muto}, {Mayama},
  {McElwain}, {Fukagawa}, {Maaskant}, {Min}, {Russell}, {Kudo}, {Kusakabe},
  {Hashimoto}, {Abe}, {Akiyama}, {Brandner}, {Brandt}, {Carson}, {Currie},
  {Egner}, {Feldt}, {Goto}, {Guyon}, {Hayano}, {Hayashi}, {Hayashi}, {Henning},
  {Hodapp}, {Ishii}, {Iye}, {Janson}, {Kandori}, {Knapp}, {Kuzuhara}, {Kwon},
  {Matsuo}, {Miyama}, {Morino}, {Moro-Martin}, {Nishimura}, {Pyo}, {Serabyn},
  {Suenaga}, {Suto}, {Suzuki}, {Takahashi}, {Takato}, {Terada}, {Thalmann},
  {Tomono}, {Turner}, {Watanabe}, {Wisniewski}, {Yamada}, {Takami}, {Usuda}, \&
  {Tamura}}]{follette15}
{Follette}, K.~B., {Grady}, C.~A., {Swearingen}, J.~R., {et~al.} 2015, \apj,
  798, 132

\bibitem[{{Follette} {et~al.}(2017){Follette}, {Rameau}, {Dong}, {Pueyo},
  {Close}, {Duch{\^e}ne}, {Fung}, {Leonard}, {Macintosh}, {Males}, {Marois},
  {Millar-Blanchaer}, {Morzinski}, {Mullen}, {Perrin}, {Spiro}, {Wang},
  {Ammons}, {Bailey}, {Barman}, {Bulger}, {Chilcote}, {Cotten}, {De Rosa},
  {Doyon}, {Fitzgerald}, {Goodsell}, {Graham}, {Greenbaum}, {Hibon}, {Hung},
  {Ingraham}, {Kalas}, {Konopacky}, {Larkin}, {Maire}, {Marchis}, {Metchev},
  {Nielsen}, {Oppenheimer}, {Palmer}, {Patience}, {Poyneer}, {Rajan},
  {Rantakyr{\"o}}, {Savransky}, {Schneider}, {Sivaramakrishnan}, {Song},
  {Soummer}, {Thomas}, {Vega}, {Wallace}, {Ward-Duong}, {Wiktorowicz}, \&
  {Wolff}}]{follette17}
{Follette}, K.~B., {Rameau}, J., {Dong}, R., {et~al.} 2017, \aj, 153, 264

\bibitem[{{Folsom} {et~al.}(2012){Folsom}, {Bagnulo}, {Wade}, {Alecian},
  {Landstreet}, {Marsden}, \& {Waite}}]{folsom12}
{Folsom}, C.~P., {Bagnulo}, S., {Wade}, G.~A., {et~al.} 2012, \mnras, 422, 2072

\bibitem[{{Forgan} \& {Rice}(2013)}]{forgan13l1527}
{Forgan}, D., \& {Rice}, K. 2013, \mnras, 433, 1796

\bibitem[{{Fortney} {et~al.}(2008){Fortney}, {Marley}, {Saumon}, \&
  {Lodders}}]{fortney08}
{Fortney}, J.~J., {Marley}, M.~S., {Saumon}, D., \& {Lodders}, K. 2008, \apj,
  683, 1104

\bibitem[{{Fossati} {et~al.}(2009){Fossati}, {Ryabchikova}, {Bagnulo},
  {Alecian}, {Grunhut}, {Kochukhov}, \& {Wade}}]{fossati09}
{Fossati}, L., {Ryabchikova}, T., {Bagnulo}, S., {et~al.} 2009, \aap, 503, 945

\bibitem[{{Fukagawa} {et~al.}(2003){Fukagawa}, {Tamura}, {Itoh}, {Hayashi}, \&
  {Oasa}}]{fukagawa03}
{Fukagawa}, M., {Tamura}, M., {Itoh}, Y., {Hayashi}, S.~S., \& {Oasa}, Y. 2003,
  \apjl, 590, L49

\bibitem[{{Fukagawa} {et~al.}(2006){Fukagawa}, {Tamura}, {Itoh}, {Kudo},
  {Imaeda}, {Oasa}, {Hayashi}, \& {Hayashi}}]{fukagawa06}
{Fukagawa}, M., {Tamura}, M., {Itoh}, Y., {et~al.} 2006, \apjl, 636, L153

\bibitem[{{Fukagawa} {et~al.}(2004){Fukagawa}, {Hayashi}, {Tamura}, {Itoh},
  {Hayashi}, {Oasa}, {Takeuchi}, {Morino}, {Murakawa}, {Oya}, {Yamashita},
  {Suto}, {Mayama}, {Naoi}, {Ishii}, {Pyo}, {Nishikawa}, {Takato}, {Usuda},
  {Ando}, {Iye}, {Miyama}, \& {Kaifu}}]{fukagawa04}
{Fukagawa}, M., {Hayashi}, M., {Tamura}, M., {et~al.} 2004, \apjl, 605, L53

\bibitem[{{Fung} \& {Dong}(2015)}]{fung15}
{Fung}, J., \& {Dong}, R. 2015, \apjl, 815, L21

\bibitem[{{Gaia Collaboration} {et~al.}(2016){Gaia Collaboration}, {Brown},
  {Vallenari}, {Prusti}, {de Bruijne}, {Mignard}, {Drimmel}, {Babusiaux},
  {Bailer-Jones}, {Bastian}, \& et~al.}]{gaia16}
{Gaia Collaboration}, {Brown}, A.~G.~A., {Vallenari}, A., {et~al.} 2016, \aap,
  595, A2

\bibitem[{{Gammie}(2001)}]{gammie01}
{Gammie}, C.~F. 2001, \apj, 553, 174

\bibitem[{{Garcia Lopez} {et~al.}(2006){Garcia Lopez}, {Natta}, {Testi}, \&
  {Habart}}]{garcialopez06}
{Garcia Lopez}, R., {Natta}, A., {Testi}, L., \& {Habart}, E. 2006, \aap, 459,
  837

\bibitem[{{Garufi} {et~al.}(2014){Garufi}, {Quanz}, {Schmid}, {Avenhaus},
  {Buenzli}, \& {Wolf}}]{garufi14}
{Garufi}, A., {Quanz}, S.~P., {Schmid}, H.~M., {et~al.} 2014, \aap, 568, A40

\bibitem[{{Garufi} {et~al.}(2013){Garufi}, {Quanz}, {Avenhaus}, {Buenzli},
  {Dominik}, {Meru}, {Meyer}, {Pinilla}, {Schmid}, \& {Wolf}}]{garufi13}
{Garufi}, A., {Quanz}, S.~P., {Avenhaus}, H., {et~al.} 2013, \aap, 560, A105

\bibitem[{{Garufi} {et~al.}(2016){Garufi}, {Quanz}, {Schmid}, {Mulders},
  {Avenhaus}, {Boccaletti}, {Ginski}, {Langlois}, {Stolker}, {Augereau},
  {Benisty}, {Lopez}, {Dominik}, {Gratton}, {Henning}, {Janson}, {M{\'e}nard},
  {Meyer}, {Pinte}, {Sissa}, {Vigan}, {Zurlo}, {Bazzon}, {Buenzli}, {Bonnefoy},
  {Brandner}, {Chauvin}, {Cheetham}, {Cudel}, {Desidera}, {Feldt}, {Galicher},
  {Kasper}, {Lagrange}, {Lannier}, {Maire}, {Mesa}, {Mouillet}, {Peretti},
  {Perrot}, {Salter}, \& {Wildi}}]{garufi16}
{Garufi}, A., {Quanz}, S.~P., {Schmid}, H.~M., {et~al.} 2016, \aap, 588, A8

\bibitem[{{Garufi} {et~al.}(2017){Garufi}, {Meeus}, {Benisty}, {Quanz},
  {Banzatti}, {Kama}, {Canovas}, {Eiroa}, {Schmid}, {Stolker}, {Pohl},
  {Rigliaco}, {M{\'e}nard}, {Meyer}, {van Boekel}, \& {Dominik}}]{garufi17}
{Garufi}, A., {Meeus}, G., {Benisty}, M., {et~al.} 2017, \aap, 603, A21

\bibitem[{{Ginski} {et~al.}(2016){Ginski}, {Stolker}, {Pinilla}, {Dominik},
  {Boccaletti}, {de Boer}, {Benisty}, {Biller}, {Feldt}, {Garufi}, {Keller},
  {Kenworthy}, {Maire}, {M{\'e}nard}, {Mesa}, {Milli}, {Min}, {Pinte}, {Quanz},
  {van Boekel}, {Bonnefoy}, {Chauvin}, {Desidera}, {Gratton}, {Girard},
  {Keppler}, {Kopytova}, {Lagrange}, {Langlois}, {Rouan}, \&
  {Vigan}}]{ginski16}
{Ginski}, C., {Stolker}, T., {Pinilla}, P., {et~al.} 2016, \aap, 595, A112

\bibitem[{{Goldreich} \& {Lynden-Bell}(1965)}]{goldreich65}
{Goldreich}, P., \& {Lynden-Bell}, D. 1965, \mnras, 130, 97

\bibitem[{{Goldreich} \& {Tremaine}(1979)}]{goldreich79}
{Goldreich}, P., \& {Tremaine}, S. 1979, \apj, 233, 857

\bibitem[{{Grady} {et~al.}(2001){Grady}, {Polomski}, {Henning}, {Stecklum},
  {Woodgate}, {Telesco}, {Pi{\~n}a}, {Gull}, {Boggess}, {Bowers}, {Bruhweiler},
  {Clampin}, {Danks}, {Green}, {Heap}, {Hutchings}, {Jenkins}, {Joseph},
  {Kaiser}, {Kimble}, {Kraemer}, {Lindler}, {Linsky}, {Maran}, {Moos}, {Plait},
  {Roesler}, {Timothy}, \& {Weistrop}}]{grady01}
{Grady}, C.~A., {Polomski}, E.~F., {Henning}, T., {et~al.} 2001, \aj, 122, 3396

\bibitem[{{Grady} {et~al.}(2007){Grady}, {Schneider}, {Hamaguchi}, {Sitko},
  {Carpenter}, {Hines}, {Collins}, {Williger}, {Woodgate}, {Henning},
  {M{\'e}nard}, {Wilner}, {Petre}, {Palunas}, {Quirrenbach}, {Nuth},
  {Silverstone}, \& {Kim}}]{grady07}
{Grady}, C.~A., {Schneider}, G., {Hamaguchi}, K., {et~al.} 2007, \apj, 665,
  1391

\bibitem[{{Grady} {et~al.}(2009){Grady}, {Schneider}, {Sitko}, {Williger},
  {Hamaguchi}, {Brittain}, {Ablordeppey}, {Apai}, {Beerman}, {Carpenter},
  {Collins}, {Fukagawa}, {Hammel}, {Henning}, {Hines}, {Kimes}, {Lynch},
  {M{\'e}nard}, {Pearson}, {Russell}, {Silverstone}, {Smith}, {Troutman},
  {Wilner}, {Woodgate}, \& {Clampin}}]{grady09}
{Grady}, C.~A., {Schneider}, G., {Sitko}, M.~L., {et~al.} 2009, \apj, 699, 1822

\bibitem[{{Grady} {et~al.}(2010){Grady}, {Hamaguchi}, {Schneider}, {Stecklum},
  {Woodgate}, {McCleary}, {Williger}, {Sitko}, {M{\'e}nard}, {Henning},
  {Brittain}, {Troutmann}, {Donehew}, {Hines}, {Wisniewski}, {Lynch},
  {Russell}, {Rudy}, {Day}, {Shenoy}, {Wilner}, {Silverstone}, {Bouret},
  {Meusinger}, {Clampin}, {Kim}, {Petre}, {Sahu}, {Endres}, \&
  {Collins}}]{grady10}
{Grady}, C.~A., {Hamaguchi}, K., {Schneider}, G., {et~al.} 2010, \apj, 719,
  1565

\bibitem[{{Grady} {et~al.}(2013){Grady}, {Muto}, {Hashimoto}, {Fukagawa},
  {Currie}, {Biller}, {Thalmann}, {Sitko}, {Russell}, {Wisniewski}, {Dong},
  {Kwon}, {Sai}, {Hornbeck}, {Schneider}, {Hines}, {Moro Mart{\'{\i}}n},
  {Feldt}, {Henning}, {Pott}, {Bonnefoy}, {Bouwman}, {Lacour}, {Mueller},
  {Juh{\'a}sz}, {Crida}, {Chauvin}, {Andrews}, {Wilner}, {Kraus}, {Dahm},
  {Robitaille}, {Jang-Condell}, {Abe}, {Akiyama}, {Brandner}, {Brandt},
  {Carson}, {Egner}, {Follette}, {Goto}, {Guyon}, {Hayano}, {Hayashi},
  {Hayashi}, {Hodapp}, {Ishii}, {Iye}, {Janson}, {Kandori}, {Knapp}, {Kudo},
  {Kusakabe}, {Kuzuhara}, {Mayama}, {McElwain}, {Matsuo}, {Miyama}, {Morino},
  {Nishimura}, {Pyo}, {Serabyn}, {Suto}, {Suzuki}, {Takami}, {Takato},
  {Terada}, {Tomono}, {Turner}, {Watanabe}, {Yamada}, {Takami}, {Usuda}, \&
  {Tamura}}]{grady13}
{Grady}, C.~A., {Muto}, T., {Hashimoto}, J., {et~al.} 2013, \apj, 762, 48

\bibitem[{{Greene} \& {Meyer}(1995)}]{greene95}
{Greene}, T.~P., \& {Meyer}, M.~R. 1995, \apj, 450, 233

\bibitem[{{Guimar{\~a}es} {et~al.}(2006){Guimar{\~a}es}, {Alencar}, {Corradi},
  \& {Vieira}}]{guimaraes06}
{Guimar{\~a}es}, M.~M., {Alencar}, S.~H.~P., {Corradi}, W.~J.~B., \& {Vieira},
  S.~L.~A. 2006, \aap, 457, 581

\bibitem[{{Hall} {et~al.}(2016){Hall}, {Forgan}, {Rice}, {Harries}, {Klaassen},
  \& {Biller}}]{hall16}
{Hall}, C., {Forgan}, D., {Rice}, K., {et~al.} 2016, \mnras, 458, 306

\bibitem[{{Hamidouche}(2010)}]{hamidouche10}
{Hamidouche}, M. 2010, \apj, 722, 204

\bibitem[{{Hartmann} {et~al.}(1998){Hartmann}, {Calvet}, {Gullbring}, \&
  {D'Alessio}}]{hartmann98}
{Hartmann}, L., {Calvet}, N., {Gullbring}, E., \& {D'Alessio}, P. 1998, \apj,
  495, 385

\bibitem[{{Hashimoto} {et~al.}(2011){Hashimoto}, {Tamura}, {Muto}, {Kudo},
  {Fukagawa}, {Fukue}, {Goto}, {Grady}, {Henning}, {Hodapp}, {Honda},
  {Inutsuka}, {Kokubo}, {Knapp}, {McElwain}, {Momose}, {Ohashi}, {Okamoto},
  {Takami}, {Turner}, {Wisniewski}, {Janson}, {Abe}, {Brandner}, {Carson},
  {Egner}, {Feldt}, {Golota}, {Guyon}, {Hayano}, {Hayashi}, {Hayashi}, {Ishii},
  {Kandori}, {Kusakabe}, {Matsuo}, {Mayama}, {Miyama}, {Morino}, {Moro-Martin},
  {Nishimura}, {Pyo}, {Suto}, {Suzuki}, {Takato}, {Terada}, {Thalmann},
  {Tomono}, {Watanabe}, {Yamada}, {Takami}, \& {Usuda}}]{hashimoto11}
{Hashimoto}, J., {Tamura}, M., {Muto}, T., {et~al.} 2011, \apjl, 729, L17

\bibitem[{{Hashimoto} {et~al.}(2012){Hashimoto}, {Dong}, {Kudo}, {Honda},
  {McClure}, {Zhu}, {Muto}, {Wisniewski}, {Abe}, {Brandner}, {Brandt},
  {Carson}, {Egner}, {Feldt}, {Fukagawa}, {Goto}, {Grady}, {Guyon}, {Hayano},
  {Hayashi}, {Hayashi}, {Henning}, {Hodapp}, {Ishii}, {Iye}, {Janson},
  {Kandori}, {Knapp}, {Kusakabe}, {Kuzuhara}, {Kwon}, {Matsuo}, {Mayama},
  {McElwain}, {Miyama}, {Morino}, {Moro-Martin}, {Nishimura}, {Pyo}, {Serabyn},
  {Suenaga}, {Suto}, {Suzuki}, {Takahashi}, {Takami}, {Takato}, {Terada},
  {Thalmann}, {Tomono}, {Turner}, {Watanabe}, {Yamada}, {Takami}, {Usuda}, \&
  {Tamura}}]{hashimoto12}
{Hashimoto}, J., {Dong}, R., {Kudo}, T., {et~al.} 2012, \apjl, 758, L19

\bibitem[{{Hashimoto} {et~al.}(2015){Hashimoto}, {Tsukagoshi}, {Brown}, {Dong},
  {Muto}, {Zhu}, {Wisniewski}, {Ohashi}, {kudo}, {Kusakabe}, {Abe}, {Akiyama},
  {Brandner}, {Brandt}, {Carson}, {Currie}, {Egner}, {Feldt}, {Grady}, {Guyon},
  {Hayano}, {Hayashi}, {Hayashi}, {Henning}, {Hodapp}, {Ishii}, {Iye},
  {Janson}, {Kandori}, {Knapp}, {Kuzuhara}, {Kwon}, {Matsuo}, {McElwain},
  {Mayama}, {Mede}, {Miyama}, {Morino}, {Moro-Martin}, {Nishimura}, {Pyo},
  {Serabyn}, {Suenaga}, {Suto}, {Suzuki}, {Takahashi}, {Takami}, {Takato},
  {Terada}, {Thalmann}, {Tomono}, {Turner}, {Watanabe}, {Yamada}, {Takami},
  {Usuda}, \& {Tamura}}]{hashimoto15}
{Hashimoto}, J., {Tsukagoshi}, T., {Brown}, J.~M., {et~al.} 2015, \apj, 799, 43

\bibitem[{{Hayashi}(1981)}]{hayashi81}
{Hayashi}, C. 1981, Progress of Theoretical Physics Supplement, 70, 35

\bibitem[{{Henning} {et~al.}(1994){Henning}, {Launhardt}, {Steinacker}, \&
  {Thamm}}]{henning94}
{Henning}, T., {Launhardt}, R., {Steinacker}, J., \& {Thamm}, E. 1994, \aap,
  291, 546

\bibitem[{{Hern{\'a}ndez} {et~al.}(2004){Hern{\'a}ndez}, {Calvet},
  {Brice{\~n}o}, {Hartmann}, \& {Berlind}}]{hernandez04}
{Hern{\'a}ndez}, J., {Calvet}, N., {Brice{\~n}o}, C., {Hartmann}, L., \&
  {Berlind}, P. 2004, \aj, 127, 1682

\bibitem[{{Hern{\'a}ndez} {et~al.}(2005){Hern{\'a}ndez}, {Calvet}, {Hartmann},
  {Brice{\~n}o}, {Sicilia-Aguilar}, \& {Berlind}}]{hernandez05}
{Hern{\'a}ndez}, J., {Calvet}, N., {Hartmann}, L., {et~al.} 2005, \aj, 129, 856

\bibitem[{{Hern{\'a}ndez} {et~al.}(2008){Hern{\'a}ndez}, {Hartmann}, {Calvet},
  {Jeffries}, {Gutermuth}, {Muzerolle}, \& {Stauffer}}]{hernandez08}
{Hern{\'a}ndez}, J., {Hartmann}, L., {Calvet}, N., {et~al.} 2008, \apj, 686,
  1195

\bibitem[{{H{\o}g} {et~al.}(2000){H{\o}g}, {Fabricius}, {Makarov}, {Urban},
  {Corbin}, {Wycoff}, {Bastian}, {Schwekendiek}, \& {Wicenec}}]{hog00}
{H{\o}g}, E., {Fabricius}, C., {Makarov}, V.~V., {et~al.} 2000, \aap, 355, L27

\bibitem[{{Honda} {et~al.}(2009){Honda}, {Inoue}, {Fukagawa}, {Oka},
  {Nakamoto}, {Ishii}, {Terada}, {Takato}, {Kawakita}, {Okamoto}, {Shibai},
  {Tamura}, {Kudo}, \& {Itoh}}]{honda09}
{Honda}, M., {Inoue}, A.~K., {Fukagawa}, M., {et~al.} 2009, \apjl, 690, L110

\bibitem[{{Houk} \& {Cowley}(1975)}]{houk75}
{Houk}, N., \& {Cowley}, A.~P. 1975, {University of Michigan Catalogue of
  two-dimensional spectral types for the HD stars. Volume I. Declinations -90
  to -53 degree}

\bibitem[{{Huang} {et~al.}(2018){Huang}, {Andrews}, {Cleeves}, {{\"O}berg},
  {Wilner}, {Bai}, {Birnstiel}, {Carpenter}, {Hughes}, {Isella}, {P{\'e}rez},
  {Ricci}, \& {Zhu}}]{huang18}
{Huang}, J., {Andrews}, S.~M., {Cleeves}, L.~I., {et~al.} 2018, \apj, 852, 122

\bibitem[{{Isella} {et~al.}(2009){Isella}, {Carpenter}, \&
  {Sargent}}]{isella09}
{Isella}, A., {Carpenter}, J.~M., \& {Sargent}, A.~I. 2009, \apj, 701, 260

\bibitem[{{Isella} \& {Turner}(2018)}]{isella18}
{Isella}, A., \& {Turner}, N.~J. 2018, \apj, 860, 27

\bibitem[{{Isella} {et~al.}(2016){Isella}, {Guidi}, {Testi}, {Liu}, {Li}, {Li},
  {Weaver}, {Boehler}, {Carperter}, {De Gregorio-Monsalvo}, {Manara}, {Natta},
  {P{\'e}rez}, {Ricci}, {Sargent}, {Tazzari}, \& {Turner}}]{isella16hd163296}
{Isella}, A., {Guidi}, G., {Testi}, L., {et~al.} 2016, Physical Review Letters,
  117, 251101

\bibitem[{{Itoh} {et~al.}(2002){Itoh}, {Tamura}, {Hayashi}, {Oasa}, {Fukagawa},
  {Kaifu}, {Suto}, {Murakawa}, {Doi}, {Ebizuka}, {Naoi}, {Takami}, {Takato},
  {Gaessler}, {Kanzawa}, {Hayano}, {Kamata}, {Saint-Jacques}, \&
  {Iye}}]{itoh02}
{Itoh}, Y., {Tamura}, M., {Hayashi}, S.~S., {et~al.} 2002, \pasj, 54, 963

\bibitem[{{Itoh} {et~al.}(2014){Itoh}, {Oasa}, {Kudo}, {Kusakabe}, {Hashimoto},
  {Abe}, {Brandner}, {Brandt}, {Carson}, {Egner}, {Feldt}, {Grady}, {Guyon},
  {Hayano}, {Hayashi}, {Hayashi}, {Henning}, {Hodapp}, {Ishii}, {Iye},
  {Janson}, {Kandori}, {Knapp}, {Kuzuhara}, {Kwon}, {Matsuo}, {McElwain},
  {Miyama}, {Morino}, {Moro-Martin}, {Nishimura}, {Pyo}, {Serabyn}, {Suenaga},
  {Suto}, {Suzuki}, {Takahashi}, {Takato}, {Terada}, {Thalmann}, {Tomono},
  {Turner}, {Watanabe}, {Wisniewski}, {Yamada}, {Mayama}, {Currie}, {Takami},
  {Usuda}, \& {Tamura}}]{itoh14}
{Itoh}, Y., {Oasa}, Y., {Kudo}, T., {et~al.} 2014, Research in Astronomy and
  Astrophysics, 14, 1438

\bibitem[{{Jamialahmadi} {et~al.}(2015){Jamialahmadi}, {Berio}, {Meilland},
  {Perraut}, {Mourard}, {Lopez}, {Stee}, {Nardetto}, {Pichon}, {Clausse},
  {Spang}, {McAlister}, {ten Brummelaar}, {Sturmann}, {Sturmann}, {Turner},
  {Farrington}, {Vargas}, \& {Scott}}]{jamialahmadi15}
{Jamialahmadi}, N., {Berio}, P., {Meilland}, A., {et~al.} 2015, \aap, 579, A81

\bibitem[{{Janson} {et~al.}(2016){Janson}, {Thalmann}, {Boccaletti}, {Maire},
  {Zurlo}, {Marzari}, {Meyer}, {Carson}, {Augereau}, {Garufi}, {Henning},
  {Desidera}, {Asensio-Torres}, \& {Pohl}}]{janson16}
{Janson}, M., {Thalmann}, C., {Boccaletti}, A., {et~al.} 2016, \apjl, 816, L1

\bibitem[{{Jayawardhana} {et~al.}(2002){Jayawardhana}, {Luhman}, {D'Alessio},
  \& {Stauffer}}]{jayawardhana02}
{Jayawardhana}, R., {Luhman}, K.~L., {D'Alessio}, P., \& {Stauffer}, J.~R.
  2002, \apjl, 571, L51

\bibitem[{{Juh{\'a}sz} {et~al.}(2015){Juh{\'a}sz}, {Benisty}, {Pohl},
  {Dullemond}, {Dominik}, \& {Paardekooper}}]{juhasz15}
{Juh{\'a}sz}, A., {Benisty}, M., {Pohl}, A., {et~al.} 2015, \mnras, 451, 1147

\bibitem[{{Kama} {et~al.}(2015){Kama}, {Folsom}, \& {Pinilla}}]{kama15}
{Kama}, M., {Folsom}, C.~P., \& {Pinilla}, P. 2015, \aap, 582, L10

\bibitem[{{Kasper} {et~al.}(2016){Kasper}, {Santhakumari}, {Herbst}, \&
  {K{\"o}hler}}]{kasper16}
{Kasper}, M., {Santhakumari}, K.~K.~R., {Herbst}, T.~M., \& {K{\"o}hler}, R.
  2016, \aap, 593, A50

\bibitem[{{Kley} \& {Nelson}(2012)}]{kley12}
{Kley}, W., \& {Nelson}, R.~P. 2012, \araa, 50, 211

\bibitem[{{Konishi} {et~al.}(2016){Konishi}, {Grady}, {Schneider}, {Shibai},
  {McElwain}, {Nesvold}, {Kuchner}, {Carson}, {Debes}, {Gaspar}, {Henning},
  {Hines}, {Hinz}, {Jang-Condell}, {Moro-Mart{\'{\i}}n}, {Perrin}, {Rodigas},
  {Serabyn}, {Silverstone}, {Stark}, {Tamura}, {Weinberger}, \&
  {Wisniewski}}]{konishi16}
{Konishi}, M., {Grady}, C.~A., {Schneider}, G., {et~al.} 2016, \apjl, 818, L23

\bibitem[{{Kooistra} {et~al.}(2017){Kooistra}, {Kamp}, {Fukagawa},
  {M{\'e}nard}, {Momose}, {Tsukagoshi}, {Kudo}, {Kusakabe}, {Hashimoto}, {Abe},
  {Brandner}, {Brandt}, {Carson}, {Egner}, {Feldt}, {Goto}, {Grady}, {Guyon},
  {Hayano}, {Hayashi}, {Hayashi}, {Henning}, {Hodapp}, {Ishii}, {Iye},
  {Janson}, {Kandori}, {Knapp}, {Kuzuhara}, {Kwon}, {Matsuo}, {McElwain},
  {Miyama}, {Morino}, {Moro-Martin}, {Nishimura}, {Pyo}, {Serabyn}, {Suenaga},
  {Suto}, {Suzuki}, {Takahashi}, {Takami}, {Takato}, {Terada}, {Thalmann},
  {Tomono}, {Turner}, {Watanabe}, {Wisniewski}, {Yamada}, {Takami}, {Usuda},
  {Tamura}, {Currie}, {Akiyama}, {Mayama}, {Follette}, \&
  {Nakagawa}}]{kooistra17}
{Kooistra}, R., {Kamp}, I., {Fukagawa}, M., {et~al.} 2017, \aap, 597, A132

\bibitem[{{Kratter} \& {Lodato}(2016)}]{kratter16}
{Kratter}, K., \& {Lodato}, G. 2016, \araa, 54, 271

\bibitem[{{Kratter} {et~al.}(2008){Kratter}, {Matzner}, \&
  {Krumholz}}]{kratter08}
{Kratter}, K.~M., {Matzner}, C.~D., \& {Krumholz}, M.~R. 2008, \apj, 681, 375

\bibitem[{{Kratter} {et~al.}(2010){Kratter}, {Matzner}, {Krumholz}, \&
  {Klein}}]{kratter10gidisks}
{Kratter}, K.~M., {Matzner}, C.~D., {Krumholz}, M.~R., \& {Klein}, R.~I. 2010,
  \apj, 708, 1585

\bibitem[{{Kraus} {et~al.}(2013){Kraus}, {Ireland}, {Sitko}, {Monnier},
  {Calvet}, {Espaillat}, {Grady}, {Harries}, {H{\"o}nig}, {Russell},
  {Swearingen}, {Werren}, \& {Wilner}}]{kraus13}
{Kraus}, S., {Ireland}, M.~J., {Sitko}, M.~L., {et~al.} 2013, \apj, 768, 80

\bibitem[{{Kraus} {et~al.}(2017){Kraus}, {Kreplin}, {Fukugawa}, {Muto},
  {Sitko}, {Young}, {Bate}, {Grady}, {Harries}, {Monnier}, {Willson}, \&
  {Wisniewski}}]{kraus17}
{Kraus}, S., {Kreplin}, A., {Fukugawa}, M., {et~al.} 2017, \apjl, 848, L11

\bibitem[{{Krist} {et~al.}(2000){Krist}, {Stapelfeldt}, {M{\'e}nard},
  {Padgett}, \& {Burrows}}]{krist00}
{Krist}, J.~E., {Stapelfeldt}, K.~R., {M{\'e}nard}, F., {Padgett}, D.~L., \&
  {Burrows}, C.~J. 2000, \apj, 538, 793

\bibitem[{{Krist} {et~al.}(2002){Krist}, {Stapelfeldt}, \& {Watson}}]{krist02}
{Krist}, J.~E., {Stapelfeldt}, K.~R., \& {Watson}, A.~M. 2002, \apj, 570, 785

\bibitem[{{Krist} {et~al.}(2005){Krist}, {Stapelfeldt}, {Golimowski}, {Ardila},
  {Clampin}, {Martel}, {Ford}, {Illingworth}, \& {Hartig}}]{krist05}
{Krist}, J.~E., {Stapelfeldt}, K.~R., {Golimowski}, D.~A., {et~al.} 2005, \aj,
  130, 2778

\bibitem[{{Kudo} {et~al.}(2008){Kudo}, {Tamura}, {Kitamura}, {Hayashi},
  {Kokubo}, {Fukagawa}, {Hayashi}, {Ishii}, {Itoh}, {Mayama}, {Momose},
  {Morino}, {Oasa}, {Pyo}, \& {Suto}}]{kudo08}
{Kudo}, T., {Tamura}, M., {Kitamura}, Y., {et~al.} 2008, \apjl, 673, L67

\bibitem[{{Kusakabe} {et~al.}(2012){Kusakabe}, {Grady}, {Sitko}, {Hashimoto},
  {Kudo}, {Fukagawa}, {Muto}, {Wisniewski}, {Min}, {Mayama}, {Werren}, {Day},
  {Beerman}, {Lynch}, {Russell}, {Brafford}, {Kuzuhara}, {Brandt}, {Abe},
  {Brandner}, {Carson}, {Egner}, {Feldt}, {Goto}, {Guyon}, {Hayano}, {Hayashi},
  {Hayashi}, {Henning}, {Hodapp}, {Ishii}, {Iye}, {Janson}, {Kandori}, {Knapp},
  {Matsuo}, {McElwain}, {Miyama}, {Morino}, {Moro-Martin}, {Nishimura}, {Pyo},
  {Suto}, {Suzuki}, {Takami}, {Takato}, {Terada}, {Thalmann}, {Tomono},
  {Turner}, {Watanabe}, {Yamada}, {Takami}, {Usuda}, \& {Tamura}}]{kusakabe12}
{Kusakabe}, N., {Grady}, C.~A., {Sitko}, M.~L., {et~al.} 2012, \apj, 753, 153

\bibitem[{{Lacour} {et~al.}(2016){Lacour}, {Biller}, {Cheetham}, {Greenbaum},
  {Pearce}, {Marino}, {Tuthill}, {Pueyo}, {Mamajek}, {Girard},
  {Sivaramakrishnan}, {Bonnefoy}, {Baraffe}, {Chauvin}, {Olofsson}, {Juhasz},
  {Benisty}, {Pott}, {Sicilia-Aguilar}, {Henning}, {Cardwell}, {Goodsell},
  {Graham}, {Hibon}, {Ingraham}, {Konopacky}, {Macintosh}, {Oppenheimer},
  {Perrin}, {Rantakyr{\"o}}, {Sadakuni}, \& {Thomas}}]{lacour16}
{Lacour}, S., {Biller}, B., {Cheetham}, A., {et~al.} 2016, \aap, 590, A90

\bibitem[{{Langlois} {et~al.}(2018){Langlois}, {Pohl}, {Lagrange}, {Maire},
  {Mesa}, {Boccaletti}, {Gratton}, {Denneulin}, {Klahr}, {Vigan}, {Benisty},
  {Dominik}, {Bonnefoy}, {Menard}, {Avenhaus}, {Cheetham}, {Van Boekel}, {de
  Boer}, {Chauvin}, {Desidera}, {Feldt}, {Galicher}, {Ginski}, {Girard},
  {Henning}, {Janson}, {Kopytova}, {Kral}, {Ligi}, {Messina}, {Peretti},
  {Pinte}, {Sissa}, {Stolker}, {Zurlo}, {Magnard}, {Blanchard}, {Buey},
  {Suarez}, {Cascone}, {Moller-Nilsson}, {Weber}, {Petit}, \&
  {Pragt}}]{langlois18}
{Langlois}, M., {Pohl}, A., {Lagrange}, A.-M., {et~al.} 2018, \aap, 614, A88

\bibitem[{{Lee}(2016)}]{lee16}
{Lee}, W.-K. 2016, \apj, 832, 166

\bibitem[{{Lenzen} {et~al.}(2003){Lenzen}, {Hartung}, {Brandner}, {Finger},
  {Hubin}, {Lacombe}, {Lagrange}, {Lehnert}, {Moorwood}, \&
  {Mouillet}}]{lenzen03}
{Lenzen}, R., {Hartung}, M., {Brandner}, W., {et~al.} 2003, in \procspie, Vol.
  4841, Instrument Design and Performance for Optical/Infrared Ground-based
  Telescopes, ed. M.~{Iye} \& A.~F.~M. {Moorwood}, 944--952

\bibitem[{{Ligi} {et~al.}(2018){Ligi}, {Vigan}, {Gratton}, {de Boer},
  {Benisty}, {Boccaletti}, {Quanz}, {Meyer}, {Ginski}, {Sissa}, {Gry},
  {Henning}, {Beuzit}, {Biller}, {Bonnefoy}, {Chauvin}, {Cheetham}, {Cudel},
  {Delorme}, {Desidera}, {Feldt}, {Galicher}, {Girard}, {Janson}, {Kasper},
  {Kopytova}, {Lagrange}, {Langlois}, {Lecoroller}, {Maire}, {M{\'e}nard},
  {Mesa}, {Peretti}, {Perrot}, {Pinilla}, {Pohl}, {Rouan}, {Stolker},
  {Samland}, {Wahhaj}, {Wildi}, {Zurlo}, {Buey}, {Fantinel}, {Fusco}, {Jaquet},
  {Moulin}, {Ramos}, {Suarez}, \& {Weber}}]{ligi18}
{Ligi}, R., {Vigan}, A., {Gratton}, R., {et~al.} 2018, \mnras, 473, 1774

\bibitem[{{Liu} {et~al.}(2016{\natexlab{a}}){Liu}, {Galv{\'a}n-Madrid},
  {Vorobyov}, {K{\'o}sp{\'a}l}, {Rodr{\'{\i}}guez}, {Dunham}, {Hirano},
  {Henning}, {Takami}, {Dong}, {Hashimoto}, {Hasegawa}, \&
  {Carrasco-Gonz{\'a}lez}}]{liu16}
{Liu}, H.~B., {Galv{\'a}n-Madrid}, R., {Vorobyov}, E.~I., {et~al.}
  2016{\natexlab{a}}, \apjl, 816, L29

\bibitem[{{Liu} {et~al.}(2016{\natexlab{b}}){Liu}, {Takami}, {Kudo},
  {Hashimoto}, {Dong}, {Vorobyov}, {Pyo}, {Fukagawa}, {Tamura}, {Henning},
  {Dunham}, {Karr}, {Kusakabe}, \& {Tsuribe}}]{liu16fuori}
{Liu}, H.~B., {Takami}, M., {Kudo}, T., {et~al.} 2016{\natexlab{b}}, Science
  Advances, 2, e1500875

\bibitem[{{Lodato} \& {Rice}(2004)}]{lodato04}
{Lodato}, G., \& {Rice}, W.~K.~M. 2004, \mnras, 351, 630

\bibitem[{{Lodato} \& {Rice}(2005)}]{lodato05}
---. 2005, \mnras, 358, 1489

\bibitem[{{Long} {et~al.}(2017){Long}, {Fernandes}, {Sitko}, {Wagner}, {Muto},
  {Hashimoto}, {Follette}, {Grady}, {Fukagawa}, {Hasegawa}, {Kluska}, {Kraus},
  {Mayama}, {McElwain}, {Oh}, {Tamura}, {Uyama}, {Wisniewski}, \&
  {Yang}}]{long17}
{Long}, Z.~C., {Fernandes}, R.~B., {Sitko}, M., {et~al.} 2017, \apj, 838, 62

\bibitem[{{Loomis} {et~al.}(2017){Loomis}, {{\"O}berg}, {Andrews}, \&
  {MacGregor}}]{loomis17}
{Loomis}, R.~A., {{\"O}berg}, K.~I., {Andrews}, S.~M., \& {MacGregor}, M.~A.
  2017, \apj, 840, 23

\bibitem[{{Lucas} {et~al.}(2004){Lucas}, {Fukagawa}, {Tamura}, {Beckford},
  {Itoh}, {Murakawa}, {Suto}, {Hayashi}, {Oasa}, {Naoi}, {Doi}, {Ebizuka}, \&
  {Kaifu}}]{lucas04}
{Lucas}, P.~W., {Fukagawa}, M., {Tamura}, M., {et~al.} 2004, \mnras, 352, 1347

\bibitem[{{Macintosh} {et~al.}(2008){Macintosh}, {Graham}, {Palmer}, {Doyon},
  {Dunn}, {Gavel}, {Larkin}, {Oppenheimer}, {Saddlemyer}, {Sivaramakrishnan},
  {Wallace}, {Bauman}, {Erickson}, {Marois}, {Poyneer}, \&
  {Soummer}}]{macintosh08}
{Macintosh}, B.~A., {Graham}, J.~R., {Palmer}, D.~W., {et~al.} 2008, in Society
  of Photo-Optical Instrumentation Engineers (SPIE) Conference Series, Vol.
  7015, Society of Photo-Optical Instrumentation Engineers (SPIE) Conference
  Series

\bibitem[{{Maire} {et~al.}(2017){Maire}, {Stolker}, {Messina}, {M{\"u}ller},
  {Biller}, {Currie}, {Dominik}, {Grady}, {Boccaletti}, {Bonnefoy}, {Chauvin},
  {Galicher}, {Millward}, {Pohl}, {Brandner}, {Henning}, {Lagrange},
  {Langlois}, {Meyer}, {Quanz}, {Vigan}, {Zurlo}, {van Boekel}, {Buenzli},
  {Buey}, {Desidera}, {Feldt}, {Fusco}, {Ginski}, {Giro}, {Gratton}, {Hubin},
  {Lannier}, {Le Mignant}, {Mesa}, {Peretti}, {Perrot}, {Ramos}, {Salter},
  {Samland}, {Sissa}, {Stadler}, {Thalmann}, {Udry}, \& {Weber}}]{maire17}
{Maire}, A.-L., {Stolker}, T., {Messina}, S., {et~al.} 2017, \aap, 601, A134

\bibitem[{{Malfait} {et~al.}(1998){Malfait}, {Bogaert}, \&
  {Waelkens}}]{malfait98}
{Malfait}, K., {Bogaert}, E., \& {Waelkens}, C. 1998, \aap, 331, 211

\bibitem[{{Mandy} \& {Martin}(1993)}]{mandy93}
{Mandy}, M.~E., \& {Martin}, P.~G. 1993, \apjs, 86, 199

\bibitem[{{Manoj} {et~al.}(2006){Manoj}, {Bhatt}, {Maheswar}, \&
  {Muneer}}]{manoj06}
{Manoj}, P., {Bhatt}, H.~C., {Maheswar}, G., \& {Muneer}, S. 2006, \apj, 653,
  657

\bibitem[{{Mathis}(1990)}]{mathis90}
{Mathis}, J.~S. 1990, \araa, 28, 37

\bibitem[{{Mathis} {et~al.}(1977){Mathis}, {Rumpl}, \& {Nordsieck}}]{mathis77}
{Mathis}, J.~S., {Rumpl}, W., \& {Nordsieck}, K.~H. 1977, \apj, 217, 425

\bibitem[{{Mawet} {et~al.}(2017){Mawet}, {Choquet}, {Absil}, {Huby}, {Bottom},
  {Serabyn}, {Femenia}, {Lebreton}, {Matthews}, {Gomez Gonzalez}, {Wertz},
  {Carlomagno}, {Christiaens}, {Defr{\`e}re}, {Delacroix}, {Forsberg},
  {Habraken}, {Jolivet}, {Karlsson}, {Milli}, {Pinte}, {Piron}, {Reggiani},
  {Surdej}, \& {Vargas Catalan}}]{mawet17}
{Mawet}, D., {Choquet}, {\'E}., {Absil}, O., {et~al.} 2017, \aj, 153, 44

\bibitem[{{Mayama} {et~al.}(2012){Mayama}, {Hashimoto}, {Muto}, {Tsukagoshi},
  {Kusakabe}, {Kuzuhara}, {Takahashi}, {Kudo}, {Dong}, {Fukagawa}, {Takami},
  {Momose}, {Wisniewski}, {Follette}, {Abe}, {Akiyama}, {Brandner}, {Brandt},
  {Carson}, {Egner}, {Feldt}, {Goto}, {Grady}, {Guyon}, {Hayano}, {Hayashi},
  {Hayashi}, {Henning}, {Hodapp}, {Ishii}, {Iye}, {Janson}, {Kandori}, {Kwon},
  {Knapp}, {Matsuo}, {McElwain}, {Miyama}, {Morino}, {Moro-Martin},
  {Nishimura}, {Pyo}, {Serabyn}, {Suto}, {Suzuki}, {Takato}, {Terada},
  {Thalmann}, {Tomono}, {Turner}, {Watanabe}, {Yamada}, {Takami}, {Usuda}, \&
  {Tamura}}]{mayama12}
{Mayama}, S., {Hashimoto}, J., {Muto}, T., {et~al.} 2012, \apjl, 760, L26

\bibitem[{{McCabe} {et~al.}(2002){McCabe}, {Duch{\^e}ne}, \& {Ghez}}]{mccabe02}
{McCabe}, C., {Duch{\^e}ne}, G., \& {Ghez}, A.~M. 2002, \apj, 575, 974

\bibitem[{{McClure} {et~al.}(2016){McClure}, {Bergin}, {Cleeves}, {van
  Dishoeck}, {Blake}, {Evans}, {Green}, {Henning}, {{\"O}berg}, {Pontoppidan},
  \& {Salyk}}]{mcclure16}
{McClure}, M.~K., {Bergin}, E.~A., {Cleeves}, L.~I., {et~al.} 2016, \apj, 831,
  167

\bibitem[{{Mendigut{\'{\i}}a} {et~al.}(2017{\natexlab{a}}){Mendigut{\'{\i}}a},
  {Oudmaijer}, {Mourard}, \& {Muzerolle}}]{mendigutia17}
{Mendigut{\'{\i}}a}, I., {Oudmaijer}, R.~D., {Mourard}, D., \& {Muzerolle}, J.
  2017{\natexlab{a}}, \mnras, 464, 1984

\bibitem[{{Mendigut{\'{\i}}a} {et~al.}(2017{\natexlab{b}}){Mendigut{\'{\i}}a},
  {Oudmaijer}, {Garufi}, {Lumsden}, {Hu{\'e}lamo}, {Cheetham}, {de Wit},
  {Norris}, {Olguin}, \& {Tuthill}}]{mendigutia17hd100546}
{Mendigut{\'{\i}}a}, I., {Oudmaijer}, R.~D., {Garufi}, A., {et~al.}
  2017{\natexlab{b}}, \aap, 608, A104

\bibitem[{{Meru} {et~al.}(2017){Meru}, {Juh{\'a}sz}, {Ilee}, {Clarke},
  {Rosotti}, \& {Booth}}]{meru17}
{Meru}, F., {Juh{\'a}sz}, A., {Ilee}, J.~D., {et~al.} 2017, \apjl, 839, L24

\bibitem[{{Momose} {et~al.}(2010){Momose}, {Ohashi}, {Kudo}, {Tamura}, \&
  {Kitamura}}]{momose10}
{Momose}, M., {Ohashi}, N., {Kudo}, T., {Tamura}, M., \& {Kitamura}, Y. 2010,
  \apj, 712, 397

\bibitem[{{Momose} {et~al.}(2015){Momose}, {Morita}, {Fukagawa}, {Muto},
  {Takeuchi}, {Hashimoto}, {Honda}, {Kudo}, {Okamoto}, {Kanagawa}, {Tanaka},
  {Grady}, {Sitko}, {Akiyama}, {Currie}, {Follette}, {Mayama}, {Kusakabe},
  {Abe}, {Brandner}, {Brandt}, {Carson}, {Egner}, {Feldt}, {Goto}, {Guyon},
  {Hayano}, {Hayashi}, {Hayashi}, {Henning}, {Hodapp}, {Ishii}, {Iye},
  {Janson}, {Kandori}, {Knapp}, {Kuzuhara}, {Kwon}, {Matsuo}, {McElwain},
  {Miyama}, {Morino}, {Moro-Martin}, {Nishimura}, {Pyo}, {Serabyn}, {Suenaga},
  {Suto}, {Suzuki}, {Takahashi}, {Takami}, {Takato}, {Terada}, {Thalmann},
  {Tomono}, {Turner}, {Watanabe}, {Wisniewski}, {Yamada}, {Takami}, {Usuda}, \&
  {Tamura}}]{momose15}
{Momose}, M., {Morita}, A., {Fukagawa}, M., {et~al.} 2015, \pasj, 67, 83

\bibitem[{{Monnier} {et~al.}(2017){Monnier}, {Harries}, {Aarnio}, {Adams},
  {Andrews}, {Calvet}, {Espaillat}, {Hartmann}, {Hinkley}, {Kraus}, {McClure},
  {Oppenheimer}, {Perrin}, \& {Wilner}}]{monnier17}
{Monnier}, J.~D., {Harries}, T.~J., {Aarnio}, A., {et~al.} 2017, \apj, 838, 20

\bibitem[{{Montesinos} {et~al.}(2016){Montesinos}, {Perez}, {Casassus},
  {Marino}, {Cuadra}, \& {Christiaens}}]{montesinos16}
{Montesinos}, M., {Perez}, S., {Casassus}, S., {et~al.} 2016, \apjl, 823, L8

\bibitem[{{Morzinski} {et~al.}(2014){Morzinski}, {Close}, {Males}, {Kopon},
  {Hinz}, {Esposito}, {Riccardi}, {Puglisi}, {Pinna}, {Briguglio}, {Xompero},
  {Quir{\'o}s-Pacheco}, {Bailey}, {Follette}, {Rodigas}, {Wu}, {Arcidiacono},
  {Argomedo}, {Busoni}, {Hare}, {Uomoto}, \& {Weinberger}}]{morzinski14}
{Morzinski}, K.~M., {Close}, L.~M., {Males}, J.~R., {et~al.} 2014, in
  \procspie, Vol. 9148, Adaptive Optics Systems IV, 914804

\bibitem[{{Mouillet} {et~al.}(2001){Mouillet}, {Lagrange}, {Augereau}, \&
  {M{\'e}nard}}]{mouillet01}
{Mouillet}, D., {Lagrange}, A.~M., {Augereau}, J.~C., \& {M{\'e}nard}, F. 2001,
  \aap, 372, L61

\bibitem[{{Mu{\~n}oz} \& {Lai}(2016)}]{munoz16}
{Mu{\~n}oz}, D.~J., \& {Lai}, D. 2016, \apj, 827, 43

\bibitem[{{Murakawa} {et~al.}(2008){Murakawa}, {Oya}, {Pyo}, \&
  {Ishii}}]{murakawa08}
{Murakawa}, K., {Oya}, S., {Pyo}, T.-S., \& {Ishii}, M. 2008, \aap, 492, 731

\bibitem[{{Murphy} \& {Paunzen}(2017)}]{murphy17}
{Murphy}, S.~J., \& {Paunzen}, E. 2017, \mnras, 466, 546

\bibitem[{{Muto} {et~al.}(2012){Muto}, {Grady}, {Hashimoto}, {Fukagawa},
  {Hornbeck}, {Sitko}, {Russell}, {Werren}, {Cur{\'e}}, {Currie}, {Ohashi},
  {Okamoto}, {Momose}, {Honda}, {Inutsuka}, {Takeuchi}, {Dong}, {Abe},
  {Brandner}, {Brandt}, {Carson}, {Egner}, {Feldt}, {Fukue}, {Goto}, {Guyon},
  {Hayano}, {Hayashi}, {Hayashi}, {Henning}, {Hodapp}, {Ishii}, {Iye},
  {Janson}, {Kandori}, {Knapp}, {Kudo}, {Kusakabe}, {Kuzuhara}, {Matsuo},
  {Mayama}, {McElwain}, {Miyama}, {Morino}, {Moro-Martin}, {Nishimura}, {Pyo},
  {Serabyn}, {Suto}, {Suzuki}, {Takami}, {Takato}, {Terada}, {Thalmann},
  {Tomono}, {Turner}, {Watanabe}, {Wisniewski}, {Yamada}, {Takami}, {Usuda}, \&
  {Tamura}}]{muto12}
{Muto}, T., {Grady}, C.~A., {Hashimoto}, J., {et~al.} 2012, \apjl, 748, L22

\bibitem[{{Muzerolle} {et~al.}(2004){Muzerolle}, {D'Alessio}, {Calvet}, \&
  {Hartmann}}]{muzerolle04}
{Muzerolle}, J., {D'Alessio}, P., {Calvet}, N., \& {Hartmann}, L. 2004, \apj,
  617, 406

\bibitem[{{Najita} \& {Kenyon}(2014)}]{najita14}
{Najita}, J.~R., \& {Kenyon}, S.~J. 2014, \mnras, 445, 3315

\bibitem[{{{\"O}berg} {et~al.}(2015){{\"O}berg}, {Guzm{\'a}n}, {Furuya}, {Qi},
  {Aikawa}, {Andrews}, {Loomis}, \& {Wilner}}]{oberg15}
{{\"O}berg}, K.~I., {Guzm{\'a}n}, V.~V., {Furuya}, K., {et~al.} 2015, \nat,
  520, 198

\bibitem[{{Ogilvie} \& {Lubow}(2002)}]{ogilvie02}
{Ogilvie}, G.~I., \& {Lubow}, S.~H. 2002, \mnras, 330, 950

\bibitem[{{Oh} {et~al.}(2016{\natexlab{a}}){Oh}, {Hashimoto}, {Tamura},
  {Wisniewski}, {Akiyama}, {Currie}, {Mayama}, {Takami}, {Thalmann}, {Kudo},
  {Kusakabe}, {Abe}, {Brandner}, {Brandt}, {Carson}, {Egner}, {Feldt}, {Goto},
  {Grady}, {Guyon}, {Hayano}, {Hayashi}, {Hayashi}, {Henning}, {Hodapp},
  {Ishii}, {Iye}, {Janson}, {Kandori}, {Knapp}, {Kuzuhara}, {Kwon}, {Matsuo},
  {Mcelwain}, {Miyama}, {Morino}, {Moro-Martin}, {Nishimura}, {Pyo}, {Serabyn},
  {Suenaga}, {Suto}, {Suzuki}, {Takahashi}, {Takato}, {Terada}, {Turner},
  {Watanabe}, {Yamada}, {Takami}, \& {Usuda}}]{oh16lkca15}
{Oh}, D., {Hashimoto}, J., {Tamura}, M., {et~al.} 2016{\natexlab{a}}, \pasj,
  68, L3

\bibitem[{{Oh} {et~al.}(2016{\natexlab{b}}){Oh}, {Hashimoto}, {Carson},
  {Janson}, {Kwon}, {Nakagawa}, {Mayama}, {Uyama}, {Yang}, {Kudo}, {Kusakabe},
  {Abe}, {Akiyama}, {Brandner}, {Brandt}, {Currie}, {Feldt}, {Goto}, {Grady},
  {Guyon}, {Hayano}, {Hayashi}, {Hayashi}, {Henning}, {Hodapp}, {Ishii}, {Iye},
  {Kandori}, {Knapp}, {Kuzuhara}, {Matsuo}, {Mcelwain}, {Miyama}, {Morino},
  {Moro-Martin}, {Nishimura}, {Pyo}, {Serabyn}, {Suenaga}, {Suto}, {Suzuki},
  {Takahashi}, {Takato}, {Terada}, {Thalmann}, {Turner}, {Watanabe}, {Yamada},
  {Takami}, {Usuda}, \& {Tamura}}]{oh16gmaur}
{Oh}, D., {Hashimoto}, J., {Carson}, J.~C., {et~al.} 2016{\natexlab{b}}, \apjl,
  831, L7

\bibitem[{{Ohta} {et~al.}(2016){Ohta}, {Fukagawa}, {Sitko}, {Muto}, {Kraus},
  {Grady}, {Wisniewski}, {Swearingen}, {Shibai}, {Sumi}, {Hashimoto}, {Kudo},
  {Kusakabe}, {Momose}, {Okamoto}, {Kotani}, {Takami}, {Currie}, {Thalmann},
  {Janson}, {Akiyama}, {Follette}, {Mayama}, {Abe}, {Brandner}, {Brandt},
  {Carson}, {Egner}, {Feldt}, {Goto}, {Guyon}, {Hayano}, {Hayashi}, {Hayashi},
  {Henning}, {Hodapp}, {Ishii}, {Iye}, {Kandori}, {Knapp}, {Kuzuhara}, {Kwon},
  {Matsuo}, {McElwain}, {Miyama}, {Morino}, {Moro-Mart{\'{\i}}n}, {Nishimura},
  {Pyo}, {Serabyn}, {Suenaga}, {Suto}, {Suzuki}, {Takahashi}, {Takami},
  {Takato}, {Terada}, {Tomono}, {Turner}, {Usuda}, {Watanabe}, {Yamada}, \&
  {Tamura}}]{ohta16}
{Ohta}, Y., {Fukagawa}, M., {Sitko}, M.~L., {et~al.} 2016, \pasj, 68, 53

\bibitem[{{Pascucci} {et~al.}(2016){Pascucci}, {Testi}, {Herczeg}, {Long},
  {Manara}, {Hendler}, {Mulders}, {Krijt}, {Ciesla}, {Henning}, {Mohanty},
  {Drabek-Maunder}, {Apai}, {Sz{\H u}cs}, {Sacco}, \& {Olofsson}}]{pascucci16}
{Pascucci}, I., {Testi}, L., {Herczeg}, G.~J., {et~al.} 2016, \apj, 831, 125

\bibitem[{{Pecaut} \& {Mamajek}(2013)}]{pecaut13}
{Pecaut}, M.~J., \& {Mamajek}, E.~E. 2013, \apjs, 208, 9

\bibitem[{{P{\'e}rez} {et~al.}(2016){P{\'e}rez}, {Carpenter}, {Andrews},
  {Ricci}, {Isella}, {Linz}, {Sargent}, {Wilner}, {Henning}, {Deller},
  {Chandler}, {Dullemond}, {Lazio}, {Menten}, {Corder}, {Storm}, {Testi},
  {Tazzari}, {Kwon}, {Calvet}, {Greaves}, {Harris}, \& {Mundy}}]{perez16}
{P{\'e}rez}, L.~M., {Carpenter}, J.~M., {Andrews}, S.~M., {et~al.} 2016,
  Science, 353, 1519

\bibitem[{{Perrin} {et~al.}(2006){Perrin}, {Duch{\^e}ne}, {Kalas}, \&
  {Graham}}]{perrin06}
{Perrin}, M.~D., {Duch{\^e}ne}, G., {Kalas}, P., \& {Graham}, J.~R. 2006, \apj,
  645, 1272

\bibitem[{{Perrin} {et~al.}(2009){Perrin}, {Schneider}, {Duchene}, {Pinte},
  {Grady}, {Wisniewski}, \& {Hines}}]{perrin09}
{Perrin}, M.~D., {Schneider}, G., {Duchene}, G., {et~al.} 2009, \apjl, 707,
  L132

\bibitem[{{Perrot} {et~al.}(2016){Perrot}, {Boccaletti}, {Pantin}, {Augereau},
  {Lagrange}, {Galicher}, {Maire}, {Mazoyer}, {Milli}, {Rousset}, {Gratton},
  {Bonnefoy}, {Brandner}, {Buenzli}, {Langlois}, {Lannier}, {Mesa}, {Peretti},
  {Salter}, {Sissa}, {Chauvin}, {Desidera}, {Feldt}, {Vigan}, {Di Folco},
  {Dutrey}, {P{\'e}ricaud}, {Baudoz}, {Benisty}, {De Boer}, {Garufi}, {Girard},
  {Menard}, {Olofsson}, {Quanz}, {Mouillet}, {Christiaens}, {Casassus},
  {Beuzit}, {Blanchard}, {Carle}, {Fusco}, {Giro}, {Hubin}, {Maurel},
  {Moeller-Nilsson}, {Sevin}, \& {Weber}}]{perrot16}
{Perrot}, C., {Boccaletti}, A., {Pantin}, E., {et~al.} 2016, \aap, 590, L7

\bibitem[{{Pfalzner}(2003)}]{pfalzner03}
{Pfalzner}, S. 2003, \apj, 592, 986

\bibitem[{{Pinilla} {et~al.}(2012){Pinilla}, {Birnstiel}, {Ricci}, {Dullemond},
  {Uribe}, {Testi}, \& {Natta}}]{pinilla12dusttrapping}
{Pinilla}, P., {Birnstiel}, T., {Ricci}, L., {et~al.} 2012, \aap, 538, A114

\bibitem[{{Pinilla} {et~al.}(2015){Pinilla}, {de Boer}, {Benisty},
  {Juh{\'a}sz}, {de Juan Ovelar}, {Dominik}, {Avenhaus}, {Birnstiel}, {Girard},
  {Huelamo}, {Isella}, \& {Milli}}]{pinilla15j1604}
{Pinilla}, P., {de Boer}, J., {Benisty}, M., {et~al.} 2015, \aap, 584, L4

\bibitem[{{Pinte} {et~al.}(2008){Pinte}, {Padgett}, {M{\'e}nard},
  {Stapelfeldt}, {Schneider}, {Olofsson}, {Pani{\'c}}, {Augereau},
  {Duch{\^e}ne}, {Krist}, {Pontoppidan}, {Perrin}, {Grady}, {Kessler-Silacci},
  {van Dishoeck}, {Lommen}, {Silverstone}, {Hines}, {Wolf}, {Blake}, {Henning},
  \& {Stecklum}}]{pinte08}
{Pinte}, C., {Padgett}, D.~L., {M{\'e}nard}, F., {et~al.} 2008, \aap, 489, 633

\bibitem[{{Pohl} {et~al.}(2017{\natexlab{a}}){Pohl}, {Sissa}, {Langlois},
  {M{\"u}ller}, {Ginski}, {van Holstein}, {Vigan}, {Mesa}, {Maire}, {Henning},
  {Gratton}, {Olofsson}, {van Boekel}, {Benisty}, {Biller}, {Boccaletti},
  {Chauvin}, {Daemgen}, {de Boer}, {Desidera}, {Dominik}, {Garufi}, {Janson},
  {Kral}, {M{\'e}nard}, {Pinte}, {Stolker}, {Szul{\'a}gyi}, {Zurlo},
  {Bonnefoy}, {Cheetham}, {Cudel}, {Feldt}, {Kasper}, {Lagrange}, {Perrot}, \&
  {Wildi}}]{pohl17tcha}
{Pohl}, A., {Sissa}, E., {Langlois}, M., {et~al.} 2017{\natexlab{a}}, \aap,
  605, A34

\bibitem[{{Pohl} {et~al.}(2017{\natexlab{b}}){Pohl}, {Benisty}, {Pinilla},
  {Ginski}, {de Boer}, {Avenhaus}, {Henning}, {Zurlo}, {Boccaletti},
  {Augereau}, {Birnstiel}, {Dominik}, {Facchini}, {Fedele}, {Janson},
  {Keppler}, {Kral}, {Langlois}, {Ligi}, {Maire}, {M{\'e}nard}, {Meyer},
  {Pinte}, {Quanz}, {Sauvage}, {Sezestre}, {Stolker}, {Szul{\'a}gyi}, {van
  Boekel}, {van der Plas}, {Villenave}, {Baruffolo}, {Baudoz}, {Le Mignant},
  {Maurel}, {Ramos}, \& {Weber}}]{pohl17hd169142}
{Pohl}, A., {Benisty}, M., {Pinilla}, P., {et~al.} 2017{\natexlab{b}}, \apj,
  850, 52

\bibitem[{{Price} {et~al.}(2018){Price}, {Cuello}, {Pinte}, {Mentiplay},
  {Casassus}, {Christiaens}, {Kennedy}, {Cuadra}, {Sebastian Perez}, {Marino},
  {Armitage}, {Zurlo}, {Juhasz}, {Ragusa}, {Laibe}, \& {Lodato}}]{price18}
{Price}, D.~J., {Cuello}, N., {Pinte}, C., {et~al.} 2018, \mnras, 477, 1270

\bibitem[{{Quanz} {et~al.}(2013){Quanz}, {Avenhaus}, {Buenzli}, {Garufi},
  {Schmid}, \& {Wolf}}]{quanz13gap}
{Quanz}, S.~P., {Avenhaus}, H., {Buenzli}, E., {et~al.} 2013, \apjl, 766, L2

\bibitem[{{Quanz} {et~al.}(2012){Quanz}, {Birkmann}, {Apai}, {Wolf}, \&
  {Henning}}]{quanz12}
{Quanz}, S.~P., {Birkmann}, S.~M., {Apai}, D., {Wolf}, S., \& {Henning}, T.
  2012, \aap, 538, A92

\bibitem[{{Quanz} {et~al.}(2011){Quanz}, {Schmid}, {Geissler}, {Meyer},
  {Henning}, {Brandner}, \& {Wolf}}]{quanz11}
{Quanz}, S.~P., {Schmid}, H.~M., {Geissler}, K., {et~al.} 2011, \apj, 738, 23

\bibitem[{{Rameau} {et~al.}(2012){Rameau}, {Chauvin}, {Lagrange},
  {Th{\'e}bault}, {Milli}, {Girard}, \& {Bonnefoy}}]{rameau12}
{Rameau}, J., {Chauvin}, G., {Lagrange}, A.-M., {et~al.} 2012, \aap, 546, A24

\bibitem[{{Rameau} {et~al.}(2017){Rameau}, {Follette}, {Pueyo}, {Marois},
  {Macintosh}, {Millar-Blanchaer}, {Wang}, {Vega}, {Doyon}, {Lafreni{\`e}re},
  {Nielsen}, {Bailey}, {Chilcote}, {Close}, {Esposito}, {Males}, {Metchev},
  {Morzinski}, {Ruffio}, {Wolff}, {Ammons}, {Barman}, {Bulger}, {Cotten}, {De
  Rosa}, {Duchene}, {Fitzgerald}, {Goodsell}, {Graham}, {Greenbaum}, {Hibon},
  {Hung}, {Ingraham}, {Kalas}, {Konopacky}, {Larkin}, {Maire}, {Marchis},
  {Oppenheimer}, {Palmer}, {Patience}, {Perrin}, {Poyneer}, {Rajan},
  {Rantakyr{\"o}}, {Marley}, {Savransky}, {Schneider}, {Sivaramakrishnan},
  {Song}, {Soummer}, {Thomas}, {Wallace}, {Ward-Duong}, \&
  {Wiktorowicz}}]{rameau17}
{Rameau}, J., {Follette}, K.~B., {Pueyo}, L., {et~al.} 2017, \aj, 153, 244

\bibitem[{{Rapson} {et~al.}(2015{\natexlab{a}}){Rapson}, {Kastner}, {Andrews},
  {Hines}, {Macintosh}, {Millar-Blanchaer}, \& {Tamura}}]{rapson15v4046}
{Rapson}, V.~A., {Kastner}, J.~H., {Andrews}, S.~M., {et~al.}
  2015{\natexlab{a}}, \apjl, 803, L10

\bibitem[{{Rapson} {et~al.}(2015{\natexlab{b}}){Rapson}, {Kastner},
  {Millar-Blanchaer}, \& {Dong}}]{rapson15twhya}
{Rapson}, V.~A., {Kastner}, J.~H., {Millar-Blanchaer}, M.~A., \& {Dong}, R.
  2015{\natexlab{b}}, \apjl, 815, L26

\bibitem[{{Reggiani} {et~al.}(2018){Reggiani}, {Christiaens}, {Absil}, {Mawet},
  {Huby}, {Choquet}, {Gomez Gonzalez}, {Ruane}, {Femenia}, {Serabyn},
  {Matthews}, {Barraza}, {Carlomagno}, {Defr{\`e}re}, {Delacroix}, {Habraken},
  {Jolivet}, {Karlsson}, {Orban de Xivry}, {Piron}, {Surdej}, {Vargas Catalan},
  \& {Wertz}}]{reggiani18}
{Reggiani}, M., {Christiaens}, V., {Absil}, O., {et~al.} 2018, \aap, 611, A74

\bibitem[{{Ren} {et~al.}(2018){Ren}, {Dong}, {Esposito}, {Pueyo}, {Debes},
  {Poteet}, {Choquet}, {Benisty}, {Chiang}, {Grady}, {Hines}, {Schneider}, \&
  {Soummer}}]{ren18}
{Ren}, B., {Dong}, R., {Esposito}, T.~M., {et~al.} 2018, \apjl, 857, L9

\bibitem[{{Rice} \& {Armitage}(2009)}]{rice09}
{Rice}, W.~K.~M., \& {Armitage}, P.~J. 2009, \mnras, 396, 2228

\bibitem[{{Rice} {et~al.}(2003){Rice}, {Armitage}, {Bate}, \&
  {Bonnell}}]{rice03}
{Rice}, W.~K.~M., {Armitage}, P.~J., {Bate}, M.~R., \& {Bonnell}, I.~A. 2003,
  \mnras, 339, 1025

\bibitem[{{Rich} {et~al.}(2015){Rich}, {Wisniewski}, {Mayama}, {Brandt},
  {Hashimoto}, {Kudo}, {Kusakabe}, {Espaillat}, {Abe}, {Akiyama}, {Brandner},
  {Carson}, {Currie}, {Egner}, {Feldt}, {Follette}, {Goto}, {Grady}, {Guyon},
  {Hayano}, {Hayashi}, {Hayashi}, {Henning}, {Hodapp}, {Ishii}, {Iye},
  {Janson}, {Kandori}, {Knapp}, {Kuzuhara}, {Kwon}, {Matsuo}, {McElwain},
  {Miyama}, {Morino}, {Moro-Martin}, {Nishimura}, {Pyo}, {Qi}, {Serabyn},
  {Suenaga}, {Suto}, {Suzuki}, {Takahashi}, {Takami}, {Takato}, {Terada},
  {Thalmann}, {Tomono}, {Turner}, {Watanabe}, {Yamada}, {Takami}, {Usuda}, \&
  {Tamura}}]{rich15}
{Rich}, E.~A., {Wisniewski}, J.~P., {Mayama}, S., {et~al.} 2015, \aj, 150, 86

\bibitem[{{Richert} {et~al.}(2018){Richert}, {Getman}, {Feigelson}, {Kuhn},
  {Broos}, {Povich}, {Bate}, \& {Garmire}}]{richert18}
{Richert}, A.~J.~W., {Getman}, K.~V., {Feigelson}, E.~D., {et~al.} 2018,
  \mnras, 477, 5191

\bibitem[{{Rodgers}(2001)}]{rodgers01}
{Rodgers}, B.~M. 2001, PhD thesis, UNIVERSITY OF WASHINGTON

\bibitem[{{Rodigas} {et~al.}(2014){Rodigas}, {Follette}, {Weinberger}, {Close},
  \& {Hines}}]{rodigas14}
{Rodigas}, T.~J., {Follette}, K.~B., {Weinberger}, A., {Close}, L., \& {Hines},
  D.~C. 2014, \apjl, 791, L37

\bibitem[{{Rosenfeld} {et~al.}(2013){Rosenfeld}, {Andrews}, {Wilner},
  {Kastner}, \& {McClure}}]{rosenfeld13}
{Rosenfeld}, K.~A., {Andrews}, S.~M., {Wilner}, D.~J., {Kastner}, J.~H., \&
  {McClure}, M.~K. 2013, \apj, 775, 136

\bibitem[{{Rosotti} {et~al.}(2017){Rosotti}, {Clarke}, {Manara}, \&
  {Facchini}}]{rosotti17}
{Rosotti}, G.~P., {Clarke}, C.~J., {Manara}, C.~F., \& {Facchini}, S. 2017,
  \mnras, 468, 1631

\bibitem[{{Sandell} {et~al.}(2011){Sandell}, {Weintraub}, \&
  {Hamidouche}}]{sandell11}
{Sandell}, G., {Weintraub}, D.~A., \& {Hamidouche}, M. 2011, \apj, 727, 26

\bibitem[{{Sartori} {et~al.}(2010){Sartori}, {Gregorio-Hetem}, {Rodrigues},
  {Hetem}, \& {Batalha}}]{sartori10}
{Sartori}, M.~J., {Gregorio-Hetem}, J., {Rodrigues}, C.~V., {Hetem}, Jr., A.,
  \& {Batalha}, C. 2010, \aj, 139, 27

\bibitem[{{Schneider} {et~al.}(2014){Schneider}, {Grady}, {Hines}, {Stark},
  {Debes}, {Carson}, {Kuchner}, {Perrin}, {Weinberger}, {Wisniewski},
  {Silverstone}, {Jang-Condell}, {Henning}, {Woodgate}, {Serabyn},
  {Moro-Martin}, {Tamura}, {Hinz}, \& {Rodigas}}]{schneider14}
{Schneider}, G., {Grady}, C.~A., {Hines}, D.~C., {et~al.} 2014, \aj, 148, 59

\bibitem[{{Sicilia-Aguilar} {et~al.}(2010){Sicilia-Aguilar}, {Henning}, \&
  {Hartmann}}]{siciliaaguilar10}
{Sicilia-Aguilar}, A., {Henning}, T., \& {Hartmann}, L.~W. 2010, \apj, 710, 597

\bibitem[{{Siess} {et~al.}(2000){Siess}, {Dufour}, \& {Forestini}}]{siess00}
{Siess}, L., {Dufour}, E., \& {Forestini}, M. 2000, \aap, 358, 593

\bibitem[{{Silber} {et~al.}(2000){Silber}, {Gledhill}, {Duch{\^e}ne}, \&
  {M{\'e}nard}}]{silber00}
{Silber}, J., {Gledhill}, T., {Duch{\^e}ne}, G., \& {M{\'e}nard}, F. 2000,
  \apjl, 536, L89

\bibitem[{{Spezzi} {et~al.}(2008){Spezzi}, {Alcal{\'a}}, {Covino}, {Frasca},
  {Gandolfi}, {Oliveira}, {Chapman}, {Evans}, {Huard}, {J{\o}rgensen},
  {Mer{\'{\i}}n}, \& {Stapelfeldt}}]{spezzi08}
{Spezzi}, L., {Alcal{\'a}}, J.~M., {Covino}, E., {et~al.} 2008, \apj, 680, 1295

\bibitem[{{Spiegel} \& {Burrows}(2012)}]{spiegel12}
{Spiegel}, D.~S., \& {Burrows}, A. 2012, \apj, 745, 174

\bibitem[{{Stamatellos} \& {Whitworth}(2008)}]{stamatellos08}
{Stamatellos}, D., \& {Whitworth}, A.~P. 2008, \aap, 480, 879

\bibitem[{{Stapelfeldt} {et~al.}(1998){Stapelfeldt}, {Krist}, {M{\'e}nard},
  {Bouvier}, {Padgett}, \& {Burrows}}]{stapelfeldt98}
{Stapelfeldt}, K.~R., {Krist}, J.~E., {M{\'e}nard}, F., {et~al.} 1998, \apjl,
  502, L65

\bibitem[{{Stapelfeldt} {et~al.}(2003){Stapelfeldt}, {M{\'e}nard}, {Watson},
  {Krist}, {Dougados}, {Padgett}, \& {Brandner}}]{stapelfeldt03}
{Stapelfeldt}, K.~R., {M{\'e}nard}, F., {Watson}, A.~M., {et~al.} 2003, \apj,
  589, 410

\bibitem[{{Stapelfeldt} {et~al.}(1999){Stapelfeldt}, {Watson}, {Krist},
  {Burrows}, {Crisp}, {Ballester}, {Clarke}, {Evans}, {Gallagher}, {Griffiths},
  {Hester}, {Hoessel}, {Holtzman}, {Mould}, {Scowen}, \&
  {Trauger}}]{stapelfeldt99}
{Stapelfeldt}, K.~R., {Watson}, A.~M., {Krist}, J.~E., {et~al.} 1999, \apjl,
  516, L95

\bibitem[{{Stolker} {et~al.}(2016){Stolker}, {Dominik}, {Avenhaus}, {Min}, {de
  Boer}, {Ginski}, {Schmid}, {Juhasz}, {Bazzon}, {Waters}, {Garufi},
  {Augereau}, {Benisty}, {Boccaletti}, {Henning}, {Langlois}, {Maire},
  {M{\'e}nard}, {Meyer}, {Pinte}, {Quanz}, {Thalmann}, {Beuzit}, {Carbillet},
  {Costille}, {Dohlen}, {Feldt}, {Gisler}, {Mouillet}, {Pavlov}, {Perret},
  {Petit}, {Pragt}, {Rochat}, {Roelfsema}, {Salasnich}, {Soenke}, \&
  {Wildi}}]{stolker16sao206462}
{Stolker}, T., {Dominik}, C., {Avenhaus}, H., {et~al.} 2016, \aap, 595, A113

\bibitem[{{Stolker} {et~al.}(2017){Stolker}, {Sitko}, {Lazareff}, {Benisty},
  {Dominik}, {Waters}, {Min}, {Perez}, {Milli}, {Garufi}, {de Boer}, {Ginski},
  {Kraus}, {Berger}, \& {Avenhaus}}]{stolker17}
{Stolker}, T., {Sitko}, M., {Lazareff}, B., {et~al.} 2017, \apj, 849, 143

\bibitem[{{Takami} {et~al.}(2013){Takami}, {Karr}, {Hashimoto}, {Kim},
  {Wisniewski}, {Henning}, {Grady}, {Kandori}, {Hodapp}, {Kudo}, {Kusakabe},
  {Chou}, {Itoh}, {Momose}, {Mayama}, {Currie}, {Follette}, {Kwon}, {Abe},
  {Brandner}, {Brandt}, {Carson}, {Egner}, {Feldt}, {Guyon}, {Hayano},
  {Hayashi}, {Hayashi}, {Ishii}, {Iye}, {Janson}, {Knapp}, {Kuzuhara},
  {McElwain}, {Matsuo}, {Miyama}, {Morino}, {Moro-Martin}, {Nishimura}, {Pyo},
  {Serabyn}, {Suto}, {Suzuki}, {Takato}, {Terada}, {Thalmann}, {Tomono},
  {Turner}, {Watanabe}, {Yamada}, {Takami}, {Usuda}, \& {Tamura}}]{takami13}
{Takami}, M., {Karr}, J.~L., {Hashimoto}, J., {et~al.} 2013, \apj, 772, 145

\bibitem[{{Tamura}(2009)}]{tamura09}
{Tamura}, M. 2009, in American Institute of Physics Conference Series, Vol.
  1158, American Institute of Physics Conference Series, ed. T.~{Usuda},
  M.~{Tamura}, \& M.~{Ishii}, 11--16

\bibitem[{{Tamura} {et~al.}(2000){Tamura}, {Suto}, {Itoh}, {Ebizuka}, {Doi},
  {Murakawa}, {Hayashi}, {Oasa}, {Takami}, \& {Kaifu}}]{tamura00}
{Tamura}, M., {Suto}, H., {Itoh}, Y., {et~al.} 2000, in \procspie, Vol. 4008,
  Optical and IR Telescope Instrumentation and Detectors, ed. M.~{Iye} \& A.~F.
  {Moorwood}, 1153--1161

\bibitem[{{Tamura} {et~al.}(2006){Tamura}, {Hodapp}, {Takami}, {Abe}, {Suto},
  {Guyon}, {Jacobson}, {Kandori}, {Morino}, {Murakami}, {Stahlberger},
  {Suzuki}, {Tavrov}, {Yamada}, {Nishikawa}, {Ukita}, {Hashimoto}, {Izumiura},
  {Hayashi}, {Nakajima}, \& {Nishimura}}]{tamura06}
{Tamura}, M., {Hodapp}, K., {Takami}, H., {et~al.} 2006, in \procspie, Vol.
  6269, Society of Photo-Optical Instrumentation Engineers (SPIE) Conference
  Series, 62690V

\bibitem[{{Tang} {et~al.}(2012){Tang}, {Guilloteau}, {Pi{\'e}tu}, {Dutrey},
  {Ohashi}, \& {Ho}}]{tang12}
{Tang}, Y.-W., {Guilloteau}, S., {Pi{\'e}tu}, V., {et~al.} 2012, \aap, 547, A84

\bibitem[{{Tang} {et~al.}(2017){Tang}, {Guilloteau}, {Dutrey}, {Muto}, {Shen},
  {Gu}, {Inutsuka}, {Momose}, {Pietu}, {Fukagawa}, {Chapillon}, {Ho}, {di
  Folco}, {Corder}, {Ohashi}, \& {Hashimoto}}]{tang17}
{Tang}, Y.-W., {Guilloteau}, S., {Dutrey}, A., {et~al.} 2017, \apj, 840, 32

\bibitem[{{Tanii} {et~al.}(2012){Tanii}, {Itoh}, {Kudo}, {Hioki}, {Oasa},
  {Gupta}, {Sen}, {Wisniewski}, {Muto}, {Grady}, {Hashimoto}, {Fukagawa},
  {Mayama}, {Hornbeck}, {Sitko}, {Russell}, {Werren}, {Cur{\'e}}, {Currie},
  {Ohashi}, {Okamoto}, {Momose}, {Honda}, {Inutsuka}, {Takeuchi}, {Dong},
  {Abe}, {Brandner}, {Brandt}, {Carson}, {Egner}, {Feldt}, {Fukue}, {Goto},
  {Guyon}, {Hayano}, {Hayashi}, {Hayashi}, {Henning}, {Hodapp}, {Ishii}, {Iye},
  {Janson}, {Kandori}, {Knapp}, {Kusakabe}, {Kuzuhara}, {Matsuo}, {McElwain},
  {Miyama}, {Morino}, {Moro-Mart{\'{\i}}n}, {Nishimura}, {Pyo}, {Serabyn},
  {Suto}, {Suzuki}, {Takami}, {Takato}, {Terada}, {Thalmann}, {Tomono},
  {Turner}, {Watanabe}, {Yamada}, {Takami}, {Usuda}, \& {Tamura}}]{tanii12}
{Tanii}, R., {Itoh}, Y., {Kudo}, T., {et~al.} 2012, \pasj, 64, 124

\bibitem[{{Thalmann} {et~al.}(2010){Thalmann}, {Grady}, {Goto}, {Wisniewski},
  {Janson}, {Henning}, {Fukagawa}, {Honda}, {Mulders}, {Min},
  {Moro-Mart{\'{\i}}n}, {McElwain}, {Hodapp}, {Carson}, {Abe}, {Brandner},
  {Egner}, {Feldt}, {Fukue}, {Golota}, {Guyon}, {Hashimoto}, {Hayano},
  {Hayashi}, {Hayashi}, {Ishii}, {Kandori}, {Knapp}, {Kudo}, {Kusakabe},
  {Kuzuhara}, {Matsuo}, {Miyama}, {Morino}, {Nishimura}, {Pyo}, {Serabyn},
  {Shibai}, {Suto}, {Suzuki}, {Takami}, {Takato}, {Terada}, {Tomono}, {Turner},
  {Watanabe}, {Yamada}, {Takami}, {Usuda}, \& {Tamura}}]{thalmann10}
{Thalmann}, C., {Grady}, C.~A., {Goto}, M., {et~al.} 2010, \apjl, 718, L87

\bibitem[{{Thalmann} {et~al.}(2014){Thalmann}, {Mulders}, {Hodapp}, {Janson},
  {Grady}, {Min}, {de Juan Ovelar}, {Carson}, {Brandt}, {Bonnefoy}, {McElwain},
  {Leisenring}, {Dominik}, {Henning}, \& {Tamura}}]{thalmann14}
{Thalmann}, C., {Mulders}, G.~D., {Hodapp}, K., {et~al.} 2014, \aap, 566, A51

\bibitem[{{Thalmann} {et~al.}(2015){Thalmann}, {Mulders}, {Janson}, {Olofsson},
  {Benisty}, {Avenhaus}, {Quanz}, {Schmid}, {Henning}, {Buenzli}, {M{\'e}nard},
  {Carson}, {Garufi}, {Messina}, {Dominik}, {Leisenring}, {Chauvin}, \&
  {Meyer}}]{thalmann15}
{Thalmann}, C., {Mulders}, G.~D., {Janson}, M., {et~al.} 2015, \apjl, 808, L41

\bibitem[{{Thalmann} {et~al.}(2016){Thalmann}, {Janson}, {Garufi},
  {Boccaletti}, {Quanz}, {Sissa}, {Gratton}, {Salter}, {Benisty}, {Bonnefoy},
  {Chauvin}, {Daemgen}, {Desidera}, {Dominik}, {Engler}, {Feldt}, {Henning},
  {Lagrange}, {Langlois}, {Lannier}, {Le Coroller}, {Ligi}, {M{\'e}nard},
  {Mesa}, {Meyer}, {Mulders}, {Olofsson}, {Pinte}, {Schmid}, {Vigan}, \&
  {Zurlo}}]{thalmann16}
{Thalmann}, C., {Janson}, M., {Garufi}, A., {et~al.} 2016, \apjl, 828, L17

\bibitem[{{The} {et~al.}(1994){The}, {de Winter}, \& {Perez}}]{the94}
{The}, P.~S., {de Winter}, D., \& {Perez}, M.~R. 1994, \aaps, 104, 315

\bibitem[{{Thi} {et~al.}(2013){Thi}, {M{\'e}nard}, {Meeus}, {Carmona},
  {Riviere-Marichalar}, {Augereau}, {Kamp}, {Woitke}, {Pinte},
  {Mendigut{\'{\i}}a}, {Eiroa}, {Montesinos}, {Britain}, \& {Dent}}]{thi13}
{Thi}, W.~F., {M{\'e}nard}, F., {Meeus}, G., {et~al.} 2013, \aap, 557, A111

\bibitem[{{Tobin} {et~al.}(2016){Tobin}, {Kratter}, {Persson}, {Looney},
  {Dunham}, {Segura-Cox}, {Li}, {Chandler}, {Sadavoy}, {Harris}, {Melis}, \&
  {P{\'e}rez}}]{tobin16}
{Tobin}, J.~J., {Kratter}, K.~M., {Persson}, M.~V., {et~al.} 2016, \nat, 538,
  483

\bibitem[{{Tomida} {et~al.}(2017){Tomida}, {Machida}, {Hosokawa}, {Sakurai}, \&
  {Lin}}]{tomida17}
{Tomida}, K., {Machida}, M.~N., {Hosokawa}, T., {Sakurai}, Y., \& {Lin}, C.~H.
  2017, \apjl, 835, L11

\bibitem[{{Toomre}(1964)}]{toomre64}
{Toomre}, A. 1964, \apj, 139, 1217

\bibitem[{{Tsukagoshi} {et~al.}(2014){Tsukagoshi}, {Momose}, {Hashimoto},
  {Kudo}, {Andrews}, {Saito}, {Kitamura}, {Ohashi}, {Wilner}, {Kawabe}, {Abe},
  {Akiyama}, {Brandner}, {Brandt}, {Carson}, {Currie}, {Egner}, {Goto},
  {Grady}, {Guyon}, {Hayano}, {Hayashi}, {Hayashi}, {Henning}, {Hodapp},
  {Ishii}, {Iye}, {Janson}, {Kandori}, {Knapp}, {Kusakabe}, {Kuzuhara}, {Kwon},
  {McElwain}, {Matsuo}, {Mayama}, {Miyama}, {Morino}, {Moro-Mart{\'{\i}}n},
  {Nishimura}, {Pyo}, {Serabyn}, {Suenaga}, {Suto}, {Suzuki}, {Takahashi},
  {Takami}, {Takami}, {Takato}, {Terada}, {Thalmann}, {Tomono}, {Turner},
  {Usuda}, {Watanabe}, {Wisniewski}, {Yamada}, \& {Tamura}}]{tsukagoshi14}
{Tsukagoshi}, T., {Momose}, M., {Hashimoto}, J., {et~al.} 2014, \apj, 783, 90

\bibitem[{{Uyama} {et~al.}(2017){Uyama}, {Hashimoto}, {Kuzuhara}, {Mayama},
  {Akiyama}, {Currie}, {Livingston}, {Kudo}, {Kusakabe}, {Abe}, {Brandner},
  {Brandt}, {Carson}, {Egner}, {Feldt}, {Goto}, {Grady}, {Guyon}, {Hayano},
  {Hayashi}, {Hayashi}, {Henning}, {Hodapp}, {Ishii}, {Iye}, {Janson},
  {Kandori}, {Knapp}, {Kwon}, {Matsuo}, {Mcelwain}, {Miyama}, {Morino},
  {Moro-Martin}, {Nishimura}, {Pyo}, {Serabyn}, {Suenaga}, {Suto}, {Suzuki},
  {Takahashi}, {Takami}, {Takato}, {Terada}, {Thalmann}, {Turner}, {Watanabe},
  {Wisniewski}, {Yamada}, {Takami}, {Usuda}, \& {Tamura}}]{uyama17}
{Uyama}, T., {Hashimoto}, J., {Kuzuhara}, M., {et~al.} 2017, \aj, 153, 106

\bibitem[{{Uyama} {et~al.}(2018){Uyama}, {Hashimoto}, {Muto}, {Akiyama},
  {Dong}, {de Leon}, {Sakon}, {Kudo}, {Kusakabe}, {Kuzuhara}, {Bonnefoy},
  {Abe}, {Brandner}, {Brandt}, {Carson}, {Currie}, {Egner}, {Feldt}, {Fung},
  {Goto}, {Grady}, {Guyon}, {Hayano}, {Hayashi}, {Hayashi}, {Henning},
  {Hodapp}, {Ishii}, {Iye}, {Janson}, {Kandori}, {Knapp}, {Kwon}, {Matsuo},
  {Mayama}, {Mcelwain}, {Miyama}, {Morino}, {Moro-Martin}, {Nishimura}, {Pyo},
  {Serabyn}, {Sitko}, {Suenaga}, {Suto}, {Suzuki}, {Takahashi}, {Takami},
  {Takato}, {Terada}, {Thalmann}, {Turner}, {Watanabe}, {Wisniewski}, {Yamada},
  {Yang}, {Takami}, {Usuda}, \& {Tamura}}]{uyama18}
{Uyama}, T., {Hashimoto}, J., {Muto}, T., {et~al.} 2018, ArXiv e-prints,
  arXiv:1804.05934

\bibitem[{{Valdes} {et~al.}(2004){Valdes}, {Gupta}, {Rose}, {Singh}, \&
  {Bell}}]{valdes04}
{Valdes}, F., {Gupta}, R., {Rose}, J.~A., {Singh}, H.~P., \& {Bell}, D.~J.
  2004, \apjs, 152, 251

\bibitem[{{Valenti} {et~al.}(2003){Valenti}, {Fallon}, \&
  {Johns-Krull}}]{valenti03}
{Valenti}, J.~A., {Fallon}, A.~A., \& {Johns-Krull}, C.~M. 2003, \apjs, 147,
  305

\bibitem[{{van Boekel} {et~al.}(2017){van Boekel}, {Henning}, {Menu}, {de
  Boer}, {Langlois}, {M{\"u}ller}, {Avenhaus}, {Boccaletti}, {Schmid},
  {Thalmann}, {Benisty}, {Dominik}, {Ginski}, {Girard}, {Gisler}, {Lobo Gomes},
  {Menard}, {Min}, {Pavlov}, {Pohl}, {Quanz}, {Rabou}, {Roelfsema}, {Sauvage},
  {Teague}, {Wildi}, \& {Zurlo}}]{vanboekel17}
{van Boekel}, R., {Henning}, T., {Menu}, J., {et~al.} 2017, \apj, 837, 132

\bibitem[{{van den Ancker} {et~al.}(1996){van den Ancker}, {The}, \& {de
  Winter}}]{vandenancker96}
{van den Ancker}, M.~E., {The}, P.~S., \& {de Winter}, D. 1996, \aap, 309, 809

\bibitem[{{van der Marel} {et~al.}(2016){van der Marel}, {Verhaar}, {van
  Terwisga}, {Mer{\'{\i}}n}, {Herczeg}, {Ligterink}, \& {van
  Dishoeck}}]{vandermarel16whole}
{van der Marel}, N., {Verhaar}, B.~W., {van Terwisga}, S., {et~al.} 2016, \aap,
  592, A126

\bibitem[{{van der Marel} {et~al.}(2013){van der Marel}, {van Dishoeck},
  {Bruderer}, {Birnstiel}, {Pinilla}, {Dullemond}, {van Kempen}, {Schmalzl},
  {Brown}, {Herczeg}, {Mathews}, \& {Geers}}]{vandermarel13}
{van der Marel}, N., {van Dishoeck}, E.~F., {Bruderer}, S., {et~al.} 2013,
  Science, 340, 1199

\bibitem[{{Vieira} {et~al.}(2003){Vieira}, {Corradi}, {Alencar}, {Mendes},
  {Torres}, {Quast}, {Guimar{\~a}es}, \& {da Silva}}]{vieira03}
{Vieira}, S.~L.~A., {Corradi}, W.~J.~B., {Alencar}, S.~H.~P., {et~al.} 2003,
  \aj, 126, 2971

\bibitem[{{Vigan} {et~al.}(2017){Vigan}, {Bonavita}, {Biller}, {Forgan},
  {Rice}, {Chauvin}, {Desidera}, {Meunier}, {Delorme}, {Schlieder}, {Bonnefoy},
  {Carson}, {Covino}, {Hagelberg}, {Henning}, {Janson}, {Lagrange}, {Quanz},
  {Zurlo}, {Beuzit}, {Boccaletti}, {Buenzli}, {Feldt}, {Girard}, {Gratton},
  {Kasper}, {Le Coroller}, {Mesa}, {Messina}, {Meyer}, {Montagnier},
  {Mordasini}, {Mouillet}, {Moutou}, {Reggiani}, {Segransan}, \&
  {Thalmann}}]{vigan17}
{Vigan}, A., {Bonavita}, M., {Biller}, B., {et~al.} 2017, \aap, 603, A3

\bibitem[{{Vorobyov} \& {Basu}(2005)}]{vorobyov05}
{Vorobyov}, E.~I., \& {Basu}, S. 2005, \apjl, 633, L137

\bibitem[{{Vorobyov} \& {Basu}(2010)}]{vorobyov10burst}
---. 2010, \apj, 719, 1896

\bibitem[{{Wagner} {et~al.}(2015){Wagner}, {Apai}, {Kasper}, \&
  {Robberto}}]{wagner15hd100453}
{Wagner}, K., {Apai}, D., {Kasper}, M., \& {Robberto}, M. 2015, \apjl, 813, L2

\bibitem[{{Wagner} {et~al.}(2018){Wagner}, {Dong}, {Sheehan}, {Apai}, {Kasper},
  {McClure}, {Morzinski}, {Close}, {Males}, {Hinz}, {Quanz}, \&
  {Fung}}]{wagner18}
{Wagner}, K., {Dong}, R., {Sheehan}, P., {et~al.} 2018, \apj, 854, 130

\bibitem[{{Wahhaj} {et~al.}(2015){Wahhaj}, {Cieza}, {Mawet}, {Yang}, {Canovas},
  {de Boer}, {Casassus}, {M{\'e}nard}, {Schreiber}, {Liu}, {Biller}, {Nielsen},
  \& {Hayward}}]{wahhaj15}
{Wahhaj}, Z., {Cieza}, L.~A., {Mawet}, D., {et~al.} 2015, \aap, 581, A24

\bibitem[{{Walker} \& {Wolstencroft}(1988)}]{walker88}
{Walker}, H.~J., \& {Wolstencroft}, R.~D. 1988, \pasp, 100, 1509

\bibitem[{{Walsh} {et~al.}(2016){Walsh}, {Juh{\'a}sz}, {Meeus}, {Dent}, {Maud},
  {Aikawa}, {Millar}, \& {Nomura}}]{walsh16}
{Walsh}, C., {Juh{\'a}sz}, A., {Meeus}, G., {et~al.} 2016, \apj, 831, 200

\bibitem[{{Walsh} {et~al.}(2014){Walsh}, {Juh{\'a}sz}, {Pinilla}, {Harsono},
  {Mathews}, {Dent}, {Hogerheijde}, {Birnstiel}, {Meeus}, {Nomura}, {Aikawa},
  {Millar}, \& {Sandell}}]{walsh14}
{Walsh}, C., {Juh{\'a}sz}, A., {Pinilla}, P., {et~al.} 2014, \apjl, 791, L6

\bibitem[{{Watson} \& {Stapelfeldt}(2004)}]{watson04}
{Watson}, A.~M., \& {Stapelfeldt}, K.~R. 2004, \apj, 602, 860

\bibitem[{{Weinberger} {et~al.}(1999){Weinberger}, {Becklin}, {Schneider},
  {Smith}, {Lowrance}, {Silverstone}, {Zuckerman}, \& {Terrile}}]{weinberger99}
{Weinberger}, A.~J., {Becklin}, E.~E., {Schneider}, G., {et~al.} 1999, \apjl,
  525, L53

\bibitem[{{Weinberger} {et~al.}(2002){Weinberger}, {Becklin}, {Schneider},
  {Chiang}, {Lowrance}, {Silverstone}, {Zuckerman}, {Hines}, \&
  {Smith}}]{weinberger02}
---. 2002, \apj, 566, 409

\bibitem[{{White} {et~al.}(2016){White}, {Boley}, {Hughes}, {Flaherty}, {Ford},
  {Wilner}, {Corder}, \& {Payne}}]{white16}
{White}, J.~A., {Boley}, A.~C., {Hughes}, A.~M., {et~al.} 2016, \apj, 829, 6

\bibitem[{{Williams} \& {Best}(2014)}]{williams14}
{Williams}, J.~P., \& {Best}, W.~M.~J. 2014, \apj, 788, 59

\bibitem[{{Wolff} {et~al.}(2016){Wolff}, {Perrin}, {Millar-Blanchaer},
  {Nielsen}, {Wang}, {Cardwell}, {Chilcote}, {Dong}, {Draper}, {Duch{\^e}ne},
  {Fitzgerald}, {Goodsell}, {Grady}, {Graham}, {Greenbaum}, {Hartung}, {Hibon},
  {Hines}, {Hung}, {Kalas}, {Macintosh}, {Marchis}, {Marois}, {Pueyo},
  {Rantakyr{\"o}}, {Schneider}, {Sivaramakrishnan}, \& {Wiktorowicz}}]{wolff16}
{Wolff}, S.~G., {Perrin}, M., {Millar-Blanchaer}, M.~A., {et~al.} 2016, \apjl,
  818, L15

\bibitem[{{Yang} {et~al.}(2017){Yang}, {Hashimoto}, {Hayashi}, {Tamura},
  {Mayama}, {Rafikov}, {Akiyama}, {Carson}, {Janson}, {Kwon}, {de Leon}, {Oh},
  {Takami}, {Tang}, {Kudo}, {Kusakabe}, {Abe}, {Brandner}, {Brandt}, {Egner},
  {Feldt}, {Goto}, {Grady}, {Guyon}, {Hayano}, {Hayashi}, {Henning}, {Hodapp},
  {Ishii}, {Iye}, {Kandori}, {Knapp}, {Kuzuhara}, {Matsuo}, {Mcelwain},
  {Miyama}, {Morino}, {Moro-martin}, {Nishimura}, {Pyo}, {Serabyn}, {Suenaga},
  {Suto}, {Suzuki}, {Takahashi}, {Takato}, {Terada}, {Thalmann}, {Turner},
  {Watanabe}, {Wisniewski}, {Yamada}, {Takami}, \& {Usuda}}]{yang17}
{Yang}, Y., {Hashimoto}, J., {Hayashi}, S.~S., {et~al.} 2017, \aj, 153, 7

\bibitem[{{Zacharias} {et~al.}(2012){Zacharias}, {Finch}, {Girard}, {Henden},
  {Bartlett}, {Monet}, \& {Zacharias}}]{zacharias12}
{Zacharias}, N., {Finch}, C.~T., {Girard}, T.~M., {et~al.} 2012, VizieR Online
  Data Catalog, 1322

\bibitem[{{Zhang} {et~al.}(2014){Zhang}, {Isella}, {Carpenter}, \&
  {Blake}}]{zhang14}
{Zhang}, K., {Isella}, A., {Carpenter}, J.~M., \& {Blake}, G.~A. 2014, \apj,
  791, 42

\bibitem[{{Zhu} {et~al.}(2015){Zhu}, {Dong}, {Stone}, \&
  {Rafikov}}]{zhu15densitywaves}
{Zhu}, Z., {Dong}, R., {Stone}, J.~M., \& {Rafikov}, R.~R. 2015, \apj, 813, 88

\bibitem[{{Zhu} {et~al.}(2012){Zhu}, {Nelson}, {Dong}, {Espaillat}, \&
  {Hartmann}}]{zhu12}
{Zhu}, Z., {Nelson}, R.~P., {Dong}, R., {Espaillat}, C., \& {Hartmann}, L.
  2012, \apj, 755, 6

\end{thebibliography}
\end{document}